# Angular Momentum of Electromagnetic Radiation

Fundamental physics applied to the radio domain for innovative studies of space and and development of new concepts in wireless communications

Johan Sjöholm
Kristoffer Palmer

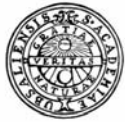

## UPPSALA UNIVERSITET

# Abstract

# Angular Momentum of Electromagnetic Radiation

*Johan Sjöholm and Kristoffer Palmer*

**Teknisk- naturvetenskaplig fakultet**
**UTH-enheten**

Besöksadress:
Ångströmlaboratoriet
Lägerhyddsvägen 1
Hus 4, Plan 0

Postadress:
Box 536
751 21 Uppsala

Telefon:
018 – 471 30 03

Telefax:
018 – 471 30 00

Hemsida:
http://www.teknat.uu.se/student

In this diploma thesis we study the characteristics of electromagnetic fields carrying orbital angular momentum (OAM) by analyzing and utilizing results achieved in optics and then apply them to the radio domain to enable innovative radio studies of space and the development of new concepts in wireless communications.

With the recent advent of fast digital converters it has become possible, over a wide radio frequency range, to manipulate not only the modulation properties of any given signal carried by a radio beam, but also the physical field vectors which make up the radio beam itself. Drawing inferences from results obtained in optics and quantum communication research during the past 10–15 years, we extract the core information about fields carrying orbital angular momentum. We show that with this information it is possible to design an array of antennas which, together with digital receivers/transmitters, can readily produce, under full software control, a radio beam that carries electromagnetic orbital angular momentum, a classical electrodynamics quantity known for a century but so far preciously little utilized in radio, if at all. This electromagnetic field is then optimized with the help of various antenna array techniques to improve the radio vector field qualities. By explicit numerical solution of the Maxwell equations from first principles, using a de facto industrial standard antenna software package, we show that the field indeed carries orbital angular momentum, and give a hint on how to detect and measure orbital angular momentum in radio beams. Finally, we discuss and give an explanation of what this can be used for and what the future might bring in this area.

# ANGULAR MOMENTUM OF ELECTROMAGNETIC RADIATION

## Fundamental physics applied to the radio domain for innovative studies of space and development of new concepts in wireless communications

JOHAN SJÖHOLM and KRISTOFFER PALMER

Uppsala School of Engineering

and

Department of Astronomy and Space Physics, Uppsala University, Sweden

# CONTENTS

# LIST OF FIGURES

# 1

## INTRODUCTION

In today's society, electromagnetic (EM) fields are used in increasingly many different contexts, ranging from fundamental research and development to communications and household appliances. However, there are still properties of the classical EM field, well known already to the pioneers of electromagnetism a century ago [43], that are not yet fully utilized, either because of actuator/sensor/detector limitations or because of lack of familiarity with the more subtle aspects of the electromagnetic field. One of the underutilized properties is the complete, instantaneous field vector (magnetic or electric) itself, meaning that all three spatial dimensions of the field vector are taken fully into account so that one can make use of the information embedded in both the magnitude and the direction of the field vector in question. Because of the technical complexities involved, sensing of electromagnetic fields, which are three-dimensional vectorial, have to this day typically been made with sensors capable of capturing only one or two of the three spatial components of the field vectors (or projections of them), *e.g.*, radio antennas erected in one or two spatial dimensions, effectively resulting in a waste of available EM information. One obvious example being the inadequacy to sense, simultaneously, the transverse part as well as the *z* component of the EM field in a beam which has a non-planar phase front.

Simple one-dimensional sensing antennas are what is typically used for picking up radio and TV broadcasts on our domestic radio and TV sets but are also used in more demanding situations. Two-dimensional sensing of the 3D vector field (*e.g.*, crossed dipole antennas) is used in many modern radio telescopes, including the LOFAR (Low Frequency Array) distributed radio telescope currently under construction in the Netherlands, Germany and France.[1] Conventional sensing of the radio field in two dimensions is also going to be used for the next generation antennas of the planned "3D" European Incoherent Scatter (EISCAT) ionospheric radar facility;[2] here "3D" does not refer to the way the EM fields of the radar are sensed but how the radar signals, once they have been sensed with a conventional 2D information wasting technique, will be used in attempts to estimate the 3D plasma dynamics in the ionosphere. The Scandinavian supplement to LOFAR, LOIS (LOFAR Outrigger in Scandinavia),[3] is the first space physics radio/radar or radio astronomy facility that utilizes the entire 3D vector information embedded in the EM fields of the radio signals. Future big Earth-based radio astronomy multi-antenna telescopes such as the Square Kilometre Array (SKA),[4] and the Long Wavelength Array (LWA),[5] are expected to benefit significantly from using the LOIS-type vector-sensing radio technology. This is even more true for space-based radio infrastructures such as the proposed Lunar Infrastructure for Exploration (LIFE) project, which aims at building a multi-antenna radio telescope on the far side of the moon [4, 37].

One particularly interesting property, used in modern physics, *e.g.*, in experiments on trapping and manipulating of atoms, molecules, and microscopic particles by the help of laser fields [16], is the electromagnetic orbital angular momentum, OAM [2, 6, 15, 23, 24, 31, 44, 45]. However, while OAM has been used to efficiently encode information for free-space communications in the optics frequency range [22], OAM has so far not been used to its full extent—if at all—in the radio domain, except for some proof-of-concept experiments in the microwave range. As will be described in this thesis, the advent of fast analog-to-digital and digital-to-analog converters has made it possible to construct combined 3D sensing antenna and tri-channel digital receiver systems which can measure coherently the instantaneous 3D field vector, a first-order quantity, of an EM signal in the radio domain. This new possibility enables the processing of EM field vectors, including OAM encoding and decoding of radio beams, with high precision and speed under full software control. This is in contrast to optics where detectors are still incoherent, capable of measuring second (and sometimes higher) order field quantities only and not the field vectors themselves. In order to cover new unexplored ground and find additional uses of electromagnetic fields, these full vectorial properties of the EM field, and their *physical* encoding, have to be explored. We think that therein lies the future for a more efficient use of electromagnetic fields and improved radio methods for research and communications [8, 20]. This is what this thesis is about.

---

[1] www.lofar.org
[2] www.eiscat.org
[3] www.lois-space.net
[4] www.telescope.org
[5] lwa.nrl.navy.mil

In optics, the properties of laser beams have been studied for a long time. Laser beams can contain several different types of laser modes, the Hermite-Gaussian modes are the most common. Two other modes are the Laguerre-Gaussian modes and the Bessel-Gaussian modes, but these are much less common than the different Hermite-Gaussian modes. Laguerre-Gaussian laser beams can be obtained by conversion of Hermite-Gaussian laser beams, this is done by a series of optical devices [48]. A Laguerre-Gaussian beam carries orbital angular momentum [3]. In theory the orbital angular momentum can have an infinite number of distinct states. The number of distinct states achievable in practice are however limited by physical issues such as the sensitivity of the devices and the spreading of the beam. If we combine the orbital angular momentum with polarization, the amount of distinct states can, within certain limits, be doubled.

Starting from the mathematical results obtained for the idealized par axial model used in optics, we translate these results into the radio domain by estimating the equivalent (complex valued) currents needed to generate OAM with the help of radio antenna arrays. In order to test our approach, we have used these current/antenna array setups to solve the Maxwell equations from first principles, using the *de facto* industrial standard antenna software package Numeric Electromagnetic Code [40]. Analyzing the radiation field vector characteristics obtained in these accurate numerical simulations we find that the fields do indeed carry orbital angular momentum (OAM). By choosing antenna arrays that can produce different OAM states and combining them we have been able to reproduce, in the radio domain, field characteristics which are very similar to those obtained in optics, thus proving the feasibility of the LOIS radio technique.

# 2

## ELECTROMAGNETIC FIELDS AND CONSERVATION LAWS

In this chapter the basic properties of electromagnetic fields will be presented. First, the conserved quantities of motion for an electromagnetic field in vacuum will be derived in the standard way from Maxwell-Lorentz equations. It is well known that conservation laws are directly related to symmetries in the dynamic equations of a physical system [29, 41] which allows for a formal, abstract approach. However, we shall derive the conservation laws by using the explicit, straightforward calculations found in most standard textbooks [15, 31]. Some of the conserved quantities derived are vectorial in nature. Hence, they effectively contain three quantities. Secondly properties of polarized electromagnetic fields will be explained.

## 2.1 Maxwell-Lorentz equations

The equations of classical electrodynamics, on a microscopic scale, are the so called Maxwell-Lorentz equations

$$\boldsymbol{\nabla} \cdot \mathbf{E} = \frac{\rho(\mathbf{x}, t)}{\epsilon_0} \tag{2.1}$$

$$\boldsymbol{\nabla} \cdot \mathbf{B} = 0 \tag{2.2}$$

$$\boldsymbol{\nabla} \times \mathbf{E} = -\frac{\partial \mathbf{B}}{\partial t} \tag{2.3}$$

$$\boldsymbol{\nabla} \times \mathbf{B} = \mu_0 \mathbf{j}(\mathbf{x}, t) + \frac{1}{c^2} \frac{\partial \mathbf{E}}{\partial t} \tag{2.4}$$

where $\epsilon_0$ is the permittivity of free space and $\mu_0$ is the permeability of vacuum, or the magnetic constant.

The Maxwell-Lorentz equations were formulated based on empirical results [31, 49]. First one derives the static electric and magnetic field from empirical force equations, Coulomb's law and Ampère's law. This results in two time-independent, uncoupled systems of equations. By introducing a time-dependent relationship for conservation of electric charge and the electric charges relation to the currents in the form of the continuity equation

$$\frac{\partial \rho(\mathbf{x}, t)}{\partial t} + \boldsymbol{\nabla} \cdot \mathbf{j}(\mathbf{x}, t) = 0 \tag{2.5}$$

one can obtain the Maxwell-Lorentz equations [49]. The Maxwell-Lorentz equations give a classical correct picture, for both macroscopic and microscopic scales. These equations are also known as Maxwell's microscopic equations.

## 2.2 Energy

The power gain of a charged particle moving with velocity $\mathbf{v}$ in an electromagnetic field $(\mathbf{E}, \mathbf{B})$, which gives rise to a Lorentz force $\mathbf{F}$ acting on the particle, is

$$\mathbf{F} \cdot \mathbf{v} = q(\mathbf{E} + \mathbf{v} \times \mathbf{B}) \cdot \mathbf{v} = q\mathbf{v} \cdot \mathbf{E} \tag{2.6}$$

If we can represent the total charge density as a continuous distribution of charges and currents within a certain volume we can write

$$\mathbf{j} = \rho \mathbf{v} \tag{2.7}$$

$$q = \int_{V'} \mathrm{d}^3 x' \, \rho(\mathbf{x}, t) \tag{2.8}$$

where $\rho$ is the charge distribution. Then equation (2.6) can then be rewritten as

$$\mathbf{F} \cdot \mathbf{v} = \int_{V'} \mathrm{d}^3 x' \, \mathbf{j} \cdot \mathbf{E} \tag{2.9}$$

This is the total work done by the field on the moving charges. It represents a conversion of electromagnetic energy into mechanical or thermal energy. As is well known, the total energy in a closed system is conserved. Therefore a change in the mechanical energy of the system has to be balanced with the corresponding rate of change in the electromagnetic field energy within the volume. Using Maxwell's equations we can rewrite (2.9) as

$$\mathbf{F} \cdot \mathbf{v} = \int_{V'} \mathrm{d}^3 x' \, \epsilon_0 \left( c^2 \mathbf{E} \cdot (\boldsymbol{\nabla} \times \mathbf{B}) - \mathbf{E} \cdot \frac{\partial \mathbf{E}}{\partial t} \right) \tag{2.10}$$

By using the vector identity

$$\frac{1}{\mu_0} \mathbf{E} \cdot (\boldsymbol{\nabla} \times \mathbf{B}) = \frac{1}{\mu_0} \mathbf{B} \cdot (\boldsymbol{\nabla} \times \mathbf{E}) - \frac{1}{\mu_0} \boldsymbol{\nabla} \cdot (\mathbf{E} \times \mathbf{B}) = -\frac{1}{\mu_0} \mathbf{B} \cdot \frac{\partial \mathbf{B}}{\partial t} - \frac{1}{\mu_0} \boldsymbol{\nabla} \cdot (\mathbf{E} \times \mathbf{B}) \tag{2.11}$$

in equation (2.10), and Gauss's theorem and then rearranging the terms, we get Poynting's theorem

$$-\int_{V'} \mathrm{d}^3 x' \, \frac{\epsilon_0}{2} \left( c^2 \mathbf{B} \cdot \frac{\partial \mathbf{B}}{\partial t} + \mathbf{E} \cdot \frac{\partial \mathbf{E}}{\partial t} \right) = \int_{V'} \mathrm{d}^3 x' \, \mathbf{j} \cdot \mathbf{E} + \oint_{S'} \mathrm{d}^2 x' \, \frac{1}{\mu_0} (\mathbf{E} \times \mathbf{B}) \cdot \hat{\boldsymbol{n}} \tag{2.12}$$

In order for us to advance any further we need to make two assumptions:
(1) The macroscopic medium is linear in its electric and magnetic properties, with negligible dispersion or losses, and (2) the total electromagnetic power for the field is represented by

$$U_{\mathrm{Field}} = \int_{V'} \mathrm{d}^3 x' \, \frac{\epsilon_0}{2} \left( \mathbf{E} \cdot \mathbf{E} + c^2 \, \mathbf{B} \cdot \mathbf{B} \right) = \int_{V'} \mathrm{d}^3 x' \, \frac{\epsilon_0}{2} \left( E^2 + c^2 B^2 \right) \tag{2.13}$$

This gives us a new version of Poynting's theorem

$$-\frac{\partial}{\partial t} \int_{V'} \mathrm{d}^3 x' \, \frac{\epsilon_0}{2} \left( c^2 B^2 + E^2 \right) = \int_{V'} \mathrm{d}^3 x' \, \mathbf{j} \cdot \mathbf{E} + \oint_{S'} \mathrm{d}^2 x' \, \frac{1}{\mu_0} (\mathbf{E} \times \mathbf{B}) \cdot \hat{\boldsymbol{n}} \qquad (2.14)$$

If we denote the electromagnetic field energy density

$$u = \frac{\epsilon_0}{2} \left( E^2 + c^2 B^2 \right) \qquad (2.15)$$

and define the Poynting vector $\mathbf{S}$ as follows

$$\mathbf{S} = \frac{1}{\mu_0} \mathbf{E} \times \mathbf{B} \qquad (2.16)$$

Poynting's theorem, equation (2.14), can be rewritten as a differential continuity equation

$$\frac{\partial u}{\partial t} + \boldsymbol{\nabla} \cdot \mathbf{S} = -\mathbf{j} \cdot \mathbf{E} \qquad (2.17)$$

Assuming that no particle leaves the volume, the total work done by the fields on the sources can be written

$$\frac{\partial U_{\text{mech}}}{\partial t} = \int_{V'} \mathrm{d}^3 x' \, \mathbf{j} \cdot \mathbf{E} \qquad (2.18)$$

Inserting this into equation (2.14), we obtain

$$\frac{\partial U_{\text{tot}}}{\partial t} = \frac{\partial}{\partial t} \left[ U_{\text{mech}} + \underbrace{\int_{V'} \mathrm{d}^3 x' \, \frac{\epsilon_0}{2} \left( c^2 B^2 + E^2 \right)}_{U_{\text{Field}}} \right] = \oint_{S'} \mathrm{d}^2 x' \, \frac{1}{\mu_0} (\mathbf{E} \times \mathbf{B}) \cdot \hat{\boldsymbol{n}} \quad (2.19)$$

If we are looking at a closed system this equation simplifies to

$$U_{\text{tot}} = \text{constant} = U_{\text{mech}} + \int_{V'} \mathrm{d}^3 x' \, \frac{\epsilon_0}{2} \left( c^2 B^2 + E^2 \right) \qquad (2.20)$$

This is the law of conservation of energy alluded to earlier in this section. It clearly shows the balance between the mechanical and electromagnetic energy and is the consequence of the invariance of the dynamic equations describing the system (equations of motion, Maxwell-Lorentz equations) with respect to changes in the time origin (homogeneity in time) [15].

### 2.2.1 Ohm's law

A current density, $\mathbf{j}$, can be produced in a conducting medium by applying an electric field.[1] If the current density is analytic it can be expressed in a Taylor series. By making the simplifying assumption that the current density in the conducting medium is proportional to the electric field impressed upon the medium, *i.e.*, can be approximated by the first term in this Taylor series:

$$\mathbf{j} = \sigma(\mathbf{E} + \mathbf{E}^{\text{EMF}})$$

This linear approximation of the relation between the current density and the electric field strength is known as Ohm's law. Inserting this into Poynting's theorem, equation (2.14), and rewriting we get

$$\underbrace{\int_{V'} \mathrm{d}^3 x' \, \mathbf{j} \cdot \mathbf{E}^{\text{EMF}}}_{\text{Applied electric power}} = \underbrace{\int_{V'} \mathrm{d}^3 x' \, \frac{j^2}{\sigma}}_{\text{Joule heat}} + \underbrace{\frac{\partial}{\partial t} \int_{V'} \mathrm{d}^3 x' \, \frac{\epsilon_0}{2} \left( c^2 B^2 + E^2 \right)}_{\text{Field energy}} + \underbrace{\oint_{S'} \mathrm{d}^2 x' \, \frac{1}{\mu_o} (\mathbf{E} \times \mathbf{B}) \cdot \hat{\boldsymbol{n}}}_{\text{Radiated power}}$$

$$(2.21)$$

## 2.3 Center of energy

In analogy with the definition for the center of mass for a mechanical system, one can define the center of energy $\mathbf{C}$ for EM field [15] by the formula [12]

$$\mathbf{C} = \mathbf{X} \int_{V'} \mathrm{d}^3 x' \, u = \int_{V'} \mathrm{d}^3 x' \, (\mathbf{x} - \mathbf{x}_0) u \tag{2.22}$$

where $\mathbf{X}$ is the coordinate vector for *the center of mass*, $u$ is the EM energy density, and $\mathbf{x}$ is the position vector. From equation (2.10) we see that the mechanical power is balanced by the power originating from the field. This means that we can integrate with respect to time and thereby obtain the total energy in the EM field. Combining this with the definition for *the center of energy*, equation (2.22), we deduce that the center of energy for the field should equal

$$-\int_{t_0}^{t} \mathrm{d}t \int_{V'} \mathrm{d}^3 x' \, (\mathbf{x} - \mathbf{x}_0) (\mathbf{j} \cdot \mathbf{E}) \tag{2.23}$$

---

[1]Electric currents may also be produced by mechanical transport of electric charges due to, *e.g.*, winds.

By integrating $(\mathbf{x} - \mathbf{x}_0)(\mathbf{j} \cdot \mathbf{E})$ and using Maxwell's equations we obtain the following expression

$$-\int_{V'} \mathrm{d}^3x' \, (\mathbf{x} - \mathbf{x}_0)(\mathbf{j} \cdot \mathbf{E}) = \int_{V'} \mathrm{d}^3x' \, \epsilon_0 (\mathbf{x} - \mathbf{x}_0) \left( c^2 \, \mathbf{E} \cdot (\boldsymbol{\nabla} \times \mathbf{B}) - \mathbf{E} \cdot \frac{\partial \mathbf{E}}{\partial t} \right) \quad (2.24)$$

Using the vector identity in equation (2.11) and Maxwell's equations, we rewrite this as

$$-\int_{V'} \mathrm{d}^3x' \, (\mathbf{x} - \mathbf{x}_0)(\mathbf{j} \cdot \mathbf{E})$$
$$= \int_{V'} \mathrm{d}^3x' \, \epsilon_0 \frac{(\mathbf{x} - \mathbf{x}_0)}{2} \frac{\partial}{\partial t} \left( c^2 B^2 + E^2 \right) + \int_{V'} \mathrm{d}^3x' \, (\mathbf{x} - \mathbf{x}_0) \frac{1}{\mu_0} \left[ \boldsymbol{\nabla} \cdot (\mathbf{E} \times \mathbf{B}) \right]$$
$$(2.25)$$

Integrating this with respect to time we obtain the equation for *the center of energy* for the field

$$\mathbf{C} = \mathbf{X} \int_{V'} \mathrm{d}^3x' \, u = \int_{V'} \mathrm{d}^3x' \, \left[ \epsilon_0 (\mathbf{x} - \mathbf{x}_0) \left( c^2 B^2 + E^2 \right) - (t - t_0)(\mathbf{E} \times \mathbf{B}) \right] \quad (2.26)$$

which is a manifestation of the equivalence of space and time.

## 2.4 Linear momentum

We start with the expression for the *Lorentz force density*, $\rho \mathbf{E} + \mathbf{j} \times \mathbf{B}$, use the Maxwell-Lorentz equations to rewrite this expression and then symmetrize it

$$\rho \mathbf{E} + \mathbf{j} \times \mathbf{B} = \epsilon_0 (\boldsymbol{\nabla} \cdot \mathbf{E}) \mathbf{E} + \epsilon_0 \left( c^2 \, \boldsymbol{\nabla} \times \mathbf{B} - \frac{\partial \mathbf{E}}{\partial t} \right) \times \mathbf{B} =$$
$$= \epsilon_0 (\boldsymbol{\nabla} \cdot \mathbf{E}) \mathbf{E} + \frac{1}{\mu_0} (\boldsymbol{\nabla} \times \mathbf{B}) \times \mathbf{B} - \epsilon_0 \frac{\partial \mathbf{E}}{\partial t} \times \mathbf{B} =$$
$$= \epsilon_0 (\boldsymbol{\nabla} \cdot \mathbf{E}) \mathbf{E} - \frac{1}{\mu_0} \mathbf{B} \times (\boldsymbol{\nabla} \times \mathbf{B}) - \epsilon_0 \frac{\partial}{\partial t} (\mathbf{E} \times \mathbf{B}) + \epsilon_0 \mathbf{E} \times \frac{\partial \mathbf{B}}{\partial t}$$
$$= \epsilon_0 \left[ \mathbf{E} (\boldsymbol{\nabla} \cdot \mathbf{E}) - \mathbf{E} \times (\boldsymbol{\nabla} \times \mathbf{E}) \right] + \frac{1}{\mu_0} \left[ \mathbf{B} \underbrace{(\boldsymbol{\nabla} \cdot \mathbf{B})}_{=0} - \mathbf{B} \times (\boldsymbol{\nabla} \times \mathbf{B}) \right] - \epsilon_0 \frac{\partial}{\partial t} (\mathbf{E} \times \mathbf{B})$$
$$(2.27)$$

The square brackets vector components can be expressed as follows

$$\left[ \mathbf{E}(\boldsymbol{\nabla} \cdot \mathbf{E}) - \mathbf{E} \times (\boldsymbol{\nabla} \times \mathbf{E}) \right]_i = \frac{1}{2}\left( \mathbf{E} \cdot \frac{\partial \mathbf{E}}{\partial x_i} - \mathbf{E} \cdot \frac{\partial \mathbf{E}}{\partial x_i} \right) + \frac{\partial}{\partial x_j}\left( E_i E_j - \frac{1}{2}\mathbf{E} \cdot \mathbf{E}\, \delta_{ij} \right) \tag{2.28}$$

and

$$\left[ \mathbf{B}(\boldsymbol{\nabla} \cdot \mathbf{B}) - \mathbf{B} \times (\boldsymbol{\nabla} \times \mathbf{B}) \right]_i = \frac{1}{2}\left( \mathbf{B} \cdot \frac{\partial \mathbf{B}}{\partial x_i} - \mathbf{B} \cdot \frac{\partial \mathbf{B}}{\partial x_i} \right) + \frac{\partial}{\partial x_j}\left( B_i B_j - \frac{1}{2}\mathbf{B} \cdot \mathbf{B}\, \delta_{ij} \right) \tag{2.29}$$

Now we use these two equations together with equation (2.27) and obtain

$$\underbrace{(\rho \mathbf{E} + \mathbf{j} \times \mathbf{B})_i - \frac{1}{2}\left[ \epsilon_0 \mathbf{E} \cdot \frac{\partial \mathbf{E}}{\partial x_i} - \epsilon_0 \mathbf{E} \cdot \frac{\partial \mathbf{E}}{\partial x_i} + \frac{1}{\mu_0}\mathbf{B} \cdot \frac{\partial \mathbf{B}}{\partial x_i} - \frac{1}{\mu_0}\mathbf{B} \cdot \frac{\partial \mathbf{B}}{\partial x_i} \right]}_{(\mathbf{F}_{ev})_i}$$

$$+ \epsilon_0 \frac{\partial}{\partial t}(\mathbf{E} \times \mathbf{B}) = \frac{\partial}{\partial x_j}\underbrace{\left( \epsilon_0 E_i E_j - \frac{\epsilon_0}{2}\mathbf{E} \cdot \mathbf{E}\, \delta_{ij} + \frac{1}{\mu_0}B_i B_j - \frac{1}{\mu_0}\frac{1}{2}\mathbf{B} \cdot \mathbf{B}\, \delta_{ij} \right)}_{T_{ij}} \tag{2.30}$$

Where $T_{ij}$ stands for the $ij$th component of Maxwell's stress tensor, $\mathbf{T}$ and $(\mathbf{F}_{ev})_i$ stand for the $i$th component of *the electric volume force*, $\mathbf{F}_{ev}$. This can now be expressed as *the force equation*

$$\left[ \mathbf{F}_{ev} + \epsilon_0 \frac{\partial}{\partial t}(\mathbf{E} \times \mathbf{B}) \right]_i = \frac{\partial T_{ij}}{\partial x_j} = (\boldsymbol{\nabla} \cdot \mathbf{T})_i \tag{2.31}$$

Rewriting this equation gives us

$$\frac{\partial T_{ij}}{\partial x_i} = \left( \mathbf{F}_{ev} + \frac{1}{c^2}\frac{\partial \mathbf{S}}{\partial t} \right)_i \tag{2.32}$$

Where $\mathbf{S}$ is the Poynting vector defined in equation (2.16). If this equation is integrated over the entire volume we retrieve

$$\int_{V'} \mathrm{d}^3 x'\, \mathbf{F}_{ev} + \frac{1}{c^2}\frac{\mathrm{d}}{\mathrm{d}t}\int_{V'} \mathrm{d}^3 x'\, \mathbf{S} = \oint_{S'} \mathrm{d}^2 x'\, \mathbf{T}_{\hat{n}} \tag{2.33}$$

This equation is called *the momentum theorem* in Maxwell's theory for fields in vacuum. By using another notation we can write this as

$$\frac{\mathrm{d}}{\mathrm{d}t}\mathbf{p}^{\text{mech}} + \frac{\mathrm{d}}{\mathrm{d}t}\mathbf{p}^{\text{field}} = \oint_{S'} \mathrm{d}^2x' \; \mathbf{T}_{\hat{n}} \tag{2.34}$$

This means that $\mathbf{p}^{\text{field}}$ can be expressed as

$$\mathbf{p}^{\text{field}} = \epsilon_0 \int_{V'} \mathrm{d}^3x' \; \mathbf{E} \times \mathbf{B} \tag{2.35}$$

for the case that we are dealing with electromagnetic fields, *e.g.*, propagating radio waves, in vacuum.

The linear momentum conservation law (2.34) is a result of the invariance of the dynamic equations with respect to changes in the coordinate origin (homogeneity in space) [15].

## 2.5 Angular momentum

In analogy with classical mechanics the electromagnetic angular momentum is defined as $\mathbf{J} = (\mathbf{x} - \mathbf{x}_0) \times \mathbf{p}$. From equation (2.35) we get the linear momentum density. By using the definition for angular momentum we derive the equation for the angular momentum density, $\mathbf{M}$, as [31]

$$\mathbf{M} = \epsilon_0 (\mathbf{x} - \mathbf{x}_0) \times (\mathbf{E} \times \mathbf{B}). \tag{2.36}$$

To retrieve the total angular momentum, $\mathbf{J}$, we have to integrate over the entire volume.

$$\mathbf{J}^{\text{field}} = \epsilon_0 \int_{V'} \mathrm{d}^3x' \; (\mathbf{x} - \mathbf{x}_0) \times (\mathbf{E} \times \mathbf{B}) \tag{2.37}$$

In the case of transverse plane waves the orbital angular momentum will be zero since the term $\mathbf{E} \times \mathbf{B}$ will have no components other than those along the axis of propagation. If $\mathbf{E} \times \mathbf{B}$ will have small non-zero components perpendicular to the axis of propagation, it could yield a net total orbital angular momentum.

Both in classical mechanics and atomic physics we know that the angular momentum can be separated into two different parts. In resemblance to Earth which rotates about its axis but also orbits the Sun and electrons can have both a orbital angular momentum and spin. It is therefore reasonable to assume that in many cases the angular momentum carried by the electromagnetic field radiation can be separated into two different parts so that

$$\mathbf{J} = \mathbf{L} + \mathbf{S}. \tag{2.38}$$

where $\mathbf{L}$ is the orbital angular momentum and $\mathbf{S}$ the spin angular momentum. According to *Jackson* [31], expressing the magnetic field in terms of the vector potential the total angular momentum can be written as

$$\mathbf{J}^{\text{field}} = \frac{1}{\mu_0 c^2} \int_V \mathrm{d}^3 x \left[ \mathbf{E} \times \mathbf{A} + \sum_{j=1}^{3} E_j (\mathbf{x} \times \boldsymbol{\nabla}) A_j \right] \tag{2.39}$$

The first term is identified as the spin angular momentum. The second term is identified as the orbital angular momentum since it contains the angular momentum operator, $\mathbf{L}^{\text{op}} = -\mathrm{i}(\mathbf{x} \times \boldsymbol{\nabla})$, well known from quantum mechanics.

Rotational invariance of the the dynamic equations (isotropy in space) manifests itself in the conservation law

$$\mathbf{J}^{\text{tot}} = \mathbf{J}^{\text{mech}} + \mathbf{J}^{\text{field}} \tag{2.40}$$

where $\mathbf{J}^{\text{mech}} = (\mathbf{x} - \mathbf{x}_0) \times \mathbf{p}^{\text{mech}}$ is the mechanical angular momentum.

## 2.6 Polarization

Wave polarization is defined as [5] "that property of an electromagnetic wave describing the time-varying direction and relative magnitude of the electric-field vector; especially, the figure traced as a function of time by the extremity of the vector at a fixed location in space, and the sense in which it is traced, as observed along the direction of propagation." So by taking the curve given by the end point of the instantaneous electric field vector we get the polarization curve. If a wave is traveling in the $\hat{\mathbf{r}}$ direction, the polarization in the $\varphi\theta$ plane is generally an ellipse. In the special case when the length of the minor axis is equal to that of the major axis the curve is a circle and we have circular polarization. When the length of the minor axis is zero, the polarization is said to be linear. For all other cases we say that the polarization is elliptic. The difference between circular and elliptic polarization is that when having circular polarization, the magnitude of the electric-field vector is constant as the electric-field vector rotates around the propagation axis, and when having elliptic polarization, it varies between the length of the minor axis and that of the major axis. The linearly polarized wave does not rotate around its axis at all, and the magnitude of the electric-field vector varies between zero and the length of the major axis.

Mathematically we use the *axial ratio*, *AR*, and *tilt angle*, $\tau$, to describe the polarization. The ratio *AR* is the length of the minor axis divided by the length of the major axis, and it attains values between 0 (linearly polarized) and 1 (circularly polarized). The tilt angle is the smallest angle between the $\theta$ axis and the major axis. This means that it falls in the interval between 0 and $\pi/2$. In general, polarized radiation is either left-hand or right-hand polarized. If **E** moves in a clockwise direction as the wave propagates the polarization is right-handed, if it moves in a counter clockwise direction the polarization is left-handed. Consider the field

$$\mathbf{E}^{\mathrm{rad}}(t, \mathbf{x}) = E_\theta(r, t)\hat{\boldsymbol{\theta}} + E_\varphi(r, t)\hat{\boldsymbol{\varphi}} \tag{2.41}$$

where

$$E_\theta(r, t) = E_{\theta 0}\cos(\omega t - kr + \phi_\theta) \tag{2.42}$$

$$E_\varphi(r, t) = E_{\varphi 0}\cos(\omega t - kr + \phi_\varphi) \tag{2.43}$$

Now, let $\Delta\phi = \phi_\theta - \phi_\varphi$. If

$$\Delta\phi = n\pi, \quad n = 0, 1, 2, 3, ... \tag{2.44}$$

the wave is linearly polarized. If instead

$$\Delta\phi = \pm\left(\frac{1}{2} + 2n\right)\pi, \quad n = 0, 1, 2, 3... \tag{2.45}$$

or

$$\Delta\phi = \pm M \neq \frac{\pi n}{2}, \quad M \in \mathbb{R} \quad n = 0, 1, 2, 3, ... \tag{2.46}$$

the wave is elliptically polarized. The plus sign corresponds to left-hand polarized waves and the minus sign to right-hand polarization. If, in the case of elliptic polarization in equation (2.45) only, $E_{\theta 0} = E_{\varphi 0}$, we have circular polarization.

The parameters $AR$ and $\tau$ can be calculated from

$$AR = \frac{\left[ E_{\varphi 0}^2 + E_{\theta 0}^2 - \left( E_{\varphi 0}^4 + E_{\theta 0}^4 + 2E_{\varphi 0}^2 E_{\theta 0}^2 \cos(2\Delta\phi) \right)^{\frac{1}{2}} \right]^{\frac{1}{2}}}{\left[ E_{\varphi 0}^2 + E_{\theta 0}^2 + \left( E_{\varphi 0}^4 + E_{\theta 0}^4 + 2E_{\varphi 0}^2 E_{\theta 0}^2 \cos(2\Delta\phi) \right)^{\frac{1}{2}} \right]^{\frac{1}{2}}} \qquad (2.47)$$

$$\tau = \frac{\pi}{2} - \frac{1}{2}\tan^{-1}\left[ \frac{2E_{\varphi 0}E_{\theta 0}}{E_{\varphi 0}^2 - E_{\theta 0}^2}\cos(\Delta\phi) \right] \qquad (2.48)$$

If $\tau$ is a multiple of $\frac{\pi}{2}$ the equation for $AR$ simplifies to

$$AR = \frac{E_{\varphi 0}}{E_{\theta 0}} \quad \text{or} \quad AR = \frac{E_{\theta 0}}{E_{\varphi 0}} \qquad (2.49)$$

where the condition that $0 \le AR \le 1$ decides which equation that is correct.

For measurements made with two crossed dipoles, a useful set of parameters that describes the state and degree of wave polarization is furnished by the Stokes parameters [31]. For the three-dimensional vectorial electromagnetic field methods described in this thesis, the generalised 3D Stokes parameters are more convenient [13, 14].

# 3

## OPTICS

It is difficult to pinpoint the first discovery of polarized light. The use of this visible manifestation of the vector nature of the electromagnetic field can be tracked down all the way to the Vikings. Erasmus Bartholinus is the one who is officially recognized as the discoverer of polarization in 1669. When looking at a picture through crystal of Icelandic spar (calcium carbonate, calcite) he saw two displaced images. After performing some experiments he wrote a memoir on the subject and this could be considered as the first scientific description of polarized light. The subject has then been studied by a number of scientist such as Huygens, Newton and Young just to mention a few.

In 1936 *Beth* [11] showed how the spin angular momentum of polarized light can exert a torque on a doubly refracting medium. With his apparatus he could detect and measure the torque and from this verify that the spin angular momentum of each photon in a beam of circularly polarized light carries the angular momentum $\hbar$. In 1950 *Kastler* [33], who was awarded the 1966 Nobel Prize in Physics, pioneered techniques for transferring electromagnetic momentum from optical beams to atoms, thereby laying the foundation for another Nobel Prize in Physics, awarded 1997 to Chu, Cohen-Tannoudji, and Phillips "for development of methods to cool and trap atoms with laser light" [16]. *Allen et al.* [3] showed that waves with helical phase fronts have a well defined orbital angular momentum. Beams of this structure are sometimes described as optical vortices [7, 10]. If the phase front of a beam is inclined, it carries orbital angular momentum which can be either an integer or a non-integer multiple of $\hbar$ [18, 47].

The mechanical effects of the orbital angular momentum of beams with an azimuthal phase term $e^{il\varphi}$ has been experimentally verified [46]. For photons, the orbital angular momentum has been experimentally verified to be $l\hbar$ [27].

## 3.1 Formation of beams

A strictly transverse wave traveling in the $z$ direction has its **E** and **B** fields transverse to the $z$ axis and therefore its linear momentum is in the $z$ direction. From this we can see that it cannot have any angular momentum, given by equation (2.37), along the $z$ axis. But if we would have a field component in the $z$ direction, it is possible to obtain angular momentum in the $z$ direction. In fact, a beam of electromagnetic radiation has this property since it consists of a superposition of many waves which do not all travel exactly parallel to the beam symmetry axis. Let us start by looking at such a beam, which is also a good approximation of a laser beam.

## 3.2 Gaussian Beams

To obtain $z$ components in the **E** and **B** fields we can start with a vector potential of the form [3]

$$\mathbf{A} = u(x, y, z) \exp(-\mathrm{i}kz)\hat{x} \tag{3.1}$$

where $u(x, y, z)$ is a function describing the field amplitude distribution. This vector potential is valid for linearly polarized waves. Since we consider monochromatic waves with frequency $\omega$, the solutions to the time-independent wave equation are also solutions to the Helmholtz equation. So to obtain $u$ we solve the Helmholtz equation

$$(\nabla^2 + k^2)u(x, y, z)\exp(-\mathrm{i}kz) = 0 \tag{3.2}$$

where $k = 2\pi/\lambda$. If we assume that the variation in $x$ and $y$ directions are larger than the variation in the $z$ direction we can neglect the $\partial^2/\partial z^2$ part. Mathematically, we write it as

$$\left| 2k\frac{\partial u}{\partial z} \right| \gg \left| \frac{\partial u^2}{\partial z^2} \right| \tag{3.3}$$

This is called the paraxial wave approximation since it basically says that the beam does not diverge much from the beam axis. In this approximation we find that

$$\nabla_t^2 u - 2ik\frac{\partial}{\partial z}u = 0 \qquad (3.4)$$

where $\nabla_t^2$ represents the transverse part of the Laplacian.

A well-known solution of the paraxial Helmholtz equation is a spherical wave given by

$$u(r) = A\frac{\exp(-ikr)}{r} \qquad (3.5)$$

where $A$ is a constant and $r$ the distance from the source. If we rewrite (3.5) in Cartesian coordinates, we obtain

$$u(x,y,z) = A\frac{\exp\left(-ikz\sqrt{1 + \frac{x^2+y^2}{z^2}}\right)}{z\sqrt{1 + \frac{x^2+y^2}{z^2}}} \approx A\frac{\exp(-ikz)\exp\left(\frac{-ik(x^2+y^2)}{2z}\right)}{z} \qquad (3.6)$$

We use this approximation and try to form our Gaussian beam with the cylindrical Ansatz [32]

$$u(\rho,\varphi,z) = A\exp[-i\times f(z)]\exp\left(-\frac{ik\rho^2}{2\times g(z)}\right) \qquad (3.7)$$

where $\rho^2 = x^2 + y^2$ and $f(z)$ and $g(z)$ are analytic functions of $z$. The amplitude function $u$ must also be a solution to the paraxial Helmholtz equation (3.4).

$$\nabla_t^2\left[\exp[-i\times f(z)]\exp\left(-\frac{ik\rho^2}{2\times g(z)}\right)\right] -$$

$$-2ik\frac{\partial}{\partial z}\left[\exp[-i\times f(z)]\exp\left(-\frac{ik\rho^2}{2\times g(z)}\right)\right] = 0 \qquad (3.8)$$

Let us start with the $\nabla_t^2$ term

$$\nabla_t^2\left[\exp[-i\times f(z)]\exp\left(-\frac{ik\rho^2}{2\times g(z)}\right)\right] =$$

$$= \exp\left[-\mathrm{i} \times f(z)\right] \frac{1}{\rho} \frac{\partial}{\partial \rho} \left[\rho \frac{\partial}{\partial \rho} \exp\left(-\frac{\mathrm{i}k\rho^2}{2 \times g(z)}\right)\right] =$$

$$= \exp\left[-\mathrm{i} \times f(z)\right] \exp\left(-\frac{\mathrm{i}k\rho^2}{2 \times g(z)}\right) \left[-\frac{2\mathrm{i}k}{g(z)} - \frac{k^2\rho^2}{g^2(z)}\right] \tag{3.9}$$

and then the $\frac{\partial}{\partial z}$ term

$$-2\mathrm{i}k \frac{\partial}{\partial z}\left[\exp\left[-\mathrm{i} \times f(z)\right] \exp\left(-\frac{\mathrm{i}k\rho^2}{2 \times g(z)}\right)\right] =$$

$$= -2\mathrm{i}k \exp\left[-\mathrm{i} \times f(z)\right] \exp\left(-\frac{\mathrm{i}k\rho^2}{2 \times g(z)}\right) \left[-\mathrm{i}\frac{\partial}{\partial z}f(z) + \frac{\mathrm{i}k\rho^2}{2 \times g^2(z)}\frac{\partial}{\partial z}g(z)\right] \tag{3.10}$$

Now we combine equations (3.9), (3.10) and (3.8) to obtain

$$\exp\left[-\mathrm{i} \times f(z)\right] \exp\left(-\frac{\mathrm{i}k\rho^2}{2 \times g(z)}\right) \times$$

$$\times \left[\left(-\frac{2\mathrm{i}k}{g(z)} - \frac{k^2\rho^2}{g^2(z)}\right) - 2\mathrm{i}k\left(-\mathrm{i}\frac{\partial}{\partial z}f(z) + \frac{\mathrm{i}k\rho^2}{2 \times g^2(z)}\frac{\partial}{\partial z}g(z)\right)\right] = 0 \Leftrightarrow$$

$$\Leftrightarrow \frac{2\mathrm{i}k}{g(z)} + 2k\frac{\partial}{\partial z}f(z) + \frac{k^2\rho^2}{g^2(z)}\left(1 - \frac{\partial}{\partial z}g(z)\right) = 0 \tag{3.11}$$

The first two terms in the left-hand member are independent of $\rho$, which means that for any $\rho$ we can separate the equation into two new ones

$$\frac{2\mathrm{i}k}{g(z)} + 2k\frac{\partial}{\partial z}f(z) = 0 \tag{3.12}$$

$$\frac{k^2\rho^2}{g^2(z)}\left(1 - \frac{\partial}{\partial z}g(z)\right) = 0 \tag{3.13}$$

which gives us

$$(3.13) \Rightarrow \frac{\partial}{\partial z}g(z) = 1 \tag{3.14}$$

$$(3.12) \Rightarrow \frac{\partial}{\partial z}f(z) = -\frac{\mathrm{i}}{g(z)} \tag{3.15}$$

After some further integration we obtain

$$g(z) = z + g_0 \tag{3.16}$$

$$f(z) = -i \ln(z + g_0) \tag{3.17}$$

where $g_0$ is a constant. Inserting the result into the Ansatz (3.7) we get the expression

$$u(\rho, \varphi, z) = A \frac{1}{z + g_0} \exp \left( -\frac{ik\rho^2}{2(z + g_0)} \right). \tag{3.18}$$

Now, we examine the characteristics of the beams, and again compare them with those of a spherical wave. The function $f(z)$ is related to a complex phase shift while $g(z)$ describes the intensity variation. First, we write $g(z)$ in terms of the *width*, $w(z)$, and *radius of curvature*, $R(z)$, measured on the beam axis [36]. The width $w(z)$ is a measure of how the electric field amplitude falls off when moving out from the beam axis. This falloff is Gaussian and $w$ is the distance from the axis where the amplitude is $1/e$ of the maximum value (which is found on the axis). So when talking about the width of a beam we do not mean the range wherein all the field distribution is located but, $1/e$ times of the field. Since the *intensity*, $I$, is proportional to $E^2$, we see that the width of the beam contains $I(1 - 1/e^2)$, or 86%, of the total intensity. At $z = 0$ the beamwidth will be minimal and we call this value the *beam waist*, $w_0$ [36].

Assume that [36]

$$\frac{1}{g(z)} = \frac{1}{R(z)} - \frac{iC}{h(w(z))} \tag{3.19}$$

where $C$ is a constant. At $z = 0$, the wave front is plane, so the radius of curvature is infinite.

$$\frac{1}{R(0)} = 0. \tag{3.20}$$

This means that at $z = 0$, we see that $1/g(z)$ is purely imaginary or, $1/g(z) = 1/g_0$. So, at $z = 0$ we have the equation

$$u(\rho, \varphi, z) = \text{constant} \times \exp \left( \frac{ik\rho^2}{2} \frac{iC}{h(w(z))} \right). \tag{3.21}$$

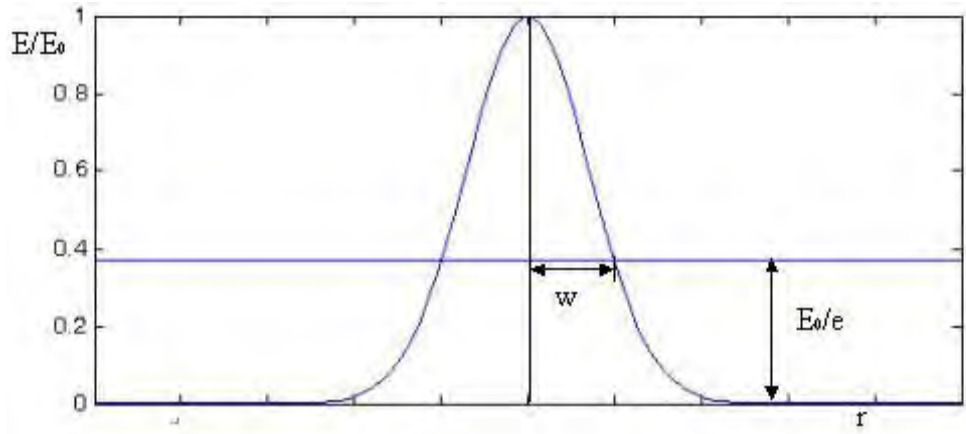

Figure 3.1: Amplitude distribution in a Gaussian beam.

Inserting $\rho = w(z)$ and requiring that the amplitude will be $1/e$ of the maximum value gives

$$\exp\left(\frac{\mathrm{i}kw(z)^2}{2}\frac{\mathrm{i}C}{h(w(z))}\right) = \exp(-1) \tag{3.22}$$

which leads to

$$h(w(z)) = w^2(z) \tag{3.23}$$

$$C = \frac{2}{k} = \frac{\lambda}{\pi} \tag{3.24}$$

$$\Rightarrow g(z) = z + \mathrm{i}\frac{\pi w_0^2}{\lambda} \tag{3.25}$$

and

$$\frac{1}{g(z)} = \frac{1}{R(z)} - \mathrm{i}\frac{\lambda}{\pi w^2(z)}. \tag{3.26}$$

Expressions for $w$ and $R$ can now be obtained

$$w(z) = w_0\sqrt{1 + \left(\frac{\lambda z}{\pi w_0^2}\right)^2} \tag{3.27}$$

$$R(z) = z \left[ 1 + \left( \frac{\pi w_0^2}{\lambda z} \right)^2 \right].$$ (3.28)

By choosing $z = \pi w_0^2 / \lambda$ in equation (3.27) we get $w(z) = \sqrt{2} w_0$ which means that at this $z$, the cross section of the beam is twice as large as the cross section at $z = 0$. We call this value of $z$ the *Rayleigh range*, or $z_R$ [36]. The Rayleigh range gives information of the spread of the beam. A short $z_R$ means that the beam diverges rapidly.

Inserting $z_R$ in equations (3.27) and (3.28) gives us

$$w(z) = w_0 \sqrt{1 + \left( \frac{z}{z_R} \right)^2}$$ (3.29)

$$R(z) = z \left[ 1 + \left( \frac{z_R}{z} \right)^2 \right].$$ (3.30)

Now, let us use these expressions for $w$ and $R$ in the equation for $f(z)$ (3.17)

$$f(z) = -\mathrm{i} \ln(z + g_0) = -\mathrm{i} \ln(z + \mathrm{i} z_R) = -\mathrm{i} \ln \left( \sqrt{z^2 + z_R^2} \times \exp \left[ \mathrm{i} \arctan \left( \frac{z_R}{z} \right) \right] \right) =$$

$$= -\mathrm{i} \ln \left( \sqrt{z^2 + z_R^2} \right) + \arctan \left( \frac{z_R}{z} \right)$$ (3.31)

so

$$\exp \left[ -\mathrm{i} f(z) \right] = \frac{\exp \left[ -\mathrm{i} \arctan \left( \frac{z_R}{z} \right) \right]}{\sqrt{z^2 + z_R^2}} = \frac{w_0}{z_R w(z)} \exp \left[ -\mathrm{i} \arctan \left( \frac{z_R}{z} \right) \right].$$ (3.32)

Now we can write equation (3.18) as

$$u_G(\rho, \varphi, z) = A \frac{w_0}{z_R w(z)} \exp \left[ -\mathrm{i} \arctan \left( \frac{z_R}{z} \right) \right] \exp \left[ -\frac{\mathrm{i} k \rho^2}{2z \left( 1 + \frac{z_R^2}{z^2} \right)} \right] \exp \left[ -\frac{\rho^2}{w^2(z)} \right]$$ (3.33)

which is the commonly used expression for a Gaussian beam. As we see in figure 3.1, the intensity of such a beam is Gaussian, hence the name. Cross sections of beams with different $w_0$ can be seen in figure 3.2.

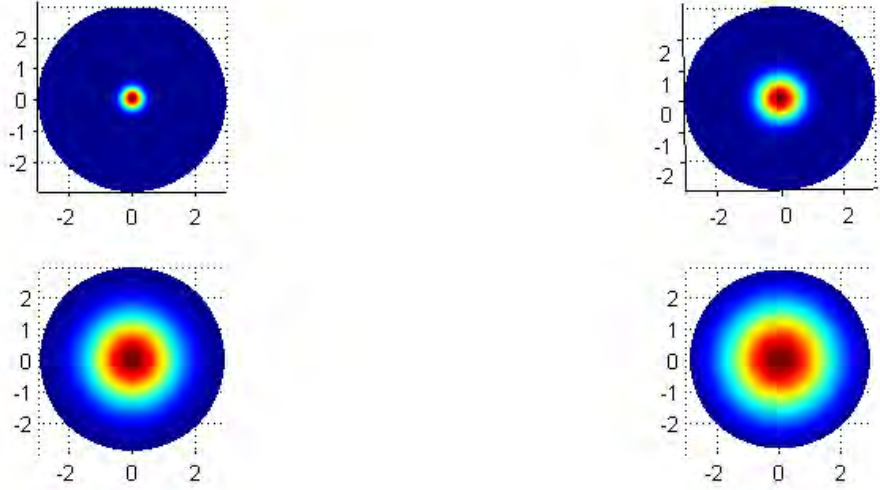

Figure 3.2: Intensities of Gaussian beams with different width parameters $w_0$. From top left to bottom right: $w_0 = 0.5\lambda$, $1\lambda$, $2\lambda$ and $2.5\lambda$. Axis lengths are in wavelengths. Blue is minimum intensity and red is maximum.

If we consider the stationary phase of the beam, in other words the imaginary parts of the exponentials of equation (3.18) and the $\exp[-\mathrm{i}kz]$ part from equation (3.1), and choose to look at planes of constant phase, we get

$$\mathrm{i}kz + \frac{\mathrm{i}z\rho^2}{z_\mathrm{R}kw^2(z)} + \mathrm{i}\arctan\left(\frac{z_\mathrm{R}}{z}\right) = \text{constant}. \tag{3.34}$$

Choosing the constant to be zero and neglecting $\arctan(z_\mathrm{R}/z)$, which is small, we get surfaces given by

$$z = -\frac{z\rho^2}{z_\mathrm{R}kw^2(z)} \tag{3.35}$$

which are parabolas looking like the one seen in figure 3.3. It is instructive to compare this to a beam consisting of parallel transverse waves which would have phase planes perpendicular to the $z$ axis.

$$kz + \phi = \text{constant} \Rightarrow z = \text{constant}. \tag{3.36}$$

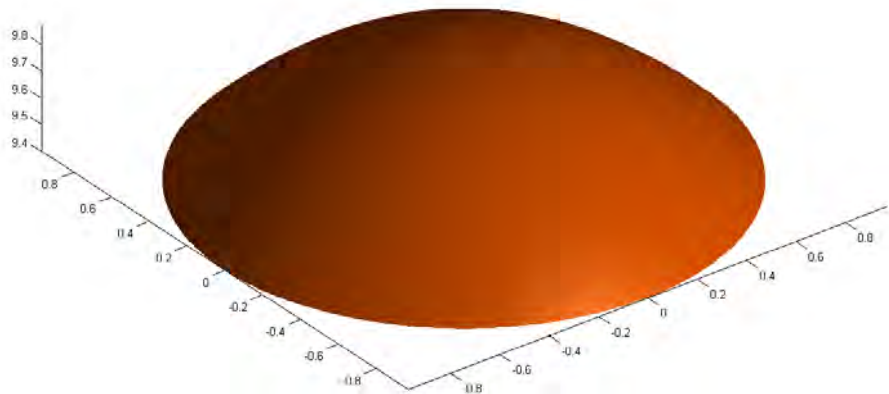

Figure 3.3: Plane of constant phase in a Gaussian beam. The values on the axes are in wavelengths.

## 3.3 Laguerre-Gaussian Beams

In order to endow our EM beam with orbital angular momentum, let us search for a solution of the Helmholtz equation with an azimuthal $e^{-il\varphi}$ dependence because, according to *Simpson et al.* [46], such a beam will carry OAM. Let us use our $u_G$ as a starting point and expand it to a trial solution of the form [36]

$$u(\rho,\varphi,z) = C \times h\left(\frac{\rho}{w}\right) \exp\left[-il\varphi\right] \exp\left[-i\phi(z)\right] u_G(\rho,\varphi,z) = F(\rho,\varphi,z)u_G(\rho,\varphi,z).$$
(3.37)

Inserting this into the Helmholtz equation (3.4), we find that

$$F\nabla_t^2 u_G + 2\left[\boldsymbol{\nabla}_t F \cdot \boldsymbol{\nabla}_t u_G\right] + u_G \nabla_t^2 F - 2iku_G\frac{\partial}{\partial z}F - 2ikF\frac{\partial}{\partial z}u_G = 0.$$
(3.38)

From equation (3.8) we know that $\nabla_t^2 u_G - 2ik(\partial/\partial z)u_G = 0$ which means that the sum of the first and last term in equation (3.38) is equal to zero, so this equation becomes

$$2\left[\boldsymbol{\nabla}_t F \cdot \boldsymbol{\nabla}_t u_G\right] + u_G\nabla_t^2 F - 2iku_G\frac{\partial}{\partial z}F = 0$$

or

$$-2\left(\frac{ik\rho}{z\left(1+\frac{z_R^2}{z^2}\right)} + \frac{2\rho}{w^2(z)}\right)\frac{\partial h}{\partial \rho} + \left(\frac{\partial h^2}{\partial \rho^2} + \frac{1}{\rho}\frac{\partial h}{\partial \rho} - \frac{l^2}{\rho^2}\right) - 2ik\left(ih\frac{\partial \phi}{\partial z} + \frac{\partial h}{\partial z}\right) = 0.$$
(3.39)

After some calculations we find that

$$h\left(\frac{\rho}{w}\right) = \left[\sqrt{2}\left(\frac{\rho}{w}\right)\right]^l \times L_p^l\left(2\frac{\rho^2}{w^2}\right)$$
(3.40)

where $L_p^l$ is a generalized Laguerre polynomial defined as regular solutions to the differential equation

$$x\frac{d^2}{dx^2}L_p^l + (l+1-x)\frac{d}{dx}L_p^l + pL_p^l = 0$$
(3.41)

where $l$ and $p$ are integers greater than $-1$. We also find that

$$\phi(z) = (2p+l)\arctan\left(\frac{z_R}{z}\right).$$
(3.42)

So now we can write equation (3.37) as

$$u_{pl}^{\text{LG}}(\rho, \varphi, z) = C_{pl} \frac{w_0}{z_R w(z)} \left[ \sqrt{2} \left( \frac{\rho}{w(z)} \right) \right]^{|l|} \times L_p^{|l|} \left( 2 \frac{\rho^2}{w^2(z)} \right) \exp\left[ -il\varphi \right] \times$$

$$\times \exp\left[ -i(2p + |l| + 1)\arctan\left( \frac{z_R}{z} \right) \right] \exp\left[ -\frac{ik\rho^2}{2z\left(1 + \frac{z_R^2}{z^2}\right)} \right] \exp\left[ -\frac{\rho^2}{w^2(z)} \right] \quad (3.43)$$

which is our final expression describing the Laguerre-Gauss (LG) beam. The reason for using the absolute value of $l$ in the Laguerre polynomial and the amplitude part is that, as stated above, the Laguerre polynomials are only defined for $l > -1$ and that having the amplitude to the power of $-l$ would result in infinite amplitude as $\rho \to 0$, which is unphysical. In the arctan part we use it for symmetry between $+l$ and $-l$. For us, the sign of $l$ is important and therefore we need to adapt our equations so we can use negative values of $l$. The coefficient $C_{pl}$ is obtained by requiring that every mode transmits the same amount of power and is given by

$$C_{pl} = A \sqrt{\frac{p!}{(p + |l|)!}} \quad (3.44)$$

where $A$ is a constant. Note that $u_{00}^{\text{LG}} = u_{\text{G}}$.

The intensity of the LG beam is proportional to the time average of the linear momentum density, $\text{Re}[\epsilon_0(\mathbf{E} \times \mathbf{B})]$, which is given by [3]

$$\frac{\epsilon_0}{2}(\mathbf{E}^* \times \mathbf{B} + \mathbf{E} \times \mathbf{B}^*) = i\omega \frac{\epsilon_0}{2}(u^* \boldsymbol{\nabla} u - u \boldsymbol{\nabla} u^*) + \omega k \epsilon_0 |u|^2 \hat{z}. \quad (3.45)$$

Depending on the values of $p$ and $l$ the intensity profile will vary. As we can see, setting $l = 0$ gives us a Gaussian intensity profile, but with higher (or lower) $l$ values the beam intensity will exhibit rings of intensity instead of spots. Letting $p$ take other values than zero will give more rings (the number of intensity peaks are $p + 1$). In figure 3.4 we can see beam cross sections with different values of $p$ and $l$.

In the same manner as for the Gaussian beam, we keep the stationary phase constant to obtain the behavior of the phase. For the LG beam it is given by

$$ikz + \frac{iz\rho^2}{z_R k w^2(z)} + il\varphi + i(2p + l + 1)\arctan\left( \frac{z_R}{z} \right) = \text{constant}. \quad (3.46)$$

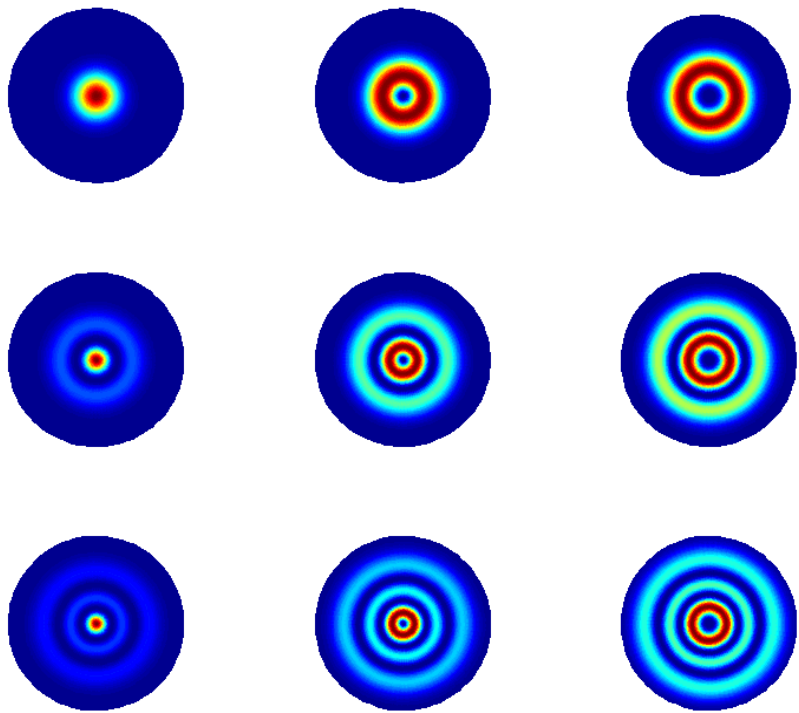

Figure 3.4: Intensities of LG beams. In the top row $p = 0$, in the middle row $p = 1$ and in the bottom row $p = 2$. In each row $l$ changes from zero to two when going from left to right.

Here the phase front appearance will depend on the value of $l$. If $l = 0$ it will resemble the Gaussian beam. For higher values of $l$ the beam will have phase fronts that look like intertwined spirals. We can also see that when moving around the axis one rotation we will cross $l$ number of planes because $l\varphi = 0 (\text{mod} 2\pi)$ will be satisfied $l$ times when $\varphi = [0, 2\pi[$. The sign of $l$ is also important since it will result in a right hand spiral for negative $l$ and left hand spirals for positive $l$. The phase front for $l = 0$ does not rotate at all. Also note that one spiral will need $l$ wavelengths to complete a rotation. This means that the rotation of the phase fronts will be higher for lower values of $l$. In figure 3.5 phase fronts for different $l$ are plotted. The phase fronts in the figure are created by choosing the constant in equation (3.46) to be zero and at the same time neglecting the second and fourth term in the left-hand member. The consequence is that the plots in figure 3.5 give the general characteristics for a beam carrying a distinct orbital angular momentum but with some restrictions as compared to the complete expression. Basically by neglecting the second term in equation (3.46) we fixate the radial curvature not to be curved like the Gaussian phase front in figure 3.3. The fourth term will generate a phase displacement but it will decrease with $z$ and always be limited.

If we examine a cross section of the beam in a plane with constant $z$ we see the $l\varphi$ phase dependence where the phase changes from zero to $2\pi\, l$ number of times when we encircle the beam. This is shown in figure 3.6.

## 3.4 Angular momentum of Laguerre-Gaussian beams

*Allen et al.* [3] use the linear momentum equation (3.45) together with the linearly polarized LG beam distribution equation (3.43) to obtain the momentum density (per unit power), which is directly related to the Poynting vector by $c^2$. For simplicity, we from now denote $u_{\text{pl}}^{\text{LG}}$ by $u$.

$$\mathbf{p}^{\text{dens}} = \frac{\mathbf{S}}{c^2} = \frac{1}{c}\left[\frac{\rho z}{z^2 + z_{\text{R}}^2}|u|^2\hat{\boldsymbol{\rho}} + \frac{l}{k\rho}|u|^2\hat{\boldsymbol{\varphi}} + |u|^2\hat{\mathbf{z}}\right]. \tag{3.47}$$

So we do not have momentum only in the $z$ direction as for a plane wave but also in the $\rho$ and $\varphi$ directions. The $\rho$ component is always positive (for positive $z$ which is what we consider) and this shows that the beam spreads. The $z$ component is the normal linear momentum of the beam and is of course also always positive. The $\varphi$ component means that momentum encircles the beam axis and thus provides a net orbital angular momentum. Depending on $l$ it is either positive or negative. From this we see that the Poynting vector spirals around the $z$ axis in a "corkscrew" fashion as the beam propagates. The time average of the angular momentum density is now given by

$$\mathbf{M} = -\frac{z}{\omega\rho}|u|^2\hat{\boldsymbol{\rho}} + \frac{\rho}{c}\left[\frac{z^2}{z^2 + z_R^2} - 1\right]|u|^2\hat{\boldsymbol{\varphi}} + \frac{l}{\omega}|u|^2\hat{\mathbf{z}}. \tag{3.48}$$

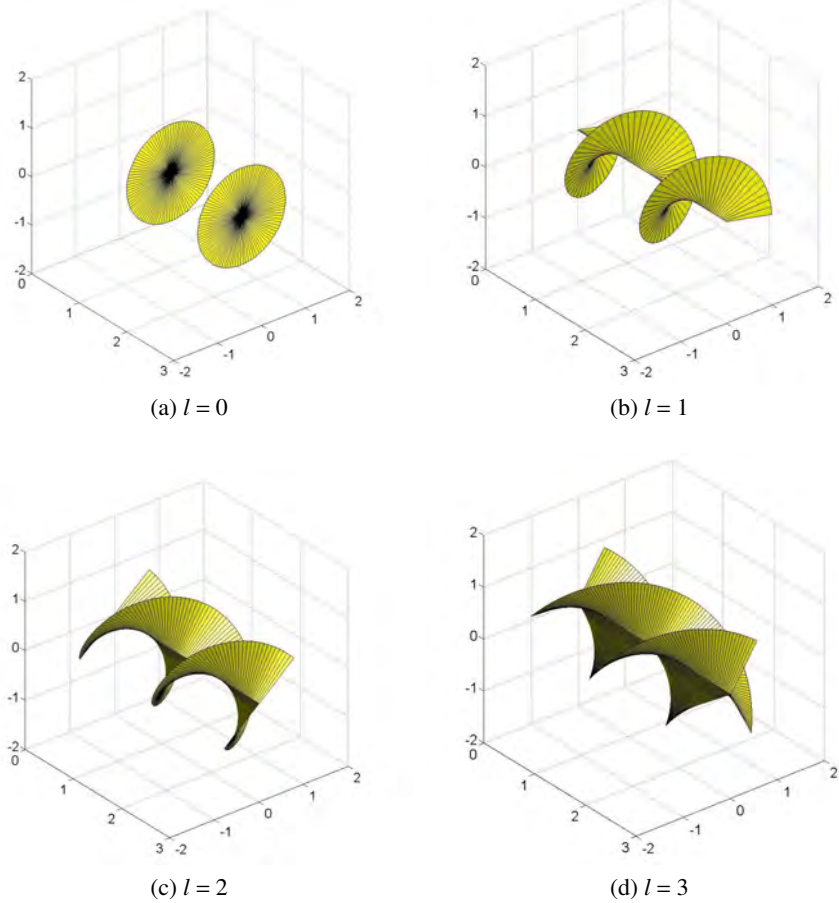

(a) $l = 0$

(b) $l = 1$

(c) $l = 2$

(d) $l = 3$

Figure 3.5: Phase fronts for beams with different orbital momentum, reaching from orbital angular momentum zero in the upper left-hand corner to an orbital angular momentum of 3 in the lower right-hand corner.

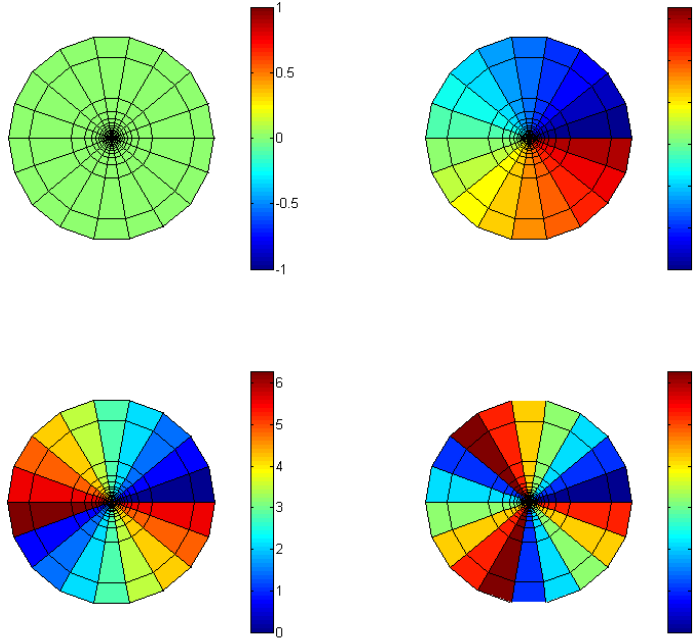

Figure 3.6: Phases in cross sections of beams with $l = 0$ (top left) to $l = 3$ (down right). Note how the phase changes from 0 (blue) to $2\pi$ (red) $l$ number of times.

When integrated over the beam, both the $\rho$ and $\varphi$ components vanish because of circular symmetry and the only angular momentum is in the $z$ direction. *Allen et al.* [3] also point out that from equations (3.47) and (3.48) one can see that the ratio of angular momentum to energy is $L/cp = l/\omega$ and the ratio of angular momentum to linear momentum is $L/p = l(\lambda/2\pi)$. This means that the orbital angular momentum is well-defined and proportional to $l$.

*Allen et al.* [3] also shows that when one has elliptic or circular polarization, equation (3.45) changes to

$$\frac{\epsilon_0}{2}(\mathbf{E}^* \times \mathbf{B} + \mathbf{E} \times \mathbf{B}^*) = i\omega\frac{\epsilon_0}{2}(u^*\boldsymbol{\nabla}u - u\boldsymbol{\nabla}u^*) + \omega k\epsilon_0|u|^2\hat{\mathbf{z}} + \omega\sigma_z\frac{\epsilon_0}{2}\frac{\partial|u|^2}{\partial\rho}\hat{\boldsymbol{\varphi}} \quad (3.49)$$

where $\sigma_z$ describes the polarization. For linearly polarized radiation $\sigma_z = 0$, and $\sigma_z = 1$ or $-1$ for left- respectively right-handed circularly polarized radiation. The $z$ component of $\mathbf{M}$ now becomes

$$M_z = \frac{l}{\omega}|u|^2 + \frac{\sigma_z r}{2\omega}\frac{\partial|u|^2}{\partial\rho}. \quad (3.50)$$

If we compare equations (3.48) and (3.50) we see that while for linearly polarized radiation the ratio of angular momentum density to energy has a well-defined unique value, the ratio of polarized radiation is dependent of the gradient of intensity in every point. But when integrated over the whole beam we get a simple result for all polarizations. For a non-linearly polarized beam the ratio of total angular momentum to energy is $J/cp = (l + \sigma_z)/\omega$.

# 4

## Angular momentum from multipoles

*Jackson* [31] shows that by using a multipole expansion of the electromagnetic fields one finds that the fields obtained contain an $l\varphi$ phase dependence and that the angular momentum is proportional to the energy in all fields which have an $l$ eigenvalue dependence. This can be compared to the results for the Laguerre-Gaussian beams obtained by *Allen et al.* [3].

## 4.1 The wave equation

Let us start with the scalar wave equation in a source-free region

$$\boldsymbol{\nabla}^2\Psi - \frac{1}{c^2}\frac{\partial^2\Psi}{\partial t^2} = 0 \tag{4.1}$$

with

$$\Psi(\mathbf{x},t) = \int_{-\infty}^{\infty} d\omega\, \Psi_\omega(\mathbf{x})\, e^{-i\omega t}. \tag{4.2}$$

Each Fourier component satisfies the Helmholtz equation

$$(\boldsymbol{\nabla}^2 + k^2)\Psi_\omega(\mathbf{x}) = 0. \tag{4.3}$$

Using spherical coordinates and separating the radial and angular variables, we make the Ansatz

$$\Psi_\omega(\mathbf{x}) = \sum_{m,l} f_{ml}(r) Y_{ml}(\theta, \varphi). \tag{4.4}$$

Starting with the radial part we find that

$$\left( \frac{d^2}{dr^2} + \frac{2}{r} \frac{d}{dr} + k^2 - \frac{m(m+1)}{r^2} \right) f_m(r) = 0 \tag{4.5}$$

where we make the substitution $f_m(r) = 1/\sqrt{r}\, u_m(r)$ we get

$$\left( \frac{d^2}{dr^2} + \frac{1}{r} \frac{d}{dr} + k^2 - \frac{(m+\frac{1}{2})^2}{r^2} \right) u_m(r) = 0 \tag{4.6}$$

This we recognize as the Bessel equation with $v = m + 1/2$. The solutions for the radial functions are therefore

$$f_{ml}(r) = \frac{A_{ml}}{r^{1/2}} J_{m+1/2}(kr) + \frac{B_{ml}}{r^{1/2}} N_{m+1/2}(kr) \tag{4.7}$$

where $A_{ml}$ and $B_{ml}$ are constants and $J_{m+1/2}(kr)$ and $N_{m+1/2}(kr)$ are Bessel functions of the first and second kind, respectively. When dealing with Bessel functions of half integer kinds it is convenient to rewrite them in terms of spherical Bessel functions or Hankel functions. The Hankel functions are given by

$$h_m^{(1,2)}(kr) = \left( \frac{\pi}{2kr} \right)^{1/2} \left[ J_{m+1/2}(kr) \pm \mathrm{i} N_{m+1/2}(kr) \right]. \tag{4.8}$$

The angular part of equation (4.4) is found from

$$-\left( \frac{1}{\sin\theta} \frac{\partial}{\partial\theta} \left( \sin\theta \frac{\partial}{\partial\theta} \right) + \frac{1}{\sin\theta} \frac{\partial^2}{\partial\varphi^2} \right) Y_{ml}(\theta, \varphi) = m(m+1) Y_{ml}(\theta, \varphi) \tag{4.9}$$

of which spherical harmonics are solutions given by

$$Y_{ml}(\theta, \varphi) = \sqrt{\frac{2m+1}{4\pi} \frac{(m-l)!}{(m+l)!}} P_m^l(\cos\theta) e^{\mathrm{i}l\varphi} \tag{4.10}$$

where $P_m^l(\cos\theta)$ are associated Legendre functions. Now we can write the general solution for $\Psi$ as

$$\Psi(\mathbf{x}) = \sum_{m,l} \left[ A_{ml}^{(1)} h_m^{(1)}(kr) + A_{ml}^{(2)} h_m^{(2)}(kr) \right] Y_{ml}(\theta, \varphi) \tag{4.11}$$

Note that equation (4.9) resembles (save a factor $\hbar$) the angular momentum operator in quantum mechanics

$$(L^{\mathrm{op}})^2 Y_{ml} = m(m+1) Y_{ml} \tag{4.12}$$

and the pseudovector operator

$$\mathbf{L}^{\mathrm{op}} = -\mathrm{i}(\mathbf{x} \times \boldsymbol{\nabla}) \tag{4.13}$$

of which

$$L_z^{\mathrm{op}} = -\mathrm{i}\frac{\partial}{\partial\varphi} \tag{4.14}$$

Note that $\mathbf{x} \cdot \mathbf{L}^{\mathrm{op}} = 0$.

## 4.2 Multipole expansion

We shall now use a multipole expansion for the electromagnetic fields and assume that we have a $e^{-\mathrm{i}\omega t}$ time dependence in a source-free region. We may then write the Maxwell equations as

$$\boldsymbol{\nabla} \cdot \mathbf{E} = 0 \tag{4.15}$$

$$\boldsymbol{\nabla} \cdot \mathbf{B} = 0 \tag{4.16}$$

$$\boldsymbol{\nabla} \times \mathbf{E} = \mathrm{i}kc\,\mathbf{B} \tag{4.17}$$

$$\boldsymbol{\nabla} \times \mathbf{B} = -\frac{\mathrm{i}k}{c}\mathbf{E}. \tag{4.18}$$

where $\omega = kc$. Combining equations (4.17) and (4.18), we can eliminate $\mathbf{E}$ and we get

$$\boldsymbol{\nabla} \cdot \mathbf{B} = 0 \tag{4.19}$$

$$(\nabla^2 + k^2)\mathbf{B} = 0. \tag{4.20}$$

From $\mathbf{B}$, the far-field $\mathbf{E}$ can be obtained by

$$\mathbf{E} = \frac{\mathrm{i}c}{k}\boldsymbol{\nabla} \times \mathbf{B} \tag{4.21}$$

Each Cartesian component satisfies the Helmholtz equation and can be written as an expansion in the form of equation (4.11) but they must also satisfy equations (4.15) and (4.16) for us to obtain a multipole field. *Jackson* [31] shows that $\mathbf{x} \cdot \mathbf{E}$ and $\mathbf{x} \cdot \mathbf{B}$ satisfy the Helmholtz equation and their solutions are also given by equation (4.11). From this *Jackson* [31] defines *a magnetic multipole field of order* $(m, l)$ by the conditions

$$\mathbf{x} \cdot \mathbf{B}_{ml}^{\mathrm{M}} = \frac{\mu_0 m(m+1)}{k} g_m(kr) Y_{ml}(\theta, \varphi) \tag{4.22}$$

$$\mathbf{x} \cdot \mathbf{E}_{ml}^{\mathrm{M}} = 0 \tag{4.23}$$

where

$$g_m(kr) = A_m^{(1)} h_m^{(1)}(kr) + A_m^{(2)} h_m^{(2)}(kr). \tag{4.24}$$

Using

$$\mathbf{B} = -\frac{\mathrm{i}}{kc}\boldsymbol{\nabla} \times \mathbf{E} \tag{4.25}$$

we relate $\mathbf{x} \cdot \mathbf{B}$ to $\mathbf{E}$ as

$$kc\,\mathbf{x} \cdot \mathbf{B} = -\mathrm{i}\,\mathbf{x} \cdot (\boldsymbol{\nabla} \times \mathbf{E}) = -\mathrm{i}(\mathbf{x} \times \boldsymbol{\nabla}) \cdot \mathbf{E} = \mathbf{L}^{\mathrm{op}} \cdot \mathbf{E} \tag{4.26}$$

where we have used equation (4.13) for $\mathbf{L}^{\mathrm{op}}$. With this we now see that the electric field from the magnetic multipole must satisfy

$$\mathbf{L}^{\mathrm{op}} \cdot \mathbf{E}_{ml}^{\mathrm{M}}(r, \theta, \varphi) = m(m+1)\mu_0 c\, g_m(kr) Y_{ml}(\theta, \varphi) \tag{4.27}$$

as well as equation (4.23). The operator $\mathbf{L}^{\mathrm{op}}$ acts on the angular variables by changing the $l$ number just as the usual shift operators, $L_+$ and $L_-$, plus the $L_z$ operator well-known from quantum mechanics. Remember that from equation 4.12 we know that our angular momentum operator is $1/\hbar$ the operator from quantum mechanics, and naturally, the same goes for the shift operators and $L_z$. However, note that in this text $l$ and $m$ are interchanged for reasons to be mentioned later. Now we can write the fields from the magnetic multipole of order $(m, l)$ as

$$\mathbf{E}_{ml}^{\mathrm{M}} = \mu_0 c\, g_l(kr)\mathbf{L}^{\mathrm{op}} Y_{ml}(\theta, \varphi) \tag{4.28}$$

and

$$\mathbf{B}_{ml}^{\mathrm{M}} = -\frac{\mathrm{i}}{kc}\boldsymbol{\nabla} \times \mathbf{E}_{ml}^{\mathrm{M}}.$$  (4.29)

In a similar way the fields from the *electric multipole of order* $(m,l)$ are found to be

$$\mathbf{B}_{ml}^{\mathrm{E}} = \mu_0 f_m(kr)\mathbf{L}^{\mathrm{op}}Y_{ml}(\theta,\varphi)$$  (4.30)

and

$$\mathbf{E}_{ml}^{\mathrm{E}} = \frac{\mathrm{i}c}{k}\boldsymbol{\nabla} \times \mathbf{B}_{ml}^{\mathrm{E}}$$  (4.31)

where $f_m(kr)$ is given by the expression for $g_m(kr)$ but with other coefficients. For convenience *Jackson* [31] rewrites $\mathbf{L}^{\mathrm{op}}Y_{ml}(\theta,\varphi)$ in a normalized form given by

$$\mathbf{X}_{ml}(\theta,\varphi) = \frac{1}{\sqrt{m(m+1)}}\mathbf{L}^{\mathrm{op}}Y_{ml}(\theta,\varphi)$$  (4.32)

with the orthogonality properties

$$\int \mathrm{d}\Omega\, \mathbf{X}_{m'l'}^* \cdot \mathbf{X}_{ml} = \delta_{ll'}\delta_{mm'}$$  (4.33)

and

$$\int \mathrm{d}\Omega\, \mathbf{X}_{m'l'}^* \cdot (\mathbf{x} \times \mathbf{X}_{ml}) = 0$$  (4.34)

for all $m, m', l$ and $l'$.

If we combine the fields from magnetic and electric multipoles, we obtain the general solution to the Maxwell equations (4.15) to (4.18)

$$\mathbf{B} = \mu_0 \sum_{m,l} \left( a_{\mathrm{E}}(m,l)f_m(kr)\mathbf{X}_{ml} - \frac{\mathrm{i}}{k}a_{\mathrm{M}}(m,l)\boldsymbol{\nabla} \times g_m(kr)\mathbf{X}_{ml} \right)$$  (4.35)

and

$$\mathbf{E} = \mu_0 c \sum_{m,l} \left( \frac{\mathrm{i}}{k}a_{\mathrm{E}}(m,l)\boldsymbol{\nabla} \times f_m(kr)\mathbf{X}_{ml} + a_{\mathrm{M}}(m,l)g_m(kr)\mathbf{X}_{ml} \right)$$  (4.36)

where $a_{\mathrm{E}}(m,l)$ and $a_{\mathrm{M}}(m,l)$ are coefficients that specify the amounts of electric and magnetic multipole fields and are determined by the sources and boundary conditions. *Jackson* [31] states that with the equations

$$\mu_0 \, a_{\mathrm{M}}(m,l) \, g_m(kr) = \frac{k}{\sqrt{m(m+1)}} \int \mathrm{d}\Omega \, Y_{ml}^* \mathbf{x} \cdot \mathbf{B} \tag{4.37}$$

and

$$\mu_0 \, c \, a_{\mathrm{E}}(m,l) \, f_m(kr) = -\frac{k}{\sqrt{m(m+1)}} \int \mathrm{d}\Omega \, Y_{ml}^* \mathbf{x} \cdot \mathbf{E} \tag{4.38}$$

and the knowledge of $\mathbf{x} \cdot \mathbf{B}$ and $\mathbf{x} \cdot \mathbf{E}$ at two different radii, $r_1$ and $r_2$ say, one can obtain the multipole fields.

If we look at $\mathbf{E}$ and $\mathbf{B}$ we see that in every term they contain the spherical harmonics, $Y_{ml}(\theta, \varphi)$, which, according to equation (4.10) include a factor $e^{il\varphi}$. According to [46] the field should therefore carry angular momentum in the $z$ direction proportional to $l$. This is the case if we consider our angular momentum operator $L_z$ given in equation (4.14). It is now also clear why we choose to interchange $m$ and $l$; it is just to avoid confusion for the reader.

## 4.3  Angular Momentum of Multipole Radiation

In the far zone the multipole fields from a localized source can be seen as spherical waves with Hankel functions as the radial function $f_m(kr)$. If we consider fields from an electric multipole of order $(m,l)$ the magnetic field goes as

$$\mathbf{B}_{ml}^{\mathrm{E}} \rightarrow (-\mathrm{i})^{m+1} \frac{e^{\mathrm{i}kr}}{\mu_0 kr} \mathbf{L}^{\mathrm{op}} Y_{ml} \tag{4.39}$$

and the electric filed is found to be

$$\mathbf{E}_{ml}^{\mathrm{E}} = c \, \mathbf{B}_{ml}^{\mathrm{E}} \times \hat{\mathbf{x}}. \tag{4.40}$$

From equations (4.28) and (4.29) we see that the fields from magnetic multipoles will be similar.

Let us consider one Fourier component from a superposition of electric multipoles of order $(m, l)$ where $m$ is a single value but $l$ may vary. According to equations (4.35) and (4.36) we may then write the fields as

$$\mathbf{B}_m = \mu_0 \sum_l a_{\mathrm{E}}(m, l) \mathbf{X}_{ml} h_m^{(1)}(kr) e^{-\mathrm{i}\omega t} \tag{4.41}$$

and

$$\mathbf{E}_m = \frac{\mathrm{i} c}{k} \boldsymbol{\nabla} \times \mathbf{B}_m. \tag{4.42}$$

The time-averaged energy density is given by

$$u = \frac{\epsilon_0}{4} \left( \mathbf{E} \cdot \mathbf{E}^* + c^2 \mathbf{B} \cdot \mathbf{B}^* \right) = \frac{\epsilon_0 c^2}{2} \mathbf{B} \cdot \mathbf{B}^* \tag{4.43}$$

where we have used that, in the far zone, the magnitude of $\mathbf{E}$ equals the magnitude of $c\,\mathbf{B}$. We can now write the differential energy in a spherical shell between $r$ and $r + \mathrm{d}r$ as

$$\mathrm{d}U = \frac{\mu_0 \, \mathrm{d}r}{2 k^2} \sum_{l,l'} a_{\mathrm{E}}^*(m, l') \, a_{\mathrm{E}}(m, l) \int \mathrm{d}\Omega \, \mathbf{X}_{ml'}^* \cdot \mathbf{X}_{ml}. \tag{4.44}$$

Using the orthogonality of $\mathbf{X}_{ml}$ from equation (4.33) we find the energy to be

$$\frac{\mathrm{d}U}{\mathrm{d}r} = \frac{\mu_0}{2 k^2} \sum_l |a_{\mathrm{E}}(m, l)|^2 \tag{4.45}$$

*Jackson* [31] points out that when one has both electric and magnetic multipoles of general order $(m, l)$ the sum changes to a sum over $m$ and $l$ and $|a_{\mathrm{E}}(m, l)|^2$ changes to $|a_{\mathrm{E}}(m, l)|^2 + |a_{\mathrm{M}}(m, l)|^2$.

The time-averaged angular momentum density is given by

$$\mathbf{M} = \frac{1}{2 \mu_0 c^2} \mathrm{Re} \left[ \mathbf{x} \times \left( \mathbf{E} \times \mathbf{B}^* \right) \right] \tag{4.46}$$

which, when considering electric multipoles, and using vector algebra together with the the relationship between $\mathbf{E}$ and $\mathbf{B}$, transforms into

$$\mathbf{M} = \frac{1}{2 \mu_0 \omega} \mathrm{Re} \left[ \mathbf{B}^* (\mathbf{L}^{\mathrm{op}} \cdot \mathbf{B}) \right] \tag{4.47}$$

For a spherical shell between between $r$ and $r + \mathrm{d}r$ the differential angular momentum from the electric multipoles is

$$\mathrm{d}\mathbf{J} = \frac{\mu_0 \, \mathrm{d}r}{2\omega k^2} \mathrm{Re} \left[ \sum_{l,l'} a_{\mathrm{E}}^*(m,l') \, a_{\mathrm{E}}(m,l) \int \mathrm{d}\Omega \, (\mathbf{L}^{\mathrm{op}} \cdot \mathbf{X}_{ml'})^* \, \mathbf{X}_{ml} \right] \qquad (4.48)$$

which, using the definition of $\mathbf{X}_{ml}$, we can write as

$$\frac{\mathrm{d}\mathbf{J}}{\mathrm{d}r} = \frac{\mu_0}{2\omega k^2} \mathrm{Re} \left[ \sum_{l,l'} a_{\mathrm{E}}^*(m,l') \, a_{\mathrm{E}}(m,l) \int \mathrm{d}\Omega \, Y_{ml'}^* \, \mathbf{L}^{\mathrm{op}} \, Y_{ml} \right]. \qquad (4.49)$$

Using equation (4.14) for how $L_z^{\mathrm{op}}$ acts on $Y_{ml}$ plus the orthogonality of the spherical harmonics we find that the $z$ component of the angular momentum can be written as

$$\frac{\mathrm{d}J_z}{\mathrm{d}r} = \frac{\mu_0}{2\omega k^2} \sum_l l \, |a_{\mathrm{E}}(m,l)|^2. \qquad (4.50)$$

Combining equation (4.45) and (4.50) we see that the energy and angular momentum is related to each other as

$$\frac{\mathrm{d}J_z}{\mathrm{d}r} = \frac{l}{\omega} \frac{\mathrm{d}U}{\mathrm{d}r} \qquad (4.51)$$

independent of $r$. In other words, the $z$ component of the angular momentum from a multipole of order $(m,l)$ is related to the EM field energy so that one photon of energy $\hbar\omega$ carries a $z$ component of angular momentum of magnitude $l\hbar$. Comparing this with the results obtained by *Allen et al.* [3] which we presented in chapter 3 we see that the results agree with each other. Furthermore, if one considers two electric dipoles oscillating along the $x$ and $y$ axis respectively with a $\pi/2$ phase difference. This corresponds to $l = \pm 1$ [31] and from equation (4.51) we see that one photon with will carry $\pm\hbar$ of angular momentum. Intuitively, we see that these crossed dipoles will create a circularly polarized beam and the result obtained agrees with what *Beth* [11] measured experimentally.

The multipole derivations are performed in more detail in [31].

# 5

## GENERATION OF AN EM BEAM
## CARRYING ANGULAR MOMENTUM

In the earlier chapters the attributes of beams carrying orbital angular momentum were established. An EM beam containing inclined phase fronts carries orbital angular momentum [18, 47]. In this chapter an array of antennas will be designed that generates a field that will have phase shifts that resembles the field of the Laguerre-Gaussian beam. Of course, radio beams generated by antennas will have somewhat different properties than Laguerre-Gaussian beams generated by lasers, but close to the axis of propagation the EM field of the two beams can be expected to be very similar, both exhibiting a relative offset in the phase of $l\varphi$.

## 5.1 The concept

In a cross-section of a Laguerre-Gaussian beam, the phase, as discussed in section 3.3, is not homogeneous. In figure 5.1 the cross-section of a Laguerre-Gaussian beam is plotted. The different colors represent different offsets in the phase and the curves represent how the phase varies over a space distance. The curves are placed in a circle of radius $\frac{2}{3}\lambda$. This will only give a schematic picture of the situation since the phases that the curves describes will be dislocated. On the other hand, it will give a nice picture of how the phase planes occur. In figure 5.1e and figure 5.1f a more accurate plot is made. It is a little less instructive, but the phase planes are clearly visible.

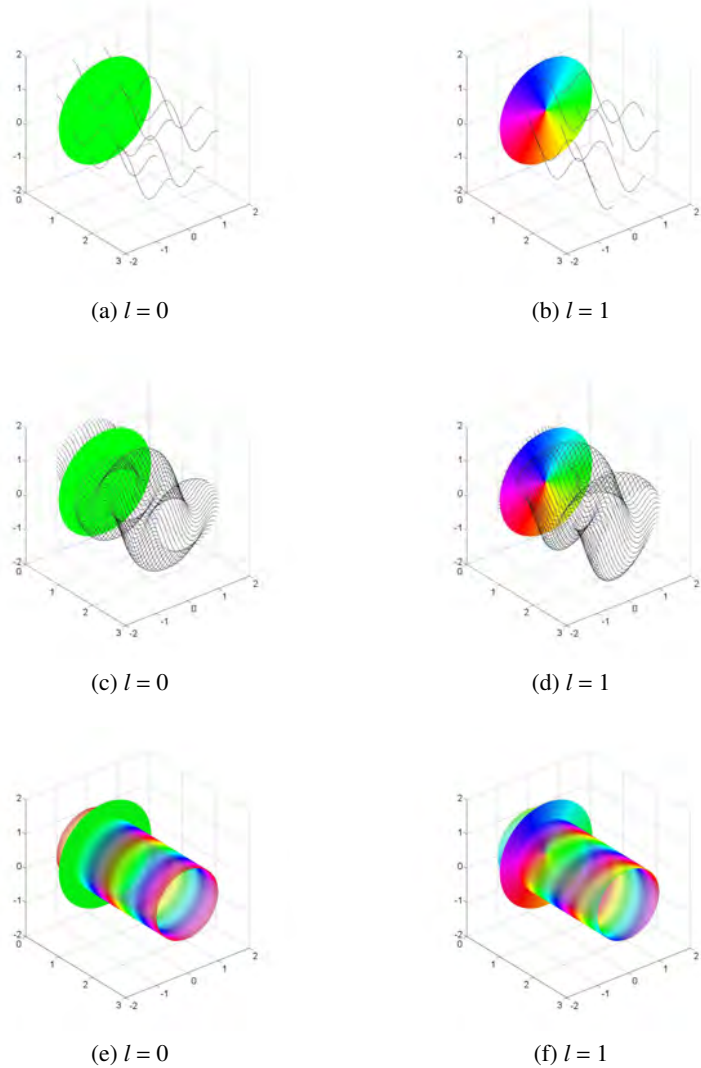

Figure 5.1: The cross section of the Laguerre-Gaussian beam and curves showing, in a schematic way, how the phase varies over a distance in space. Subfigures (a), (c) and (e) shows a plane wave, no orbital angular momentum. Subfigures (b), (d) and (f) show a beam carrying orbital angular momentum ($l = 1$).

For a Laguerre-Gaussian beam with orbital angular momentum number, $l$, the offset in the phase is described by $l\varphi$. To generate such a field we need to shift the phase for different elements in the antenna array that we have chosen. The array of antennas can have many different shapes. For simplicity, we have chosen a multiple circular array.

## 5.2 Calculation of the electric field by the use of array factors

By using multiple elements in a geometrical configuration and controlling the current and phase for each element, the electrical size of the antenna increases and so does the performance. The new antenna system created by two or more individual antennas is called an *antenna array*. Each element does not have to be identical but for simplification reasons they often are. To obtain the properties of the electric field from an array we use the array factor, $AF$.

The total field from an array can be calculated by a superposition of the fields from each element. However, with many elements this procedure is very unpractical and time consuming. By using different kinds of symmetries and identical elements in the array, it is possible to get a much simpler expression for the total field. This is achieved by calculating the so called *array factor*, $AF$, which depends on the displacement (and shape of the array), phase, current amplitude and number of elements. After calculating the array factor the total field is obtained by the *pattern multiplication* rule which is such that the total field is the product of the array factor and the field from one single element [5]

$$\mathbf{E}_{\text{total}} = \mathbf{E}_{\text{single element}} \times AF \qquad (5.1)$$

This formula is valid for all arrays consisting of identical elements. The array factor does not depend on the type of elements used, so for calculating $AF$ it is preferable to use point sources instead of the actual antennas. After calculating the $AF$, equation (5.1) is used to obtain the total field. It is only now that the actual element is being used as $\mathbf{E}_{\text{single element}}$. Arrays can be 1D (linear), 2D (planar) or 3D. In a linear array the elements are placed along a line and in a planar they are situated in a plane.

In our case where we want to have a circular symmetry over the cross section we use a circular grid. To obtain better properties we use a multiple circular grid with equal area sectors, figure 5.2.

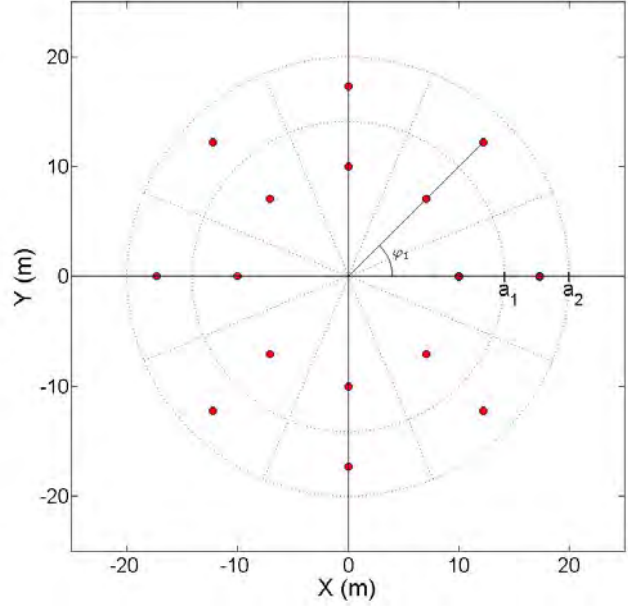

Figure 5.2: Equal area sectors circular grid (EACG).

### 5.2.1 The location of each element

The surface area for each section is $A_g$. The location of each individual element is described by $r_m$ and $\varphi_n$, where $m$ denotes which ring the element is placed in and $n$ is the location in the selected ring. $M$ is the total number of rings and $N$ is the total number of elements in each ring.

$$\frac{a_1^2 \alpha}{2} = A_g \qquad (5.2)$$

$$\frac{a_2^2 \alpha}{2} - \underbrace{\frac{a_1^2 \alpha}{2}}_{A_g} = A_g \qquad (5.3)$$

$$\frac{a_m^2 \alpha}{2} - (m-1)A_g = A_g \qquad (5.4)$$

where $\alpha$ is the angle for the area sectors. This gives us the expression for the radii of each subsection

$$a_m = \sqrt{\frac{m}{M}}\, a \tag{5.5}$$

Let $r_m$ denote the radius vector for the center of each subsection, which is the point where the antenna is located.

$$\frac{r_m^2 \alpha}{2} - \frac{a_{m-1}^2 \alpha}{2} = \frac{a_m^2 \alpha}{2} - \frac{r_m^2 \alpha}{2} \tag{5.6}$$

This gives us the following expression

$$r_m = \sqrt{\frac{(2m-1)}{2M}}\, a \tag{5.7}$$

The $\varphi_n$ for the element is given by

$$\varphi_n = \frac{2\pi n}{N} \tag{5.8}$$

### 5.2.2 Array factor

The normalized field of the array can be written as

$$E_k(r,\theta,\varphi) = \sum_{k=1}^{NM} a_k \frac{e^{-ikR_k}}{R_k} \tag{5.9}$$

where $R_k$ is the distance from the $k$th element to the observation point. Using the notations given in figure 5.2, this distance can be written,

$$R_k = R_{mn} = \left(r^2 + a_{mn}^2 + 2a_{mn}\cos\psi_{mn}\right) \tag{5.10}$$

For $r \ll a$ this expression reduces to

$$R_{mn} \simeq r - a\cos\psi_{mn} = r - a\sin\theta\cos(\varphi - \varphi_{mn}) \tag{5.11}$$

If we assume that for the amplitude $R_k \simeq r$ then we obtain the following expression for the electric field

$$E_{mn} = \frac{e^{-ikr}}{r} \sum_{m=1}^{M} \sum_{n=1}^{N} a_{mn} e^{ika_{mn}\sin\theta\cos(\varphi - \varphi_{mn})} \tag{5.12}$$

where $a_{mn}$ is the excitation coefficient and can be expressed as $I_{mn}e^{i\beta}$ which leads to

$$E_{mn} = \frac{e^{-ikr}}{r} \sum_{m=1}^{M} \sum_{n=1}^{N} I_{mn} e^{i\left( ka_{mn}\sin\theta\cos(\varphi - \varphi_{pn}) + \beta_{mn}\right)} \tag{5.13}$$

In our case $\beta$ will only be dependent on which $n$ we are considering and not the radius since the phase should be equal in "beams" from the center. The variable $\varphi$ will only be dependent on $n$, *not m*. The amplitude $I$ will be dependent of both $m$ and $n$. And finally $a$ will only depend on $m$.

$$E_{mn} = \frac{e^{-ikr}}{r} \sum_{m=1}^{M} \sum_{n=1}^{N} I_{mn} e^{i\left( ka_{m}\sin\theta\cos(\varphi - \varphi_{n}) + \beta_{n}\right)} \tag{5.14}$$

This gives us the array factor, AF

$$[\text{AF}] = \sum_{m=1}^{M} \sum_{n=1}^{N} I_{mn} e^{i\left( ka_{m}\sin\theta\cos(\varphi - \varphi_{n}) + \beta_{n}\right)} \tag{5.15}$$

### 5.2.3 The electric field

The electric field can now be calculated from equation 5.1

$$\mathbf{E} = \frac{e^{-ik|\mathbf{x} - \mathbf{x}_0|}}{|\mathbf{x} - \mathbf{x}_0|} \sum_{m=1}^{M} \sum_{n=1}^{N} I_{mn} e^{i\left( ka_{m}\sin\theta\cos(\varphi - \varphi_{n}) + \beta_{n}\right)} \hat{\mathbf{r}} \tag{5.16}$$

Taking the phase dependent part of this expression and transforming to cylindrical coordinates yields

$$\sum_{m=1}^{M} \sum_{n=1}^{N} I_{mn} e^{i\left( ka_{m}\frac{\rho}{\sqrt{\rho^2 + z^2}}\cos(\varphi - \varphi_{n}) + \beta_{n}\right)} \tag{5.17}$$

By looking at the properties for the wanted field it seems that by having a relative offset in the phase between the elements, a similar field with curved phase fronts will be generated. For a given value of $l$ the phase should shift in steps of $l2\pi/N$ between each element in each ring. The phase for the element $mn$ will then be given by $l\varphi_m$. Ideally, it should be a continuous distribution but since we are dealing with a discrete set of antennas this is not possible. Analyzing equation (5.17), with the antennas phases shifted as discussed, we confirm that a field with a offset in the phase of $l\varphi$ is indeed generated.

Using the array factor to calculate the field will give a representative picture of what the field looks like, but approximations are made along the way. Using a numerical solver for the field generated by the antennas will result in a more accurate, and realistic, picture. This is exactly what we will do in the next chapter.

# 6

## SIMULATION SETUP

To calculate the EM fields from first principles we use the numerical solver, 4nec2. This *de facto* standard program engine has proven to be highly accurate for modeling the electromagnetic response of general structures. In this chapter we will present 4nec2, the different types of antenna arrays, and the different types of elements used in this thesis.

## 6.1 The 4nec2 code

As mentioned in section 5.2, using a numerical solver to calculate the fields, a more accurate picture of the situation can be obtained. We have chosen to use the *de facto* standard program 4nec2 written by Arie Voors [30]. The program uses different versions of *Numerical Electromagnetic Code* (NEC) and we have used the Nec2dXS engine. NEC was originally developed at the Lawrence Livermore Laboratory under sponsorship of the Naval Ocean Systems Center and the Air Force Weapons Laboratory [40]. The program uses both an electric-field integral equation (EFIE) and an magnetic-field integral equation (MFIE) to model the electromagnetic response of general structures, each having their own advantages depending on the structure. The differential equations then obtained are solved numerically by a form of *the method om moments* (MOM).

   The first version of NEC was written in FORTRAN in the 1970's and since then new versions have been developed constantly, both by professionals and amateurs. Nowadays there are both free and license required public versions.

In 4nec2 the output can be viewed as patterns in two or three dimensions showing power and electric field gains. In two dimensions, information such as beamwidth and direction of maximum gain can be obtained. The output can also be given as an ASCII file which makes it possible to write post processing programs in, *e.g.*, MATLAB.

## 6.2 Grids

### 6.2.1 Equidistant radius circular grid, ERCG

In section 5.2 we used a multiple circular grid, an equal area circular grid (EACG), to estimate the properties of the field generated by the antennas. When we use 4nec2 we use a similar construction, the equidistant radius circular grid (ERCG), shown in figure 6.1a. The difference being that the radial distance between each ring is the same and not dependent on the area. This gives us quite a sparse array of antennas. The location of each element is given by $r_m$ and $\varphi_n$. The index $m$ denotes which ring the element is located in and $n$ is the number for that element in the ring. The index $n$ runs from $n = 1$ to $n = N$, where $N$ is the number of elements in each ring. The index $m$ goes from $m = 1$ to $m = M$, where $M$ is the number of rings in the array.

### 6.2.2 Equidistant circular grid, EDCG

By using a multiple circular grid array with equidistant spacing between each individual antennas in each ring, one gets a more dense array, figure 6.1b. A more dense array will lead to better sensitivity. The location of each element is given by $r_m$ and $\varphi_{mn}$, meaning that each element can be represented by $mn$. Here $m$ stands for the numbering of the ring and $n$ the numbering for the element in question in that ring. The prefix $n$ goes from $n = 1$ to $n = N_m$, where $N_m$ is the total number of elements in ring $m$. The prefix $m$ goes from $m = 1$ to $m = M$ where $M$ is the total number of rings. Using this notation for the location of the elements, means that by "equidistant spacing" we refer to the spacing between the element $mn$ and the element $m(n + 1)$.

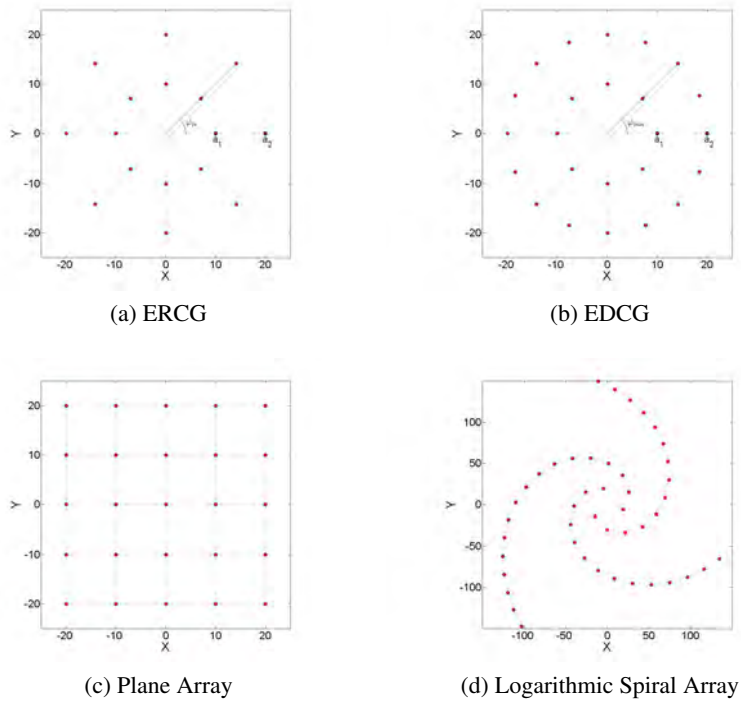

(a) ERCG

(b) EDCG

(c) Plane Array

(d) Logarithmic Spiral Array

Figure 6.1: Here various antenna arrays used and discussed in this thesis are shown.

### 6.2.3 Planar array

A planar array is an array where the elements are placed along a rectangular grid as shown in figure 6.1c. The locations of the elements are given by their Cartesian coordinates.

### 6.2.4 Logarithmic spiral array

The elements can also be placed according to a logarithmic spiral, as is the case for LOFAR [38] which can be read about in section 7.8. The location of each element is given by $r_p = ae^{b\varphi_p}$, where $p$ stands for which "branch" we are looking at. This type of antenna array is shown in figure 6.1d.

## 6.3 Elements

The types of grids presented in the previous section, section 6.2, can be used with different types of antenna elements. In this thesis we have used electric dipole, crossed electric dipole and electric tripole antennas. The crossed electric dipole antenna can be considered as a degenerate case of the electric tripole antenna.

### 6.3.1 Electric dipole antennas

The electric dipole antenna is one of the oldest, simplest and most basic antenna types. In its simplest form it is realized as a straight linear wire of high electric conductivity. Electric dipoles can be divided into different categories depending of their length compared to the wavelength; infinitesimal dipole antennas, where the length of the antenna is smaller than $\lambda/50$, small electric dipole antennas, where the length is greater than $\lambda/50$ and smaller than $\lambda/10$, finite electric dipoles, where the length is greater than $\lambda/10$, and half wave electrical dipole where the length is equal to $\lambda/2$. In this thesis we use small and finite electric dipoles.

The radiation field Fourier amplitudes for an electric dipole is given by [49, 31]

$$\mathbf{E}_\omega^{\text{rad}}(\mathbf{x}) = -\frac{1}{4\pi\epsilon_0} \frac{e^{i(k|\mathbf{x}|)}}{|\mathbf{x}|} (\mathbf{p}_\omega \times \mathbf{k}) \times \mathbf{k} \tag{6.1}$$

$$\mathbf{B}_\omega^{\text{rad}}(\mathbf{x}) = -\frac{\omega\mu_0}{4\pi} \frac{e^{i(k|\mathbf{x}|)}}{|\mathbf{x}|} (\mathbf{p}_\omega \times \mathbf{k}) \tag{6.2}$$

### 6.3.2 Tripole and crossed dipole antennas

The design of a tripole antenna is simple. Basically it comprises three equal dipole antennas perpendicular to each other, having individual feeds. The midpoints of all the three dipoles are colocated at one point i 3D space, the antenna center point. With the tripole antenna it is possible to measure the *complete* electric field vector and therefore to extract *all information available in a radio signal* [17].

In transmitting mode it is possible to transmit a beam in any direction with a desired polarization [9]. This is accomplished by controlling the current amplitude and phase individually in each of the three dipole antennas. In complex form the currents can be written [19]

$$j_k = j_0 w_k = A_k e^{i\phi_k} \tag{6.3}$$

where $A_k$ is the amplitude and $\phi_k$ the phase of the current in element $k$.

The simplest type of antenna is one where the elements are considered short compared to the wavelength, typically less than $\lambda/10$ but larger than $\lambda/50$ [5]. The Fourier amplitude of the radiated fields in the far-field approximation for such an antenna is given by [19]

$$
\begin{aligned}
\mathbf{E}_\omega^{\mathrm{rad}}(r,\varphi,\theta) = {}& i\frac{L\mu_0 ck}{8\pi}\frac{e^{ikr}}{r}\Big[ -A_1 e^{i\phi_1}(\sin\varphi\,\hat{\boldsymbol{\varphi}} - \cos\theta\cos\varphi\,\hat{\boldsymbol{\theta}}) \\
& + A_2 e^{i\phi_2}(\cos\varphi\,\hat{\boldsymbol{\varphi}} + \cos\theta\sin\varphi\,\hat{\boldsymbol{\theta}}) - A_3 e^{i\phi_3}\sin\theta\,\hat{\boldsymbol{\theta}}\Big]
\end{aligned}
\tag{6.4}
$$

and

$$
\begin{aligned}
\mathbf{B}_\omega^{\mathrm{rad}}(r,\varphi,\theta) = {}& -i\frac{L\mu_0 k}{8\pi}\frac{e^{ikr}}{r}s\Big[ -A_1 e^{i\phi_1}(\sin\varphi\,\hat{\boldsymbol{\varphi}} + \cos\theta\cos\varphi\,\hat{\boldsymbol{\theta}}) \\
& + A_2 e^{i\phi_2}(\cos\varphi\,\hat{\boldsymbol{\varphi}} - \cos\theta\sin\varphi\,\hat{\boldsymbol{\theta}}) + A_3 e^{i\phi_3}\sin\theta\,\hat{\boldsymbol{\theta}}\Big]
\end{aligned}
\tag{6.5}
$$

where $L$ is the length of the antenna. In the time domain the fields are given by [19]

$$
\begin{aligned}
\mathbf{E}^{\mathrm{rad}}(t,\mathbf{x}) = {}& -\frac{L\mu_0 ck}{8\pi r}\Bigg( \Big[ A_1 \sin(kr - \omega t + \phi_1)\cos\theta\cos\phi \\
& + A_2 \sin(kr - \omega t + \phi_2)(\cos\theta\sin\varphi - A_3\sin(kr - \omega t + \phi_3)\sin\theta\Big]\hat{\boldsymbol{\theta}} \\
& + \Big[ -A_1 \sin(kr - \omega t + \phi_1)\sin\varphi + A_2 \sin(kr - \omega t + \phi_2)\cos\varphi\Big]\hat{\boldsymbol{\varphi}} \Bigg)
\end{aligned}
\tag{6.6}
$$

and

$$
\begin{aligned}
\mathbf{B}^{\mathrm{rad}}(t,\mathbf{x}) = {}& \frac{L\mu_0 k}{8\pi r}\Bigg( \Big[ -A_1 \sin(kr - \omega t + \phi_1)\sin\varphi \\
& + A_2 \sin(kr - \omega t + \phi_2)\cos\varphi\Big]\hat{\boldsymbol{\theta}} \\
& + \Big[ -A_1 \sin(kr - \omega t + \phi_1)\cos\theta\cos\varphi \\
& - A_2 \sin(kr - \omega t + \phi_2)\cos\theta\sin\varphi + A_3 \sin(kr - \omega t + \phi_3)\sin\theta\Big]\hat{\boldsymbol{\varphi}} \Bigg)
\end{aligned}
\tag{6.7}
$$

In short we say that

$$\mathbf{E}^{\mathrm{rad}}(t,\mathbf{x}) = E_\theta \hat{\boldsymbol{\theta}} + E_\varphi \hat{\boldsymbol{\varphi}} \tag{6.8}$$

and

$$\mathbf{B}^{\mathrm{rad}}(t,\mathbf{x}) = B_\theta \hat{\boldsymbol{\theta}} + B_\varphi \hat{\boldsymbol{\varphi}} \tag{6.9}$$

Since the antenna spans 3D space, it can beam in any given direction, $(\theta_0, \varphi_0)$, and with any polarization but let us consider the case when the tilt angle, $\tau$, between the major axis of the polarization ellipse and the $\theta$ axis is a multiple of $\pi/2$. angle $\tau$ is called the tilt angle and can be anything between 0 and $\pi$. In this case the complex currents in the elements will be [19]

$$j_1 = \frac{1}{ax}\cos\theta_0\cos\varphi_0 + \mathrm{i}\sin\varphi_0 \tag{6.10}$$

$$j_2 = \frac{1}{ax}\cos\theta_0\sin\varphi_0 - \mathrm{i}\cos\varphi_0 \tag{6.11}$$

$$j_3 = -\frac{1}{ax}\sin\theta_0 \tag{6.12}$$

where $ax$ is the axial ratio including sign for right- or left-handed polarization. Its values are between $-1$ and 1. When it is $\pm 1$ the polarization circular and when it is 0 we have linear polarization. Positive $ax$ means that the polarization is left-handed and negative $ax$ means right-handed polarization. Note the similarities with $\sigma_z$ in equation (3.49). Also note that when $\theta_0 = 0$, the element in the $z$ direction should have zero current ($J_3 = 0$) and the antenna degenerates to a crossed dipole antenna.

# 7

## RESULTS

Here we present a comprehensive set of modeling results for various types of antenna arrays and different combinations of OAM. We start off by simulating fields with OAM properties and linear polarization. Then we simulate fields with OAM properties and elliptic polarization. Several types of elements are used including tripole antennas. All the simulations are performed in the far field region.

## 7.1 Electromagnetic beam with orbital angular momentum

### 7.1.1 Grid

In this section we have used the "Equidistant circular grid" (EDCG) antenna configuration, shown in figure 6.1b. We have used two rings. The first ring contains 8 elements and the second 16 elements. The elements used here are dipoles, with their orientation in the $y$ direction, according to figure 6.1b, this yields a linearly polarized field. The grid is placed 2.5 m over ground. The frequency used in this compilation is 27.3 MHz, corresponding to a vacuum wavelength of 11 m. The radius for the inner ring is 10 m and for the outer 20 m. The circle sector distance between consecutive elements in each ring is 7.85 m.

### 7.1.2 The relative offset in the phase between the elements

The field generated should have a relative offset in the phase of $l\varphi$, in resemblance with the Laguerre-Gaussian beam, where the variable $l$ stands for the orbital angular momentum number of the beam. By applying this directly to the array we will get a discrete distribution, which will introduce some physical limitations. The phase in each element will therefore be given by $l\varphi_{mn}$, for element $mn$. We have used the $x$ axis as reference for the angle $\varphi$.

### 7.1.3 Intensity

In figure 7.2 the fields obtained from the 4nec2 software package are shown. Figure 7.2a shows the field when all the elements have the same phase. In figure 7.2b - 7.2d the fields with $l = 1$, $l = 2$ and $l = 4$ are shown. In figure 3.2 the intensity profiles for Laguerre-Gaussian beams are shown for different orbital angular momenta. By comparing these profiles to the field obtained from 4nec2 we see a resemblance. The fields representing beams carrying OAM have a characteristic intensity minimum in the center of the beam and for higher values of $l$ the beam widens. This is all in agreement with the optical case. In the last figure, figure 7.2d, one weakness of this array is shown. Since the inner ring only contains 8 elements it will yield the intensity irregularities shown, there are 8 intensity spikes shown originating from the 8 inner ring antennas. To resolve this, more antennas could be added in the inner ring. With 8 antennas the maximum $l$ that could be identified with certainty would actually be 3, so in a way the antenna array is working outside its bounds.

### 7.1.4 Phase shifts in the main beam

By post-processing the output data from 4nec2, we have obtained the phase shifts over the generated electric fields. The main interest is the main lobe. The side lobes can be minimized by manipulating the arrays, for instance with more rings, different distances to ground and different radii. From figure 7.1 we obtain the beamwidth for the main lobe for each of the fields shown in figure 7.2. In figure 7.3 we can the see schematic pictures of how the phase differs in the main lobe, for different $l$. The phase is represented by the angle each arrow have, with the $x$ axis as reference. In figure 3.6 and 5.1 cross sections for Laguerre-Gaussian beams are shown, and one can see that the phase shifts in the radio beams generated here are the same. In the figures one can see that the phase in the radio beams is constant on radial rays from the center of the beam, which is also the case, as one can see in figure 3.6 and 5.1, for LG beams.

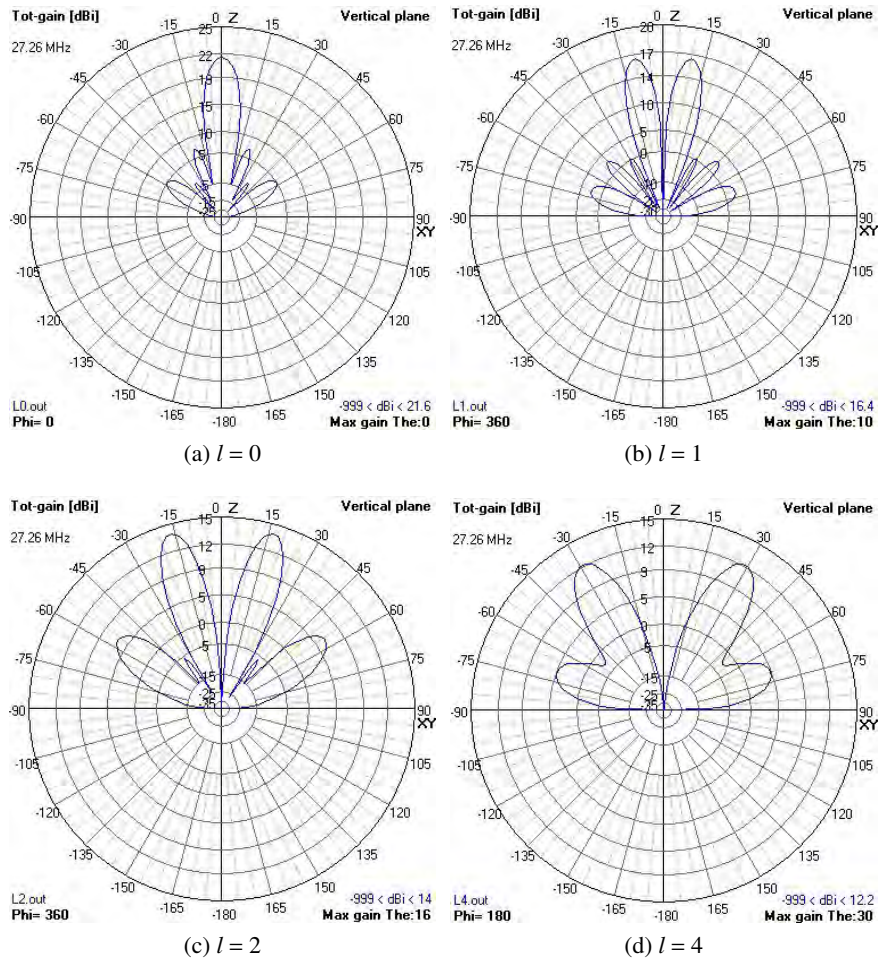

Figure 7.1: Two-dimensional field gain patterns for the fields in figure 7.2.

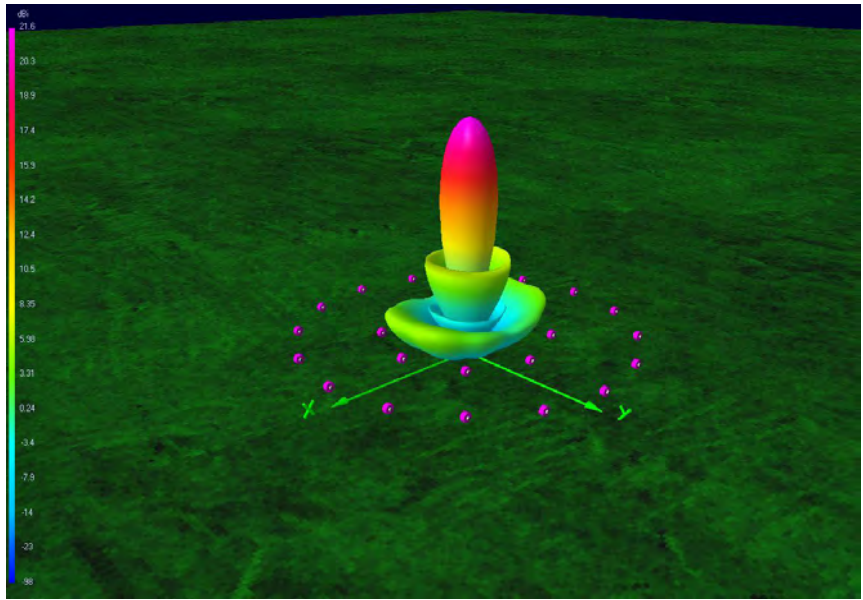

(a) $l = 0$

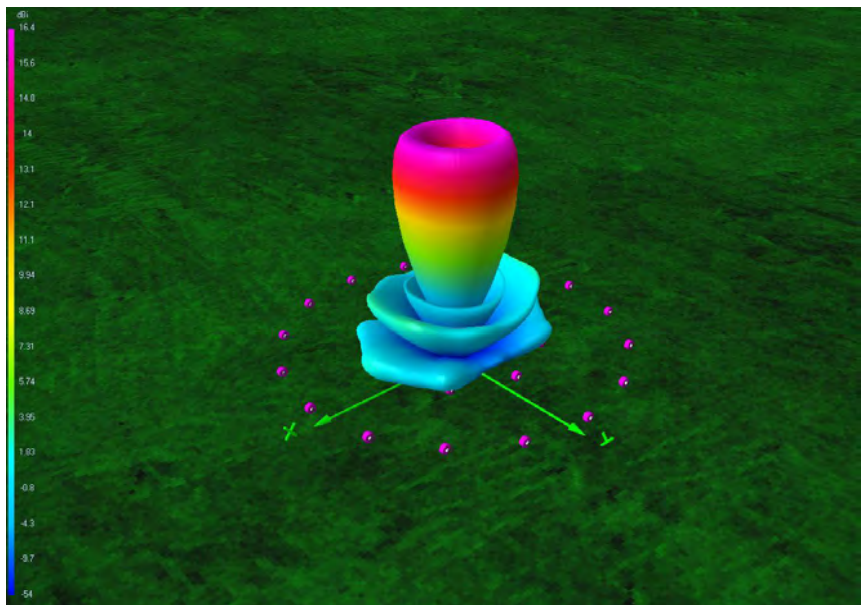

(b) $l = 1$

Figure 7.2: Field gain generated by a EDCG for different values of $l$.

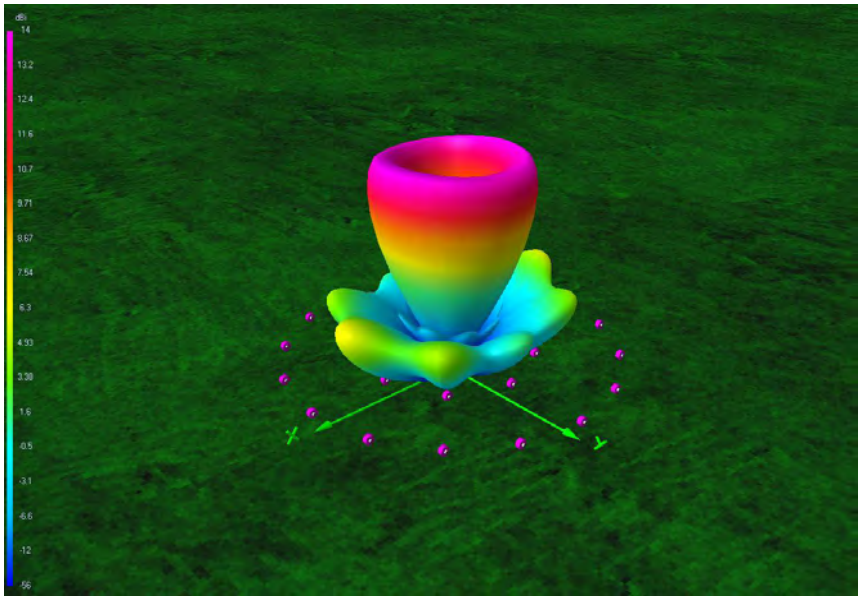

(c) $l = 2$

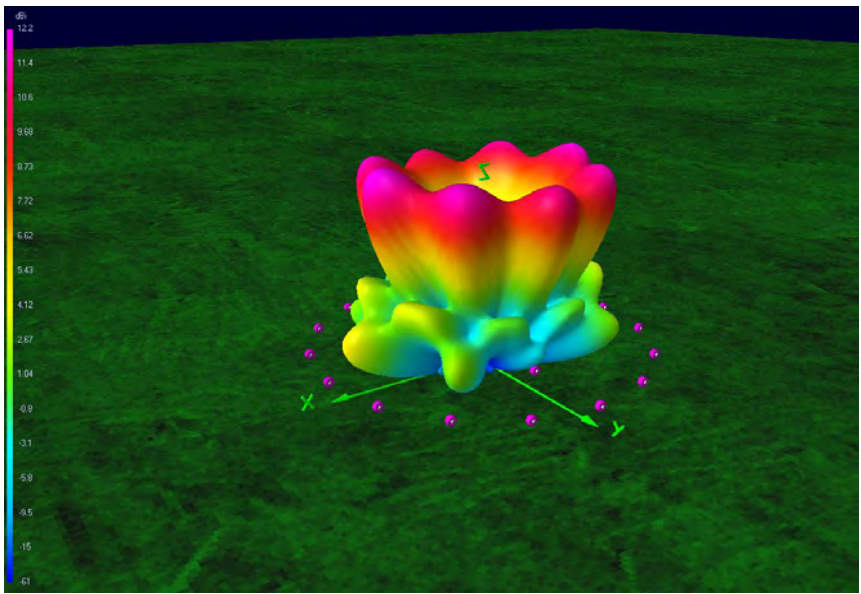

(d) $l = 4$

Figure 7.2: Field gain generated by a EDCG for different values of $l$.

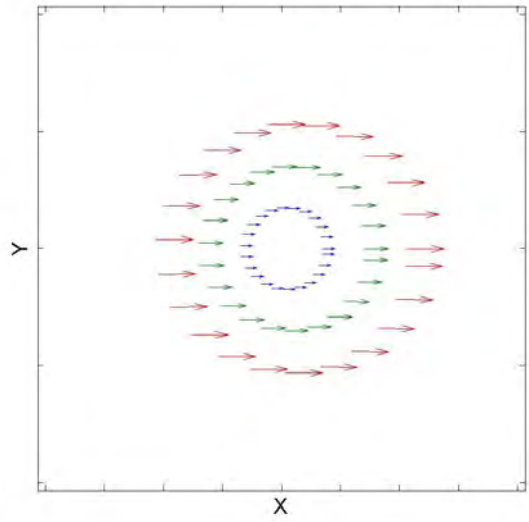

(a) $l = 0$

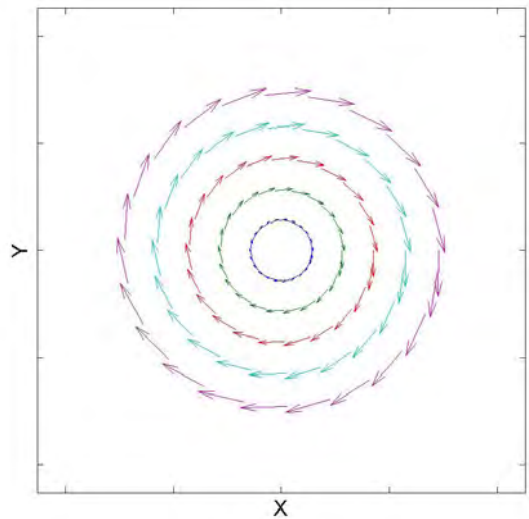

(b) $l = 1$

Figure 7.3: Phase shifts in the main lobe for fields with different $l$. The fields are those shown in figure 7.2.

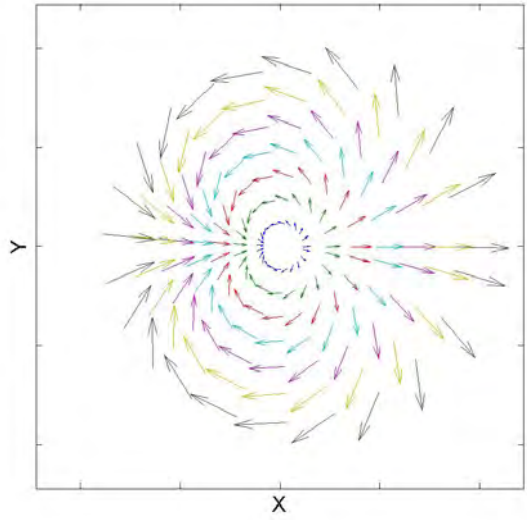

(c) $l = 2$

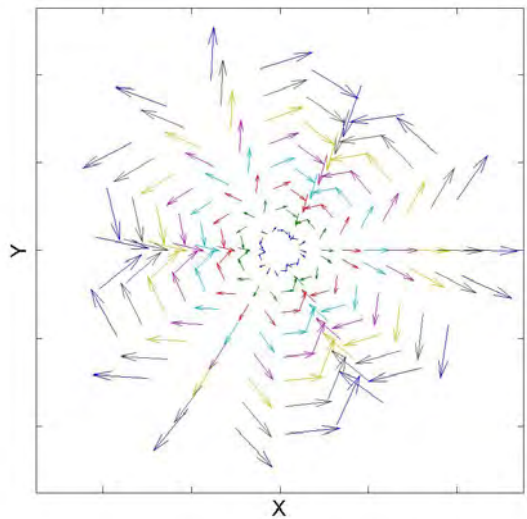

(d) $l = 4$

Figure 7.3: Phase shifts, in the main lobe, for fields with different $l$. The fields are those shown in figure 7.2.

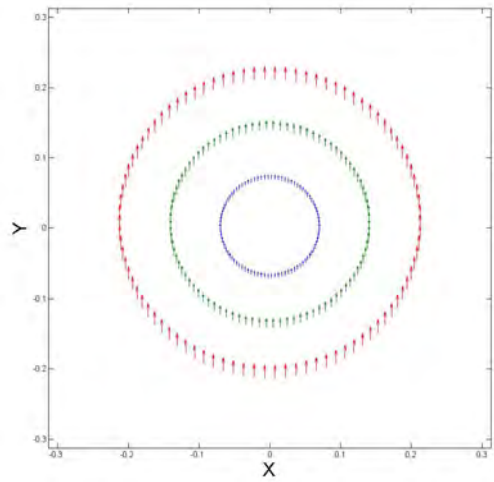

(a) $l = 0$

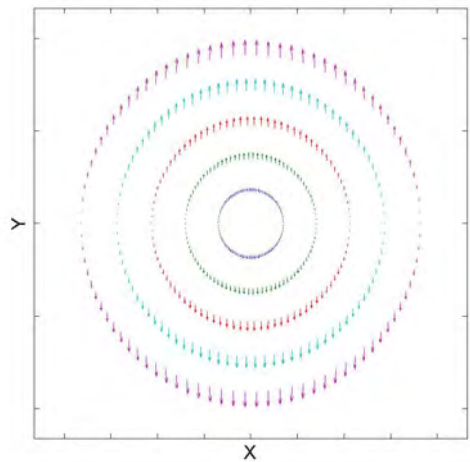

(b) $l = 1$

Figure 7.4: Electric fields in the main lobe for different $l$. Same fields as those shown in figure 7.2.

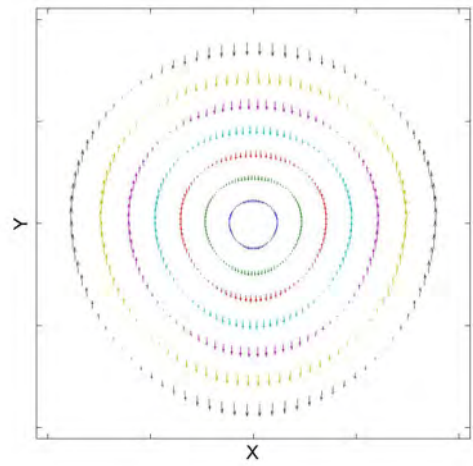

(c) $l = 2$

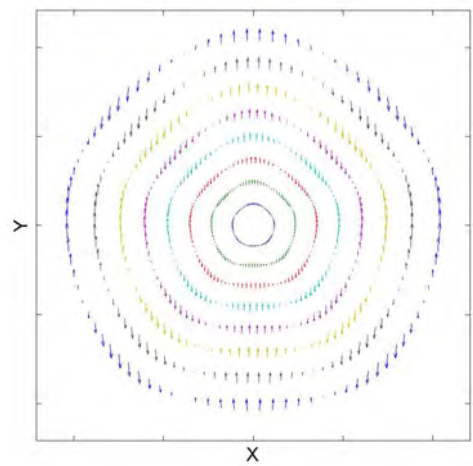

(d) $l = 4$

Figure 7.4: The electric fields, in the main lobe, for different $l$. Same fields as those shown in figure 7.2.

### 7.1.5 The electric field

From the output data one can also obtain the electric field. The fields are projected down on a plane perpendicular to the axis of propagation, 7.4. In the figure one can see the strength of the field in each location and its direction. The strength is only comparable to the electric field in the same ring, same color. As we hoped the electric field is linear, meaning that the only angular momentum in these field will originate from the phase shift; not the polarization.

## 7.2 Multiple beam interference

### 7.2.1 Grid

The same antenna configuration as in the previous section is used here, the so called EDCG, shown in figure 6.1b. The inner ring contains 8 elements and the outer 16 elements. The elements used here are dipoles with a length of 1 m,, orientated in the $y$ direction according to figure 6.1b, which should yield a linearly polarized field. The grid is placed 2.5 m over ground. The frequency used in this compilation is 15 MHz, corresponding to a vacuum wavelength, $\lambda$, of 20 m. The radius for the inner ring is 10 m and for the outer 20 m. The circle sector distance between consecutive elements in each ring is 7.85 m, corresponding to 0.3925 $\lambda$.

### 7.2.2 The relative offset in the phase between the elements

To obtain interference between beams with different values of $l$ we use an EDCG with two rings. In the first ring, $m = 1$, we have a phase shift of $l_1\varphi_{mn}$ and on the outer, $m = 2$, we have a phase shift of $l_2\varphi_{2n}$. The variables $l_1$ and $l_2$ stands for the relative phase offset factor for each ring, which corresponds to the $l$ number in the Laguerre-Gaussian beams. The $x$ axis is used as a reference for the phase shift. These patterns are shown in figure 7.5 for the different combinations of $l_1$ and $l_2$. The patterns for the interference using the Laguerre-Gaussian beams are also shown in the same figure. One can notice a distinct resemblance between the intensity for the beams generated in the optical domain and the beams generated in the radio domain. These intensity patterns would be hard to obtain without the information about beams carrying OAM. Multiple beam interference could be used to shape and direct the beam. For example the interference between $l = 2$ and $l = 4$ could be used for scanning on both sides of an object, at the same time, with one single beam.

## 7.3 Rotation of the main lobe

### 7.3.1 Grid

In this section the antenna array EDCG shown in figure 6.1b is used. We have two antenna rings. The first ring contains 8 elements, and the second 16 elements. The elements used here are dipoles with a length of 1 m, with orientation in the $y$ direction, according to figure 6.1b, which should yield a linearly polarized field. The grid is placed 2.5 m over ground. The frequency used in this compilation is 15 MHz, corresponding to a vacuum wavelength, $\lambda$, of 20 m. The radius for the inner ring is 10 m and for the outer 20 m. The circle sector distance between consecutive elements in each ring is 7.85 m, corresponding to 0.3925 $\lambda$.

### 7.3.2 The relative offset in the phase between the elements

Here the interference beam of $l_1 = 1$ and $l_2 = 2$ is used. The procedure for this was presented in the previous section. In the first ring we have a relative offset in the phase of $l_1 \varphi_{mn}$ and on the outer we have a offset of $l_2 \varphi_{mn} - \phi$. The variable $\phi$ is a rotation angle, which is varied. The variables $l_1$ and $l_2$ stands for the phase shift factor for each ring, which corresponds to the orbital angular momentum in the Laguerre-Gaussian beam. The $x$ axis is used as a reference for the angle.

### 7.3.3 Intensity in the main beam

In figure 7.6 the intensity field, given by 4nec2, is shown for different $\phi$. In these figures one can see how the main lobe rotates when $\phi$ is varied. This gives a way of rotating the beam without any mechanical devices.

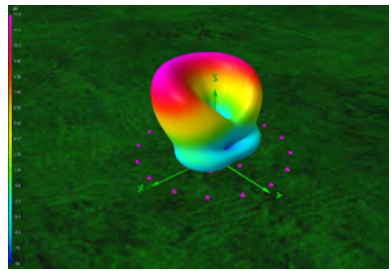
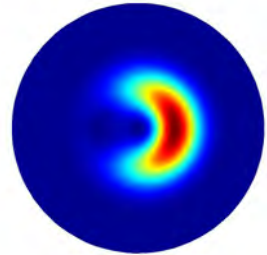

(a) $l = 1$ and $l = 2$     (b) $l = 1$ and $l = 2$

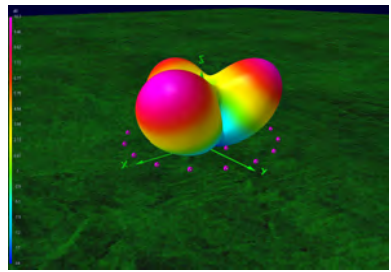
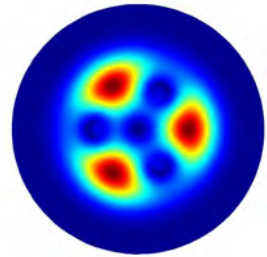

(c) $l = 1$ and $l = 4$     (d) $l = 1$ and $l = 4$

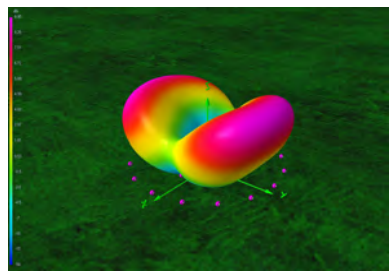
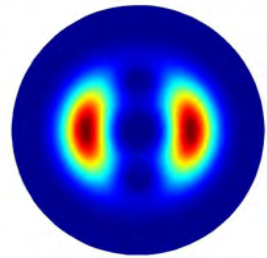

(e) $l = 2$ and $l = 4$     (f) $l = 2$ and $l = 4$

Figure 7.5: The figures show the intensity for two-beam interference. The $l$ values of the interfering beams are shown below each picture. The pictures to the left shows the intensity pattern generated by antennas and the pictures to the right shows the intensity pattern generated by Laguerre-Gaussian beams.

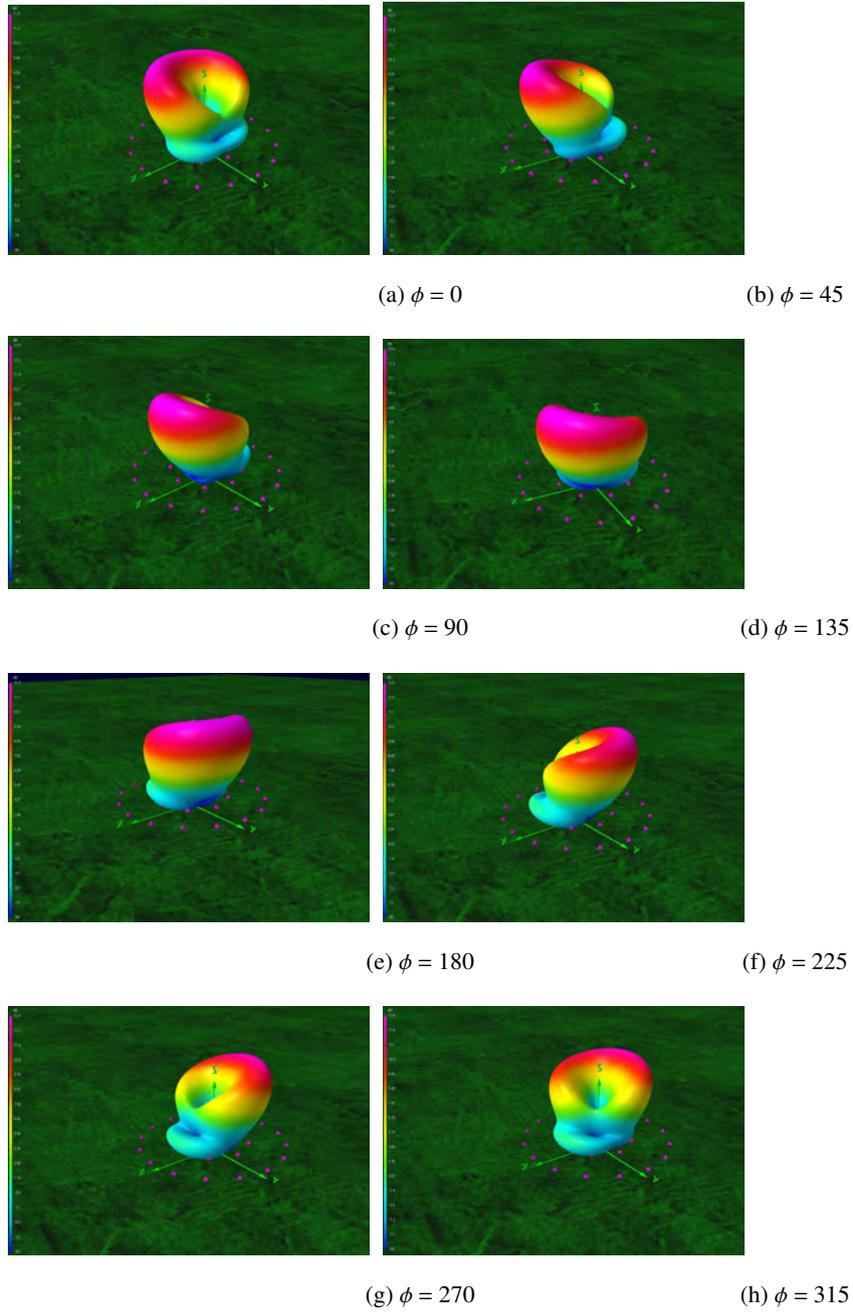

(a) $\phi = 0$      (b) $\phi = 45$

(c) $\phi = 90$      (d) $\phi = 135$

(e) $\phi = 180$      (f) $\phi = 225$

(g) $\phi = 270$      (h) $\phi = 315$

Figure 7.6: Rotation of the main lobe for the mixed state $l = 1$ and $l = 2$. The phase shift for the outer ring is given by $l_2\varphi - \phi$.

## 7.4 Five-element crossed dipole array

We start with an array with five small tripole elements operating at 300 MHz situated $\lambda/100$ over perfect ground. The radius of the array circle is $a = \lambda/5$. To transmit with the $z$ axis as symmetry axis we use equations (6.10) to (6.12) with $\theta_0 = 0$, $\varphi_0 = 0$ and $ax = \pm 1$ to calculate the currents. To do this we use the program tripole.m which, for a circular array, calculates the phases and amplitudes in any number of tripole elements for all possible $l$, directions and polarizations. The element in the $z$ direction should have zero current while the two other should have equal amplitude but be shifted $\pi/2$ in phase. Since the current in the $z$ directed element is zero we can consider the antenna as two crossed dipoles, if we neglect the small effect that the third element would still have. This effect is very small since the coupling between elements in the tripole is small [19]. When using an array with $N$ number of elements the magnitude of the highest eigenvalue , $l_{\max}$, resolvable can be calculated with the equation

$$|l_{\max}| < \frac{N}{2}. \tag{7.1}$$

When using an array with five elements, we should in other words be able to have values of $l$ between minus two and plus two, so those values are the ones we consider.

### 7.4.1 Left-hand polarization

We start with left-hand polarization *i.e.* $ax = 1$ when we calculate the currents. When plotting the electric field vector projected on a plane normal to the $z$ axis the results for $l = 1$ and $l = 2$ can be seen in figure 7.7. Each ring of vectors represents a step of $\theta = 5°$ and the outer ring is where $\theta = 60°$. Note that the vectors are scaled so that they do not intersect in the figure. This means that the lengths of the vectors are not the real length representing the electric field amplitude. However, in each individual ring, the vectors are scaled by the same constant and their lengths can be compared. The expected result is that when going around one ring the vector should complete $l$ rotations. We see that this is true for most $\theta$ and we can also see that for one $\varphi$ the vectors are in phase with each other. So it looks like the phase behavior of the beam is, as expected, depending on $l\varphi$. This can be seen more clearly if we instead plot the (spatial) phases of the electric field vectors in figure 7.8. Comparing the phase behavior to the pictures of the theoretical phase behavior in beam cross sections from chapter 3 we see that our results are indeed very promising, at least when $\theta$ is not too large. This is better seen in figure 7.9 where the phase fronts are zoomed in so the largest $\theta$ equals 65°. In these figures the expected result is that the phase should go from 0 (blue) to $2\pi$ (red) $l$ times when rotating a full circle.

We can also look at the intensity pattern in a plane of the beam and compare it to the results derived from the Laguerre-Gaussian beams from chapter 3. These are compared in figure 7.10 and 7.11. The results are very good for $l = \pm 1$ and 0 but not so good for $l = \pm 2$ where there are spots rather than rings of intensity. This is most likely because of the small number of antenna elements in the array. Just as for the LG beam the radius of the rings (if we consider the spots in the $l = 2$ beam as parts of a ring) increases with increasing $l$. The reason that the $l = 0$ beam has an intensity minimum in the middle is in the MATLAB code. In reality there is a maximum there.

As described in chapter 3, according to *Allen et al.* [3] the ratio of the $z$ component of the angular momentum, $J_z$, to the energy, $U$, is proportional to $l$, or $J_z/U = (l+ax)/\omega$. Calculating the time averaged ratio $J_z/U$ for different $l$ and plotting the results gives the points in figure 7.12. Note that our numerical calculations are not an integration in a volume but a summation on a surface. The linear dependence should however be preserved.

The result clearly shows a linear dependence, at least when we consider either positive or negative $l$. For some reason negative values of $l$ do not give negative values of $J_z/U$ as expected and therefore this requires some further investigations. The sign of $l$ seems to be undetectable. We can also see that the difference between right and left-handed polarization is very small. Looking at the equation $J_z/U = (l+ax)/\omega$ one would expect that the difference between the two polarizations would be on the same order as the difference between two consecutive $l$ values. Ideally, $ax = \pm 1$ in whole the beam but generally that is only true in the direction where the beam is directed, here $\theta, \varphi = 0$, but the magnitude of the mean value of $ax$ is still about 0.55 to 0.61 in the beam. So one could expect that for $l = 0$, the difference in $J_z/U$ between right and left-hand polarization would be $2ax$, about the same as the difference in $J_z/U$ between for instance $l = 0$ and $l = 1$ for one of the polarizations. Since this is not the case we are not sure whether the orientation of the polarization is detectable with this method. It looks like what we have is something like $J_z/U = (|l| + |ax|)/\omega$.

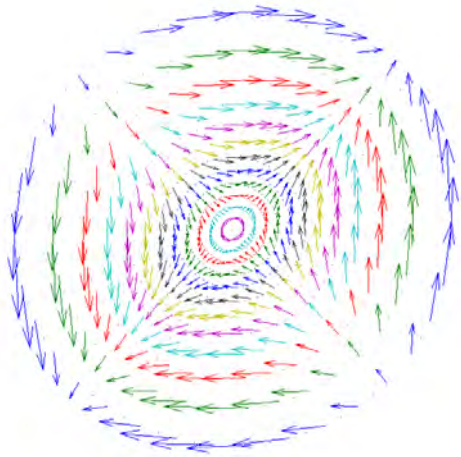

(a) $l = 1$

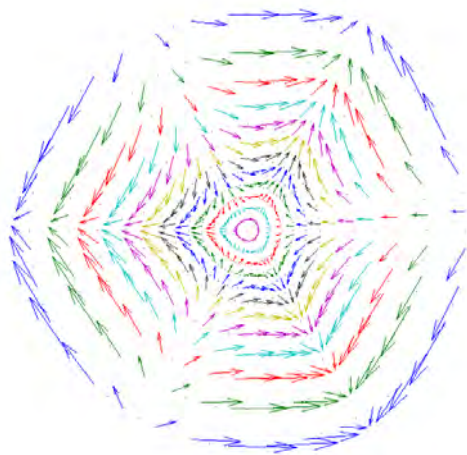

(b) $l = 2$

Figure 7.7: Electric field vectors in left-hand polarized $l = 1$ and $l = 2$ beams projected onto a plane normal to the $z$ axis. Fields created with a five-element antenna array.

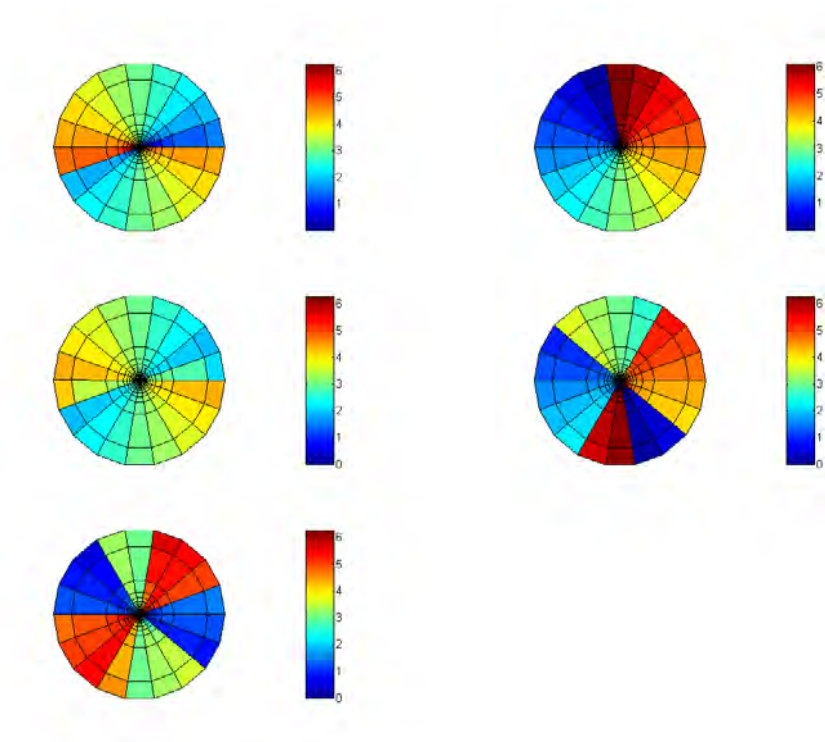

Figure 7.8: Phases of left-hand polarized beam produced by a five-element antenna array. Eigenvalues $l = -2$, $l = -1$, $l = 0$, $l = 1$ and $l = 2$ from top left. Value of phases are in radians going from 0 (blue) to $2\pi$ (red).

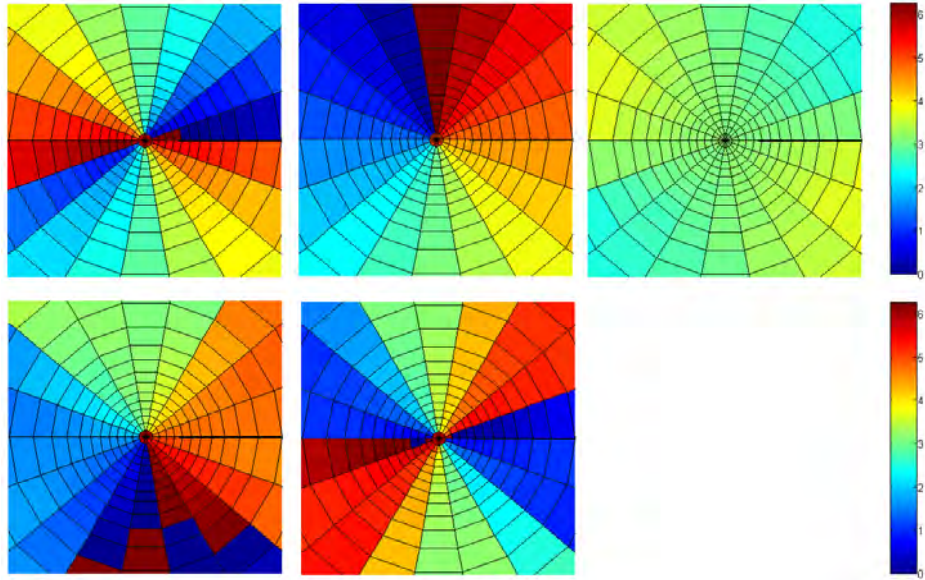

Figure 7.9: Phases of left-hand polarized beam produced by a five-element antenna array. Eigenvalues $l = -2$, $l = -1$, $l = 0$, $l = 1$ and $l = 2$ from top left. Value of phases are in radians going from 0 (blue) to $2\pi$ (red). Largest $\theta = 65°$.

## 7.4.2 Right-hand polarization

To obtain right-hand polarization we calculate the currents using $ax = -1$. We once again start with, for $l = 1$ and $l = 2$, plotting the electric field vectors in a plane normal to the $z$ axis in figure 7.13. The angular distance between the rings is as before 5° and the limits are $\theta = 60°$. Again, the length of vectors in different rings are scaled with different constants to prevent overlapping. Just like for left-handed polarization the vectors complete $l$ rotations when going around the beam but they now rotate in counterclockwise direction instead of clockwise (when moving in positive $\varphi$ direction). In figure 7.14 and 7.15 the phases are plotted. Once again we see that the phase behavior is good for not too big $\theta$. If we compare the results for left- and right-handed polarization we see that the direction of rotation for the phase fronts are opposite. For right-hand polarization the phase increases when $\varphi$ increases for positive $l$, whereas the phase decreases

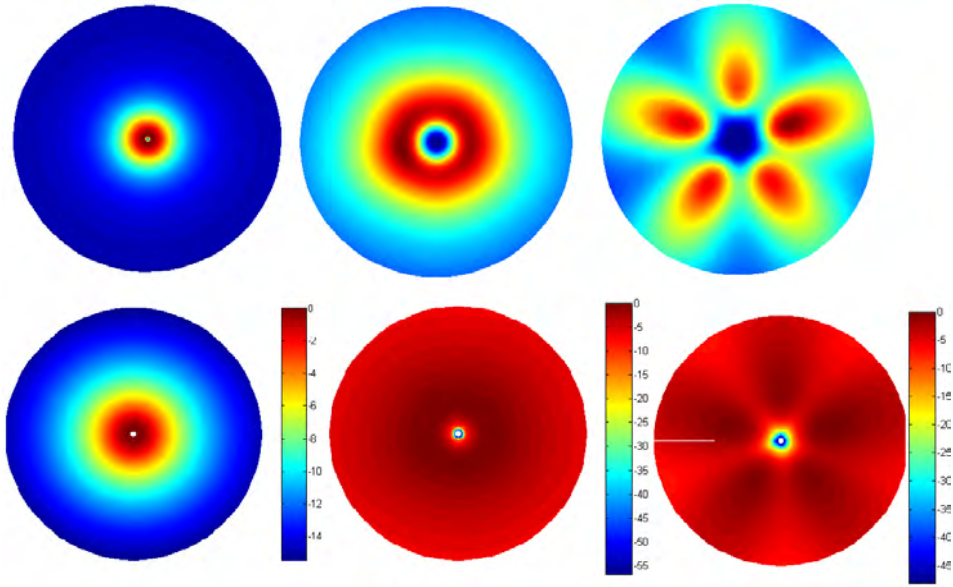

Figure 7.10: Intensity patterns for a five-element antenna array. Eigen-values $l = 0$ (left), $l = 1$ (middle) and $l = 2$ (right). Linear scale in the top row and dB scale in the bottom row.

with increasing $\varphi$ for left-hand polarization.

The intensity patterns turn out to be very similar to the left-hand polarized wave intensity patterns so they are not presented here.

The results for $J_z/U$ are presented in figure 7.12 and are discussed together with the left-hand polarized beam.

## 7.5  Ten-element crossed dipole array

By using ten instead of five elements we can, according to equation (7.1), use $l$ numbers up to four. Now we consider a circular array with radius $a = \lambda$ situated $\lambda/10$ over ground. Once again the frequency is 300 MHz and we aim the beam in the direction where $\theta$ and $\varphi$ both equals zero. We use the same equations as for the five-element array to obtain the currents and phases.

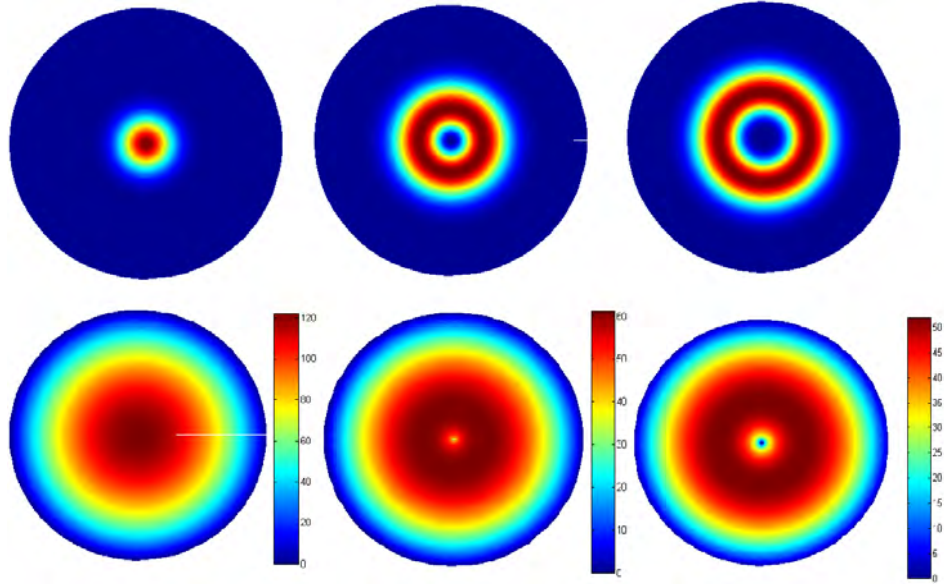

Figure 7.11: Intensities of an LG beam. Eigenvalues $l = 0$ (left), $l = 1$ (middle) and $l = 2$ (right). Linear scale in the top row and dB scale in the bottom row.

### 7.5.1 Left-hand polarization

Once again we start with $ax = 1$ to obtain a left-hand polarized beam and examine the phase behavior. The results are presented in figure 7.16. The pattern looks as expected. However, we can see that there are discontinuities in most of the figures. This is because the increase of the array radius, $a$, which has resulted in a narrower main lobe and more pronounced minor lobes. The phase in these minor lobes are, as we can see, shifted in the $\varphi$ direction.

By making the array radius smaller (here equal to $\lambda/2$) we can decrease the minor lobes and increase the beamwidth of the main lobe. In figure 7.17 we see that the phase behavior is much better. Note that these phases belong to right-hand polarized beams.

In figure 7.18 the intensity patterns of beams produced by the ten-element array are displayed. The patterns are as expected and as $l$ increases we see that the ring radii increase. We notice that the pattern for the $l = 2$ beam is much better than the one obtained from the five-element array. When $l$ increases we see that the intensities change from rings to spots. Once again, this is most likely an effect caused by the limited number of antennas.

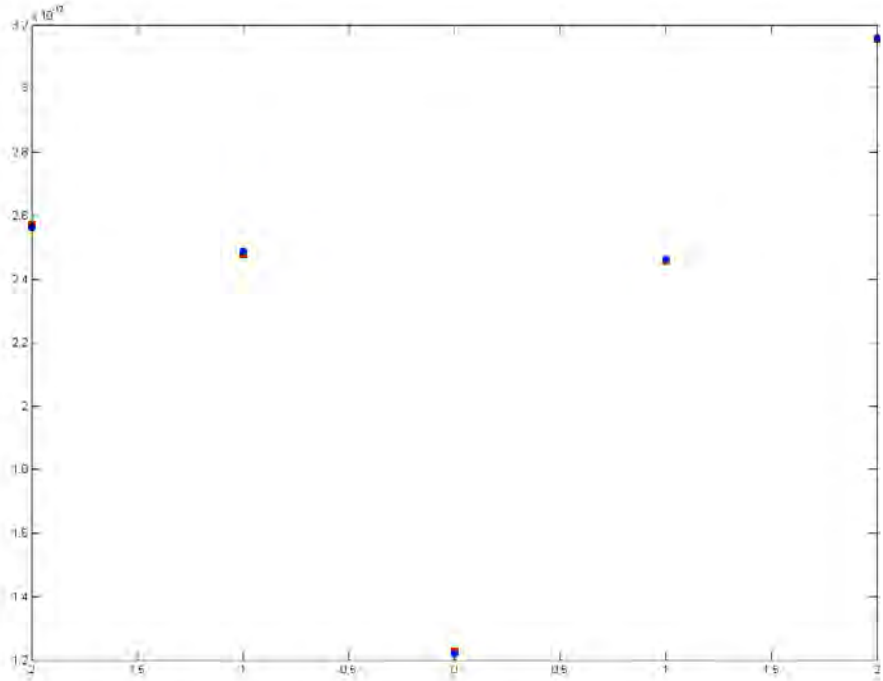

Figure 7.12: The ratio $J_z/U$ in beams with different eigenvalues $l$ produced by a five-element antenna array. The left-hand polarized wave is marked in red; right-hand polarized wave is marked in blue.

We plot the ratio $J_z/U$ in figure 7.19. Once again we see that there seems to be a linear dependence. However we can once again note that there appears to be problems separating the polarizations and the signs of $l$. To us, a dependence $J_z/U = |(|l| + |ax|)/\omega|$ seems to hold.

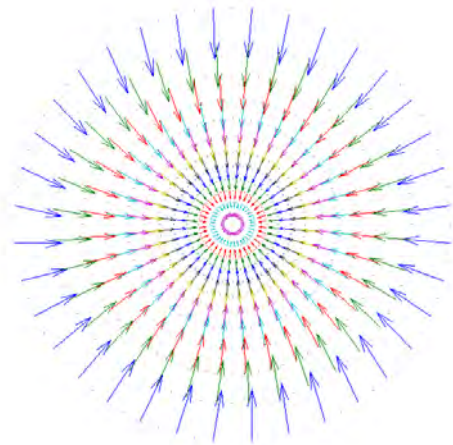

(a) $l = 1$

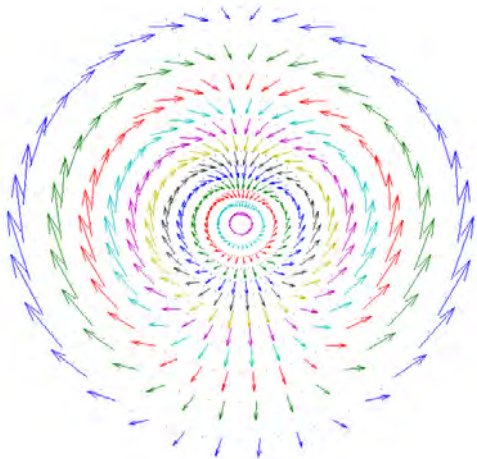

(b) $l = 2$

Figure 7.13: Electric field vectors in right-hand polarized $l = 1$ and $l = 2$ beams projected onto a plane normal to the $z$ axis. Fields produced by a five-element antenna array.

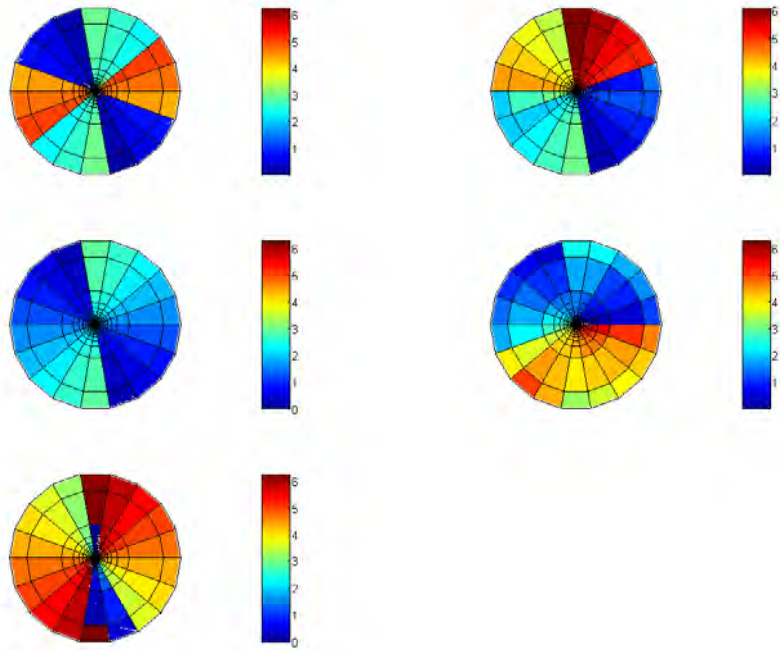

Figure 7.14: Phases of right-hand polarized beam produced by a five-element antenna array. Eigenvalues $l = -2$, $l = -1$, $l = 0$, $l = 1$ and $l = 2$ from top left. Value of phases are in radians going from 0 (blue) to $2\pi$ (red).

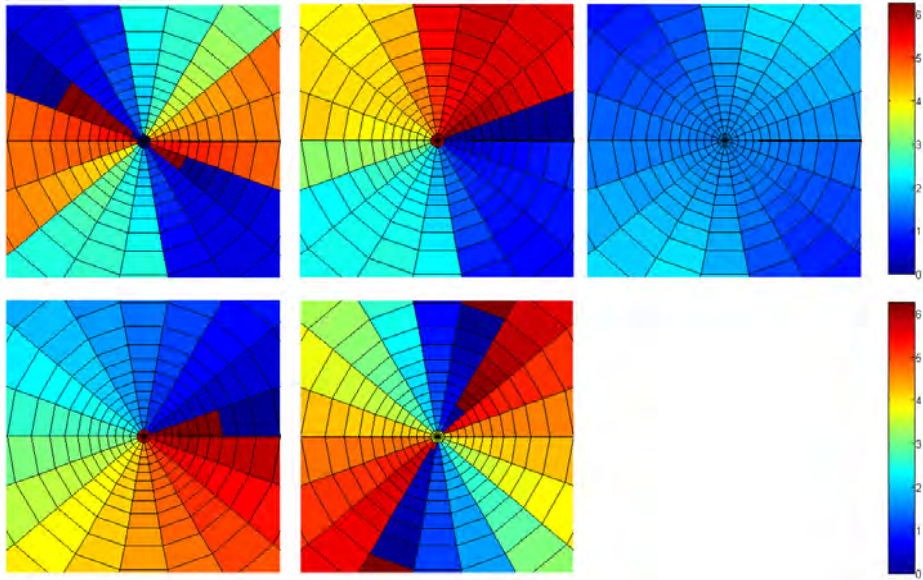

Figure 7.15: Phases of right-hand polarized beam produced by a five-element antenna array. Eigenvalues $l = -2$, $l = -1$, $l = 0$, $l = 1$ and $l = 2$ from top left. Value of phases are in radians going from 0 (blue) to $2\pi$ (red). Largest $\theta = 65°$.

### 7.5.2 Right-hand polarization

To obtain right-hand polarization, we use $ax = -1$ in equations 6.10 to 6.12 to find the currents.

As for the five-element array the results looks similar to the left-hand polarized beam (except the reverse rotations in the phase discussed in the five-element array section) and the only result we present in this report is therefore the ratio of $z$ directed angular momentum to energy found in figure 7.19 and the phase pattern from an array with radius $\lambda/2$ in figure 7.17.

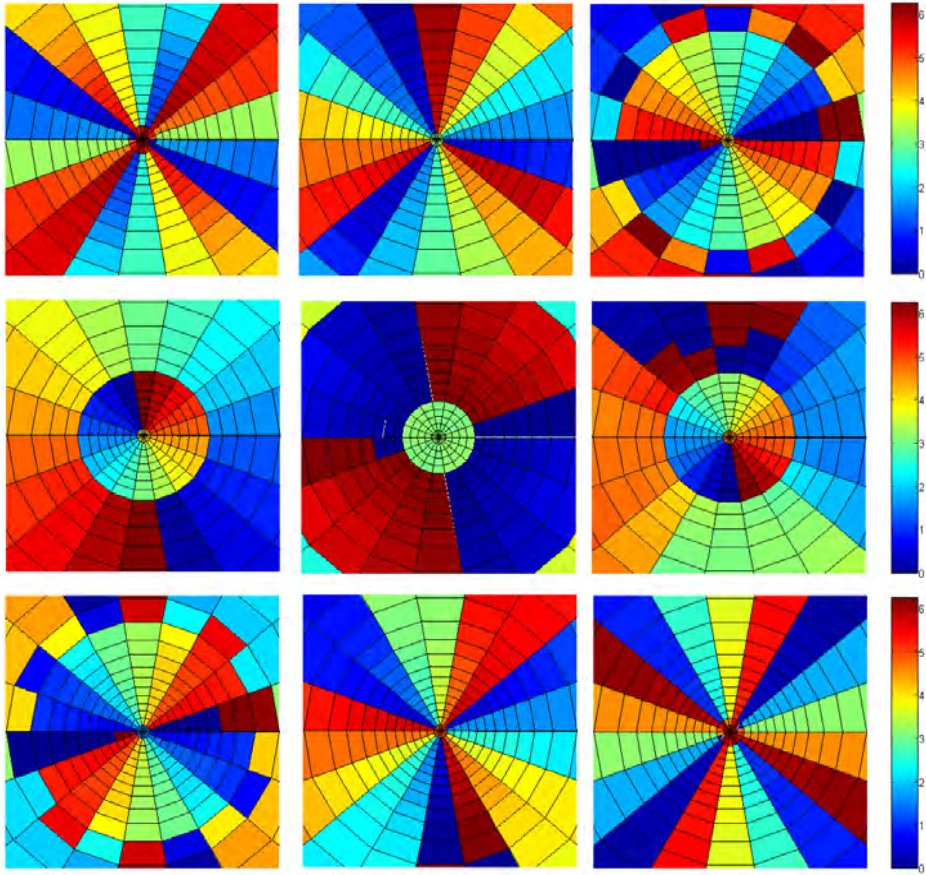

Figure 7.16: Phases of left-hand polarized beams produced by a ten-element antenna array. Eigenvalues going from $l = -4$ (top left) to $l = 4$ (bottom right). Value of phases are in radians going from 0 (blue) to $2\pi$ (red). Largest $\theta = 65°$.

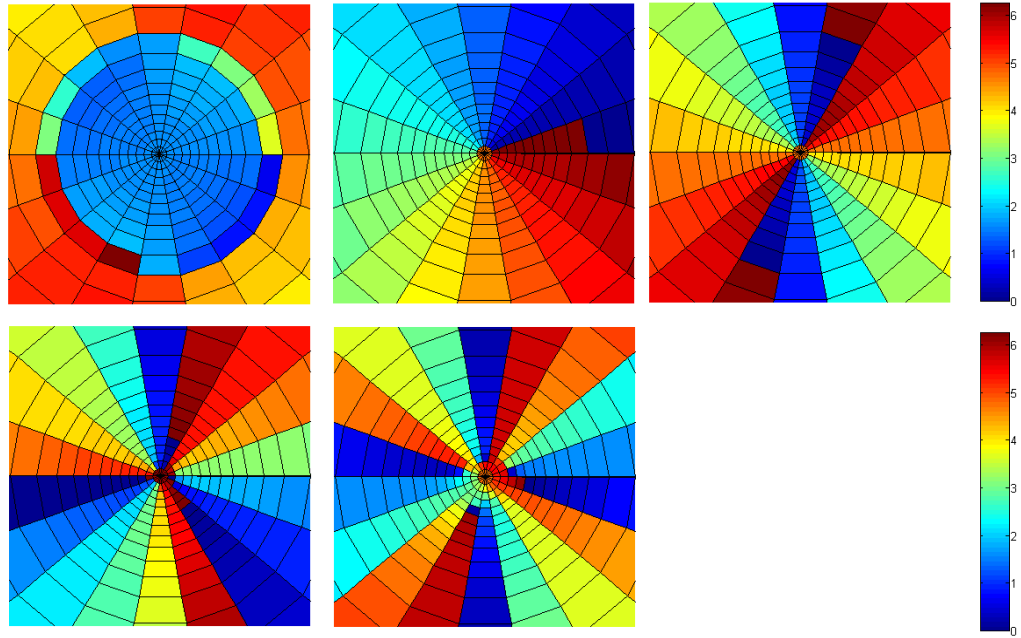

Figure 7.17: Phases in right-hand polarized beams with eigenvalues going from $l = 0$ to $l = 4$ produced by a ten-element antenna array with radius $a = \lambda/2$. The value of the phases going from 0 (blue) to $2\pi$ (red). Largest $\theta = 65°$.

## 7.6 Five-element tripole array

When one needs to point the beam in a direction other than the $z$ axis, a tripole antenna is preferred. The tripole has namely the advantage that the orthogonal antenna currents can be viewed as the $x$, $y$, and $z$ components of an antenna current vector **j**. By choosing appropriate values of these three current components, the resulting **j** can therefore be pointed along any given direction in space. Hence, a tripole antenna with short elements approximates an electric dipole moment vector **p** which can be pointed in any direction. To transmit beams with different $l$ values, the tripole is absolutely essential since we cannot control the direction of the beam with dipoles or crossed dipoles which are fixed in space along a line or in a plane and still maintain the wanted $l\varphi$ phase dependence because the phase difference between the elements is used to direct the beam! This is of course provided that we do not physically move the antennas which would be extremely unpractical. Since we want the $l\varphi$ phase dependence in the beam in a plane normal to the beam symmetry axis, we introduce a coordinate system where $z'$ is directed in the $(\theta_0, \varphi_0)$ direction and $\varphi'$ circles in a plane normal to $z'$. The desired phase dependence can then be written as $l\varphi'$.

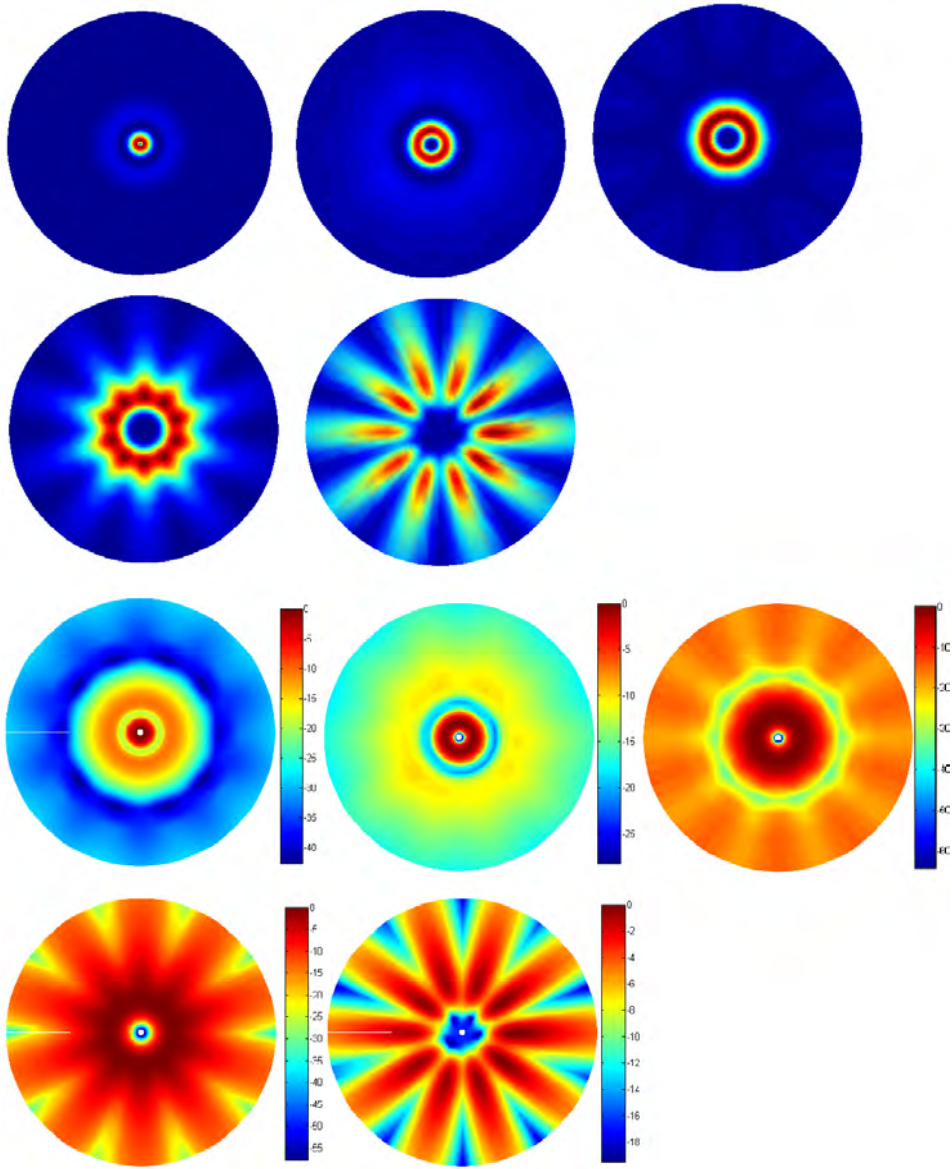

Figure 7.18: Intensity patterns produced by a ten-element antenna array. In the two top rows the scale is linear. Plotted are beams with eigenvalue $l$ going from $l = 0$ (first row, left) to $l = 4$ (second row, middle). In the two bottom rows the scale is in dB with the placing corresponding to the top rows.

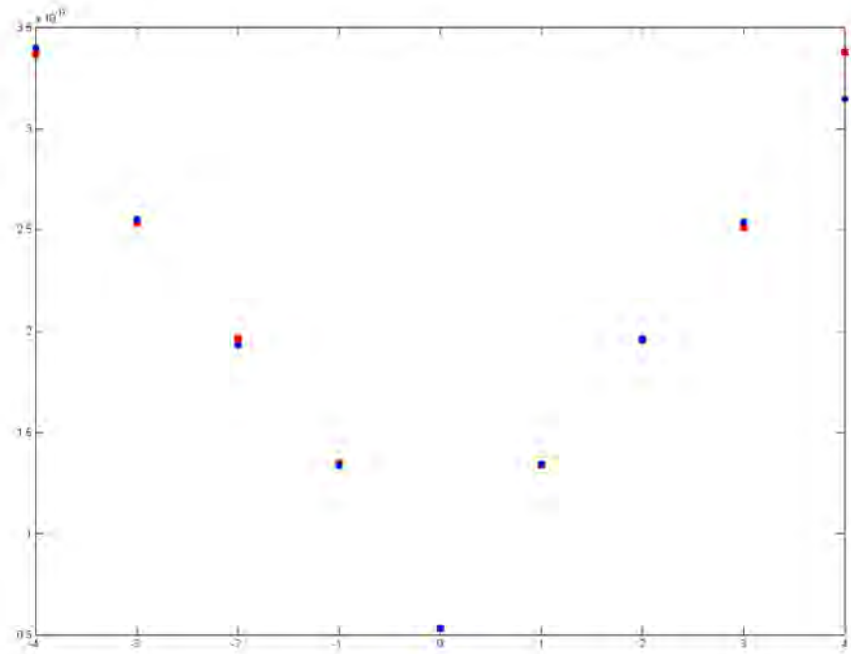

Figure 7.19: The ratio $J_z/U$ in beams with different eigenvalues $l$ produced by a ten-element antenna array. Left-hand polarized wave is marked in red; right-hand polarized wave is marked in blue.

### 7.6.1 Tripole compared to crossed dipole

We start by comparing the five-element crossed dipole array to the tripole. We choose to point a left-hand circularly polarized beam in the direction ($\theta_0 = 45°$, $\varphi_0 = 0°$). The only possible $l$ value is, as mentioned, zero. In figure 7.20 we have compared the performance of three different configurations; the five-element crossed dipole array, one single tripole antenna and an array of five tripole antennas. The beam from the crossed dipole array is pointed in the direction ($\theta_0, \varphi_0$) with the phase excitation

$$\alpha_n = -ka_n \sin\theta_0 \cos(\varphi_n - \varphi_0) \tag{7.2}$$

where $\alpha_n$ is the phase excitation in the $n$th element orientated at an angle $\varphi_n$. For the tripoles we use the same equation for the phase difference between the elements but we also point *each element* according to equations (6.10) to (6.12).

What we see is that the crossed dipole array is not even capable of producing the maximum gain in the desired $\theta$ direction! The maximum gain, which is 4.1 dB, is instead in the $\theta = 26°$ direction while in the desired direction the gain is 3.5 dB. The tripole can, as we see in figure 7.20b point the beam in the desired direction using only one element. To obtain higher gain and to be able to have an $l\varphi'$ phase dependence in the beam we use an array of several antennas. With the five-element tripole array we see that we still have maximum radiation in the desired direction and the gain is now 5.0 dB. Our conclusion is that the tripole antenna is superior to the crossed dipole when we want to point the beam in directions other than vertical ($\theta_0 = 0°$).

### 7.6.2 Beam with orbital angular momentum produced by tripole antennas

Using a five-element array with tripoles we are able to create beams with different $l$ directed in other directions than along the $z$ axis. To do this we use the program tripole.m to calculate currents and phases in the elements. The currents and phases in each tripole is first obtained by using equations (6.10) to (6.12). Then we use the equation

$$\alpha_n = l\varphi_n - k a_n \sin\theta_0 \cos(\varphi_n - \varphi_0) \tag{7.3}$$

to obtain the phase excitation in the $n$th element. We then add this phase excitation to each phase obtained with the equations (6.10) to (6.12). This will give us a beam with the symmetry axis $z'$ in the $(\theta_0, \varphi_0)$ direction and an $l\varphi'$ phase dependence where $\varphi'$ circles in a plane normal to $z'$.

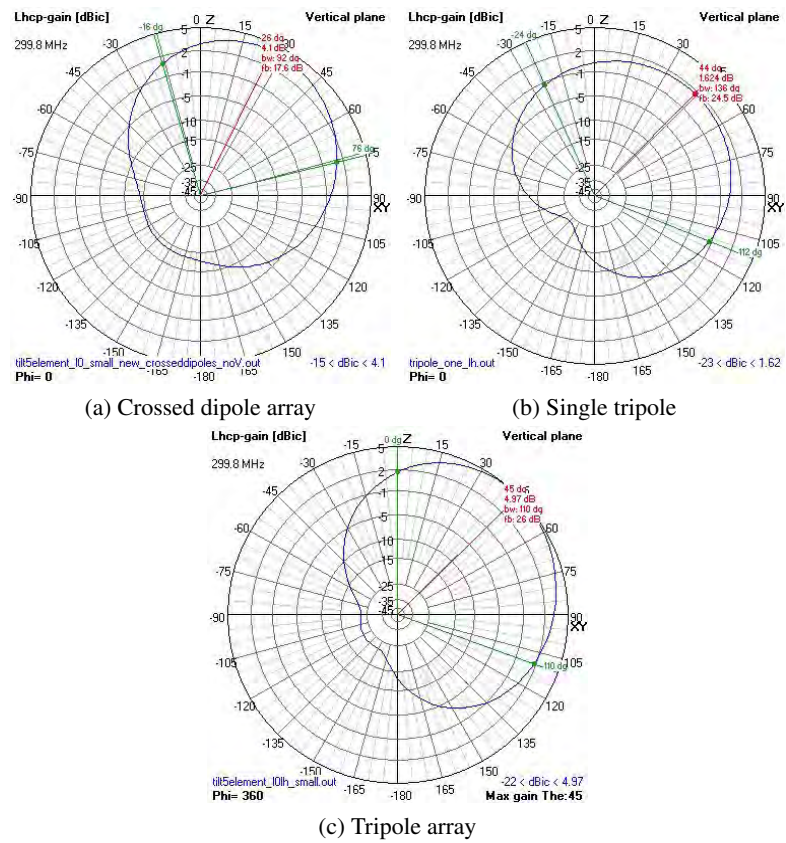

(a) Crossed dipole array

(b) Single tripole

(c) Tripole array

Figure 7.20: Left-hand gains for a five-element crossed dipole array, one single tripole and a five-element tripole array. Each configuration is pointed so that maximum radiation should occur in the ($\theta_0 = 45°$, $\varphi_0 = 0$) direction. All beams are with eigenvalue $l = 0$.

We have performed these tests for two different directions, one where $\theta_0$ is small and one with a larger $\theta_0$.

In figure 7.21 we see the results from the case where the beam symmetry axis is directed in the $(\theta_0 = 10°, \varphi_0 = 40°)$ direction. The patterns show the gain for the desired polarization. As we can see, the pattern is very similar to the patterns shown in figures 7.2a to 7.2c but the beams are directed in an other direction. Note that the difference in similarities are also because of the difference in the basic shape of the arrays, a better comparison would be with the patterns created with the five-element crossed dipole antenna but they are not presented in this report. If we compare figure 7.21c and 7.21d, we see that the pattern is more symmetric in free space than when using a ground plane.

The program for plotting the phase in the beam is designed for beams pointed along the $z$ axis. So, from a practical point of view, other directions are more complicated because of how the data in NEC output files are sorted. But since $\theta_0$ is rather small we can use the program to look at a good approximation of the phase behavior in the beams. The result is presented in figure 7.22. We find that the phase behaves just as it should in all beams and we can expect them to carry orbital angular momentum.

If we point our beam with a larger angle between $z$ and $z'$ we see that the resulting patterns become more unsymmetric, especially when using a ground plane. In figure 7.23 we can see the left-hand gain patterns for beams pointed in the $(\theta_0 = 45°, \varphi_0 = 0°)$ direction. Now we find that the ground effects are larger for all beams, and even in free space the pattern does not look good at all for the beam with eigenvalue $l = 2$.

Our conclusion is that there are two main reasons why the beams become more unsymmetric when $\theta_0$ increases. First of all, there are effects caused by the ground plane, but we have seen that even in free space the beams are deformed. If we study the array factor we find the other reason. In a beam pointed normal to the array, the array factor is, with enough elements, symmetric in each $\varphi$ plane. That means that if we move an angle $\mathrm{d}\theta$ from $\theta_0$ in positive or negative direction the array factor will be equal in those two points. In mathematical terms

$$AF(\theta_0 + \mathrm{d}\theta, \varphi) = AF(\theta_0 - \mathrm{d}\theta, \varphi), \tag{7.4}$$

for all $\varphi$.

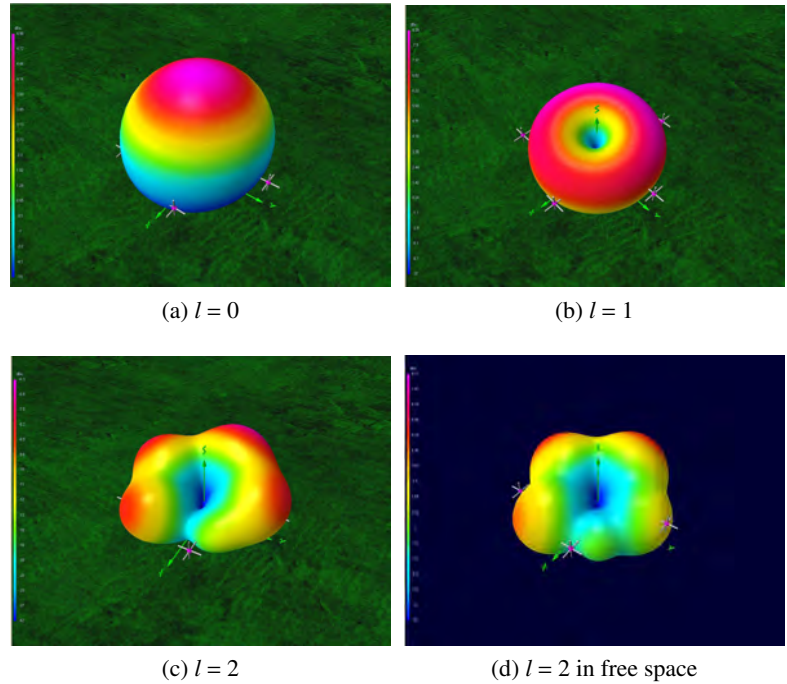

(a) $l = 0$    (b) $l = 1$

(c) $l = 2$    (d) $l = 2$ in free space

Figure 7.21: Left-hand gains of beams produced by a five-element tripole array pointed in the $(\theta_0 = 10°, \varphi_0 = 40)$ direction. The beam with eigenvalue $l = 2$ is shown with and without ground plane.

If we point the beam in any other direction (except multiples of $\theta_0 = 90°$) the symmetry condition in equation (7.4) is generally not fulfilled, and therefore the beams will not be as symmetric. The symmetry gets worse as $\theta_0$ increases. In figure 7.24 this can be seen for a five-element circular array pointed in three different $\theta_0$ directions. In figure 7.24b we can see that for $\theta_0 = 30°$ the pattern is reasonably symmetric even for large d$\theta$, but when $\theta_0 = 45°$ the beam become less symmetric. These plots are for beams without an $l\varphi$ phase dependence. With $l$ greater than zero, $AF$ is more complicated and therefore more sensitive to changes in $\theta$.

The effect of the unsymmetric array factor can be seen if we look at our beams produced by tripole antennas in figure 7.20b and 7.20c. We see that the beam from the array is not as symmetric as the one from the single tripole antenna.

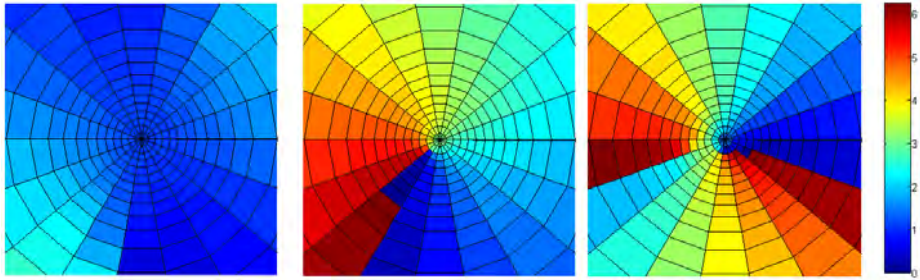

Figure 7.22: Phase pattern approximations in beams pointed in the ($\theta_0 = 10°$, $\varphi_0 = 40°$) direction. Shown are beams with eigenvalues running from $l = 0$ (left) to $l = 2$ (right). The values of the phases are in radians running from 0 (blue) to $2\pi$ (red). Largest $\theta = 65°$.

Another reason for the beam deformation is that when looking at the circular array from a direction other than its normal it changes to an ellipse and therefore the beam transmitted in that direction is not circularly symmetric. However, by using an elliptic array in our numerical simulations and examining the beams created we have not noticed any significant difference. The elliptic array had the shape of an circle projected onto a plane at an angle 45° from the circle.

## 7.7 Ten-element tripole array

Just as for the crossed dipole array we double the number of elements in the array so that we have ten instead of five. We use $ax = 1$ to obtain left-hand polarization and calculate the currents and phases using the program tripole.m. In figure 7.25 we can see the left-hand gains from beams pointed in the ($\theta_0 = 45°$, $\varphi_0 = 0°$) direction from two different ten-element arrays with different radii; one where $a = \lambda/5$ and one where $a = \lambda/2$.

As expected, we find that using more elements in the array improves the pattern. For $l = 1$, the symmetry in the beam is improved a bit compared to the beam in figure 7.23d, especially when using the larger array radius. For $l = 2$ there is a significant difference. When using the larger array, the pattern looks fairly symmetric and we find a minimum in the desired direction which was not the case in the five element array pattern shown in figure 7.23f. We even find that for $l = 3$ the pattern from the ten-element array is better than from the five element array with $l = 2$.

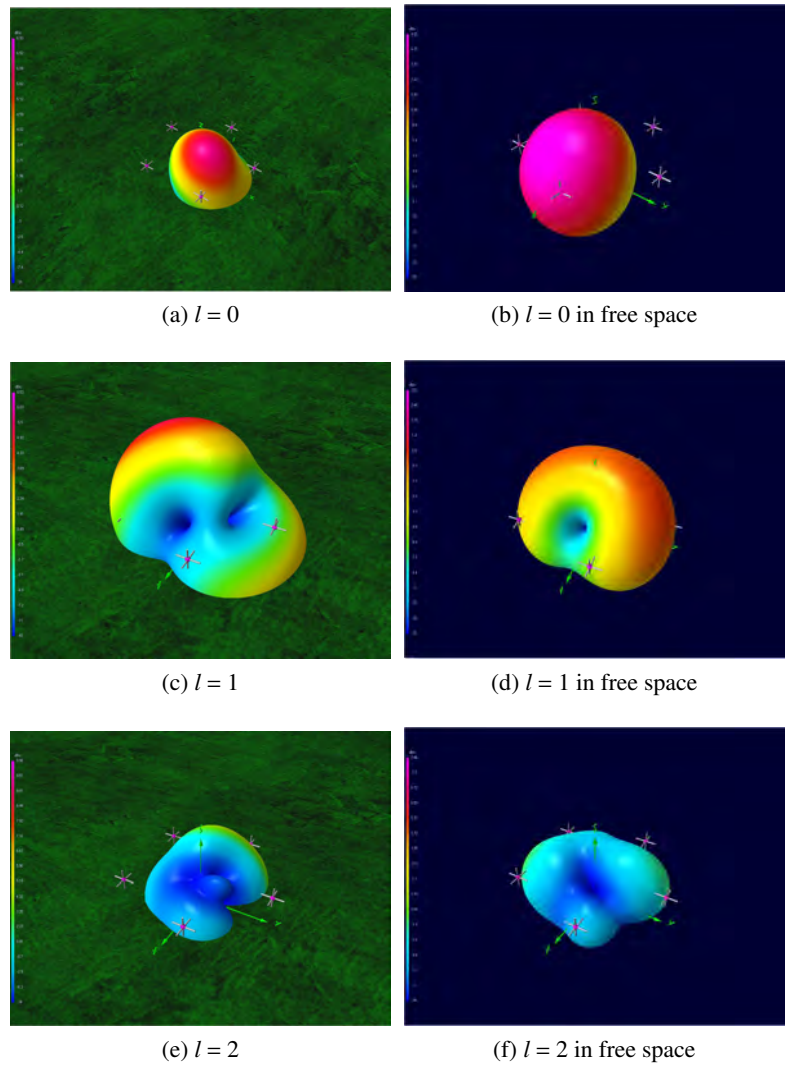

(a) $l = 0$          (b) $l = 0$ in free space

(c) $l = 1$          (d) $l = 1$ in free space

(e) $l = 2$          (f) $l = 2$ in free space

Figure 7.23: Left-hand gain of beams from a five element tripole array pointed in the ($\theta_0 = 45°$, $\varphi_0 = 0$) direction. All beams are shown with and without ground plane.

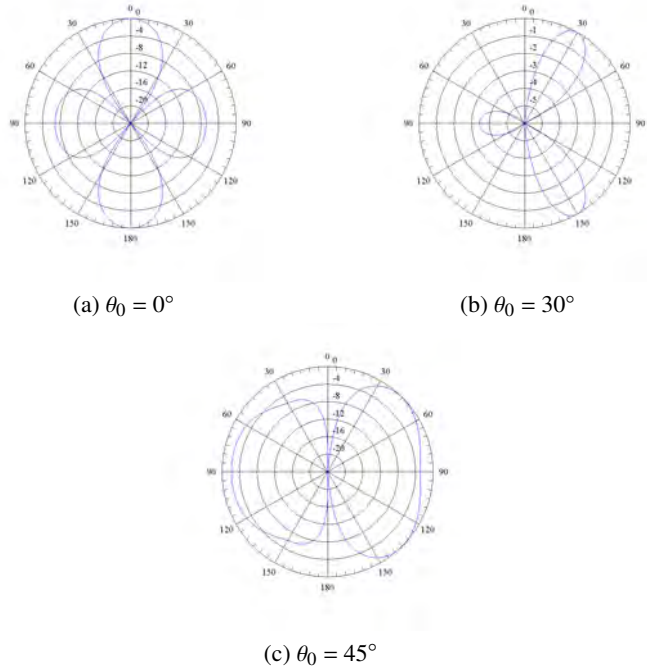

(a) $\theta_0 = 0°$

(b) $\theta_0 = 30°$

(c) $\theta_0 = 45°$

Figure 7.24: Directivity for a circular five-element array calculated for three different $\theta_0$. The angle $\varphi$ equals zero in all plots.

If we point our beam with a smaller angle $\theta_0$ the pattern should be more symmetric. In figure 7.26 the pattern for a beam with $l = 4$, which is the worst case scenario beam, is shown together with its phase pattern. Again, the program for plotting the phases are designed for beams pointed in the ($\theta_0 = 0°$, $\varphi_0 = 0°$) direction, but for small $\theta_0$ we can use it as an approximation.

What we find is that the pattern is deformed, but the basic shape is still visible and the minimum is in the right direction. From figure 7.26b we find that there is an $l\varphi$ phase dependence in the beam, which, considering that $\theta_0$ is small, indicates an $l\varphi'$ phase dependence.

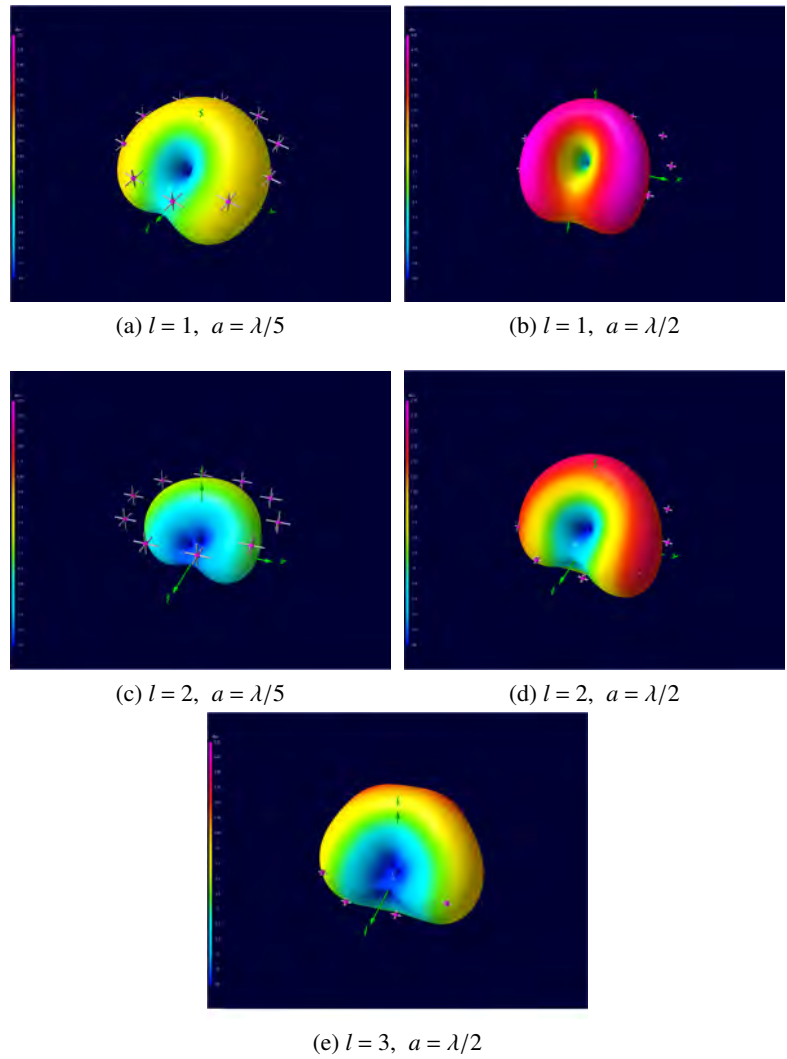

(a) $l = 1$, $a = \lambda/5$

(b) $l = 1$, $a = \lambda/2$

(c) $l = 2$, $a = \lambda/5$

(d) $l = 2$, $a = \lambda/2$

(e) $l = 3$, $a = \lambda/2$

Figure 7.25: Left-hand gain of beams from two different ten-element tripole arrays pointed in the ($\theta_0 = 45°$, $\varphi_0 = 0°$) direction. All patterns are evaluated in free-space.

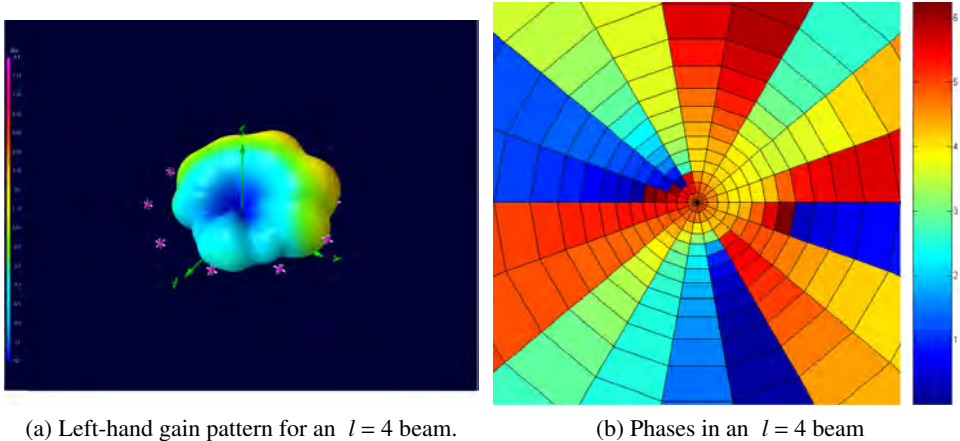

(a) Left-hand gain pattern for an $l = 4$ beam.　　(b) Phases in an $l = 4$ beam

Figure 7.26: Left-hand gain and phase pattern in a beam with $l = 4$ pointed in the ($\theta_0 = 15°$, $\varphi_0 = 0°$) direction. The beam is created using a ten-element tripole array in free-space. Values of phases are in radians going from 0 (blue) to $2\pi$ (red).

## 7.8 The LOIS Test Station

Since the 1960's the basic system for collecting radio signals has been the dish antenna. Dish antennas are large and bulky, requiring a lot of material and mechanical engineering, and to point them in different directions one has to physically rotate or move them, a slow process likely to cause loss of unexpected transient signals. This makes dish antennas unpractical and very costly. One solution to this would be to use an array of many simple antennas instead, and this is precisely what is being done in LOFAR, Low Frequency Array, the world's largest low-frequency (10–250 MHz) radio telescope [38]. The cost for LOFAR and similar radio telescopes comes mainly from electronics, and the pointing is done in software by changing the phase and amplitude of the sampled analytic signals from each individual antenna element. Another advantage, with the use of many simple elements, is that the telescope becomes modular and therefore can be expanded in an incremental manner.

LOFAR was developed to take the necessary step to increase the sensitivity for astronomical observations at radio frequencies below 250MHz. It uses simple semi-omni-directional antennas, in the form of standard crossed dipoles. The signals are digitized and sent to a central processor. There they are combined so the array will work as one huge antenna. The aggregate amount of data is around 20 Tbits/s. The array consists of a central site and 100 antenna fields, placed in a logarithmic spiral array, as the one shown in figure 6.1d. The design will, when finished in the 2008–2009 time frame, contain 25 000 antennas and have a total diameter of 350 km.

It is in connection to LOFAR that the LOIS project [39], LOFAR Outrigger in Scandinavia, started in 2000. After invitation from LOFAR, LOIS was initiated by Swedish physicists with experience within similar digital radio projects for space studies. LOIS will complement LOFAR and together they will be able to work as an advanced deep space radar. The LOIS project is planned to have up to a few thousand antennas.

There is one LOIS test station up and running in Risinge, near the university city of Växjö in southern Sweden. Additional stations are being planned in southern Sweden and in Poland. The Växjö area has a suitable geographical location with low population density, ample space for antenna arrays, little auroral and radio interference and a well developed communications infrastructure. The test station is connected with a high-bandwidth connection to the Swedish University Network and henceforth to the Internet. In the LOIS project tripole antennas are used, instead of the crossed dipoles used in the LOFAR array, giving access to the full three dimensional behavior of the field vector. The capabilities of the LOIS antennas therefore form a true superset to those of LOFAR thus allowing all types of LOFAR measurements to be made by LOIS; the *vice versa* is not true. At the time of writing this thesis the LOIS test station has six of these tripole antennas placed in one ring. The location of each element is given by the plane polar coordinates $\varphi_m = 30° + 60°m$ and $r_m$, the coordinates for each element as shown in figure 6.1a, where $m = 1$ since we consider one ring. Each tripole element is rotated as shown in figure 7.27.

To a good approximation, the three orthogonal elements of the 3D LOIS tripole constitute the $x$, $y$ and $z$ components of an electric dipole vector $\mathbf{p}$ that can, electronically, be rotated into any desirable direction. This means that the field generated by each tripole will correspond to the field generated by one "real" dipole. Technically, this is achieved by matching the amplitude and phase for each axis in each of the tripoles. Just as easy, a EM-field with polarization can be created and pointed in different directions. In figure 7.28a one can see the LOIS station's electric field gain, when we have chosen amplitude and phase, ef-fectively have a moment vector, **p**, in the $xy$ plane (as for the LOFAR antennas) directed along the $x$ axis. In figure 7.28b the same dipoles are presented but we have a relative phase offset between the elements that would correspond to $l = 1$, generating a field with the characteristic minimum along the axis of propagation, the $z$ axis.

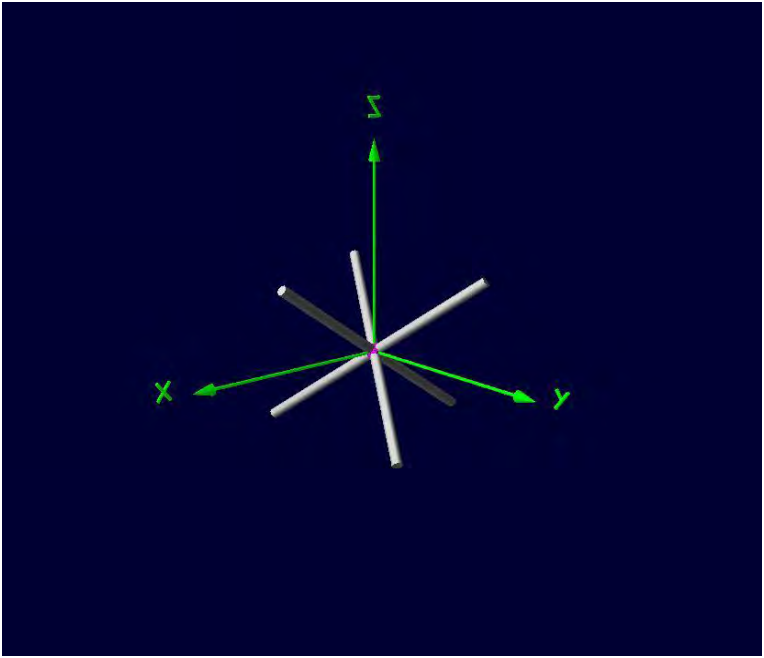

Figure 7.27: The LOIS tripole consists of three mutually orthogonal, short, active electric dipoles with co-located midpoints (feed points). Each tripole structure is mounted in a configuration that is fully symmetric relative to the ground, with the center of the structure roughly 2.5 m above ground.

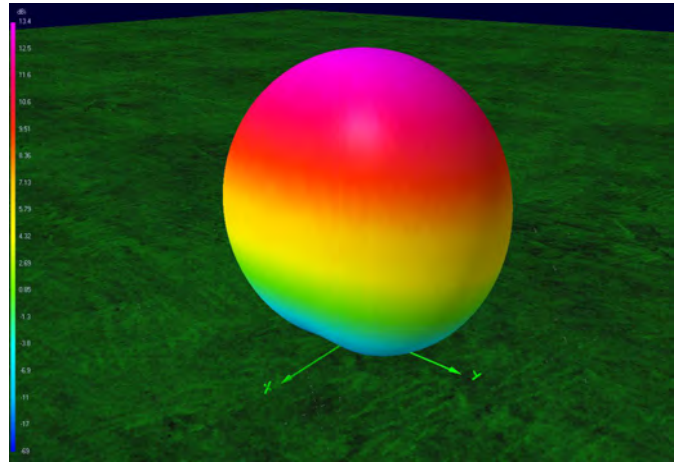

(a) $l = 0$

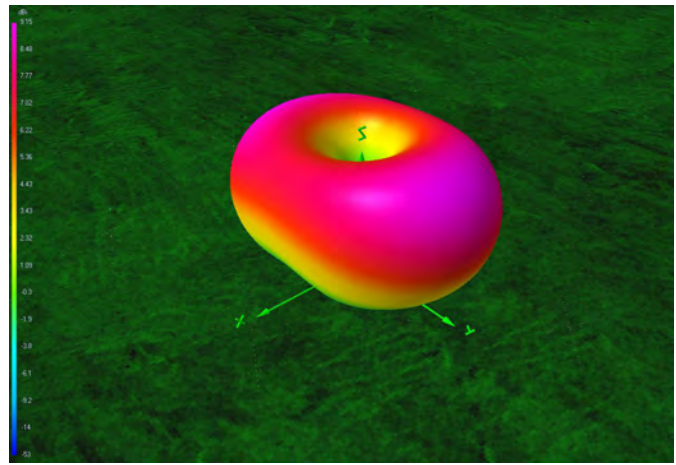

(b) $l = 1$

Figure 7.28: Antenna patterns calculated for the LOIS Risinge station, a six-element tripole test array. The elements have such a internal relative phase relationship that they comprise a single electric dipole moment vector, **p**, in the $xy$ plane, directed along the $x$ axis. In the first figure, all the elements have the same phase. In the second figure, the phase difference between the elements corresponds to $l = 1$. This generates the characteristics of a beam carrying OAM.

# 8

# DISCUSSION

## 8.1 Radio astronomy and space physics applications

The technique of applying OAM to radio beams opens up for some very interesting and powerful applications. First of all, the antenna patterns produced can be useful because of their basic shapes. If, for example, we want to observe the solar corona but not the Sun itself, the ring pattern from a beam with OAM is ideal since the intensity minimum in the center of the beam could be placed over the sun and the rest of the beam over its corona. The pattern of a beam with OAM shown in figure 7.1 has two different beamwidths. First, we have the inner width, which is characterized by the angle between two points of equal gain located inside the maximum radiation angle. The other width is the angle between two points of equal gain located outside the maximum radiation angle. Using these definitions in figure 8.1, one observes that the two half-power beamwidths, $HPBW$, are $HPBW_l^{\text{inner}} = 36°$ and $HPBW_l^{\text{outer}} = 124°$, representing inner or outer widths, respectively, for a beam eigenvalue $l \neq 0$.

In a solar radio or radar experiment [34, 35] where we want to have $HPBW_l^{\text{inner}}$ at the solar surface, we need to find an array which can produce an $l \neq 0$ beam with $HPBW_l^{\text{inner}} = 0.53°$ (if our array is situated 1 AU from the Sun). Comparing our patterns in previous chapters, we find that $HPBW_1^{\text{inner}} \approx 3/4 \times HPBW_{l=0}$. This means that the array size needed is about the same size as an array which can produce an $l = 0$ beam with $HPBW = 4/3 \times HPBW_1^{\text{inner}} = 0.71°$. If, for simplicity, we assume that the antenna array is a $M \times N$ square, we can approximate the array size with that of linear, uniform, broadside array for which the $HPBW$ can

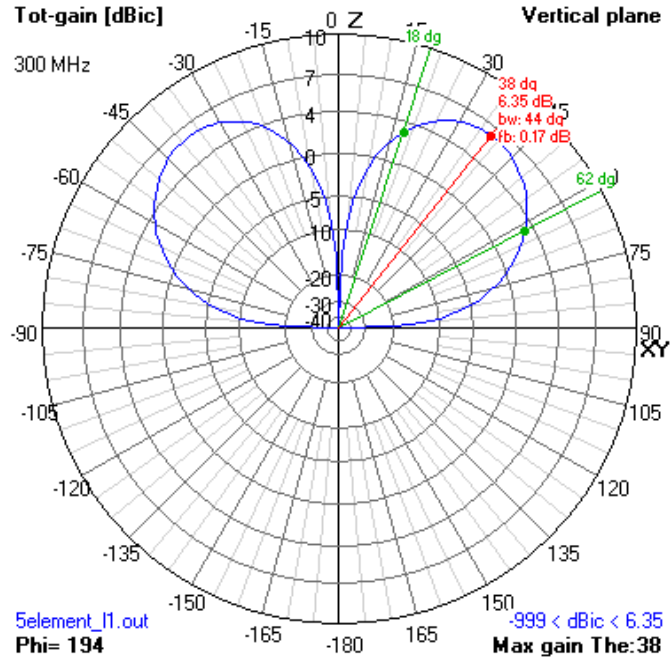

Figure 8.1: Half-power beamwidths of a beam with eigenvalue $l = 1$.

be calculated from the formula [5]

$$HPBW = 2\left[\pi/2 - \cos^{-1}\left(\frac{1.391\lambda}{\pi Nd}\right)\right] \tag{8.1}$$

where $\lambda$ is the wavelength used, $d$ the distance between elements and $N$ is the number of elements. With $HPBW = 0.71°$, we find that $Nd = 71\lambda$. In other words, the size of the array is equal to 71 times the wavelength used. This means that if we operate at the frequency $f = 10$ MHz the size of the array would have to be about 2.1 km but that for $f = 100$ MHz, an array size of 210 m will suffice. Similarly one could block out a star and search for weaker radio sources, such as exoplanets, near it.

The OAM radio technique may also be used in other astrophysical applications as suggested by *Harwit* [25, 26]. To quote *Harwit* [25]:

"Observations of the orbital angular momentum of photons — a property of electromagnetic radiation that has come to the fore in recent

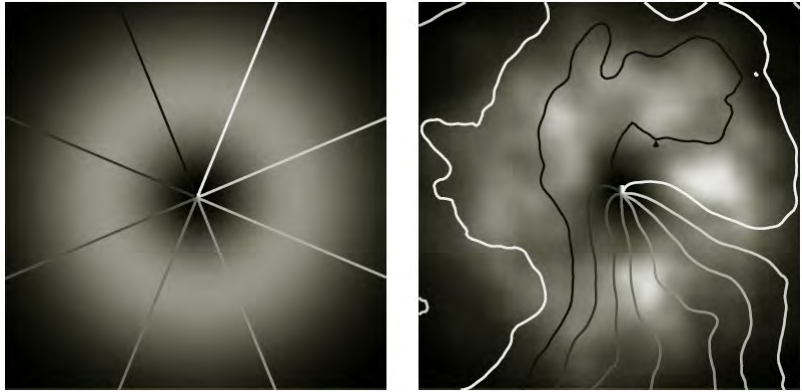

Figure 8.2: Example intensity (gray scale) and phase map ($\pi/4$-spaced contours) of a pure OAM-carrying Laguerre-Gaussian beam (left) and the same beam with aberration caused by propagation through Kolmogorov turbulence (right). This shows that OAM-carrying radio beams may be used to detect—and correct for—distortion of radio beams interacting with plasma vorticity.

years — have apparently never been attempted in astronomy. By now it is known from laboratory studies that, in addition to carrying spin angular momentum, individual photons can carry $N \gg 100$ units of orbital angular momentum $h/2\pi$. Measurements of this property of photons may have applications in several areas of astrophysical interest. Among these are: i) searches probing rapidly spinning black holes, ii) analyses of different types of discontinuities along the line of sight to astronomical masers, and iii) probes for extraterrestrial communications (where the ability of individual photons to carry $(1 + \log_2 N)$ bits of information can provide significant economies). Since photons of any wavelength may be endowed with orbital angular momentum *some of these studies will best be carried out in the radio domain*,[1] while others will be better tackled at optical or higher frequencies..."

---

[1] our emphasis.

That the radio OAM can be a sensitive detector of turbulence in the propagation medium (*e.g.*, the turbulent ionospheric plasma for radio astronomical signals) should be clear from fig. 8.2 which shows the result of a numerical simulation where a Laguerre-Gaussian beam propagated through Kolmogorov turbulence [42]. Since the total angular momentum for a given volume of the plasma (including its vorticity) and a radio beam passing through, or in any other way interacting with, this plasma volume is a conserved quantity. Vorticity in a medium is a clear signature of turbulence in the same medium. Hence, the OAM radio technique which measures the vorticity in radio signals for may be used to measure—and correct for—distortion of radio beams interacting with plasma.

### 8.1.1  Self-calibration of ionospherically aberrated radio signals

It is well-known among radio astronomers and space physicists that turbulence in the ionospheric layers can significantly alter the characteristics of radio signals that pass through them, even if the frequency of these waves (10–100 MHz) exceed the critical frequency of the ionosphere (3–10 MHz) by quite a bit. Part of the radio signal aberration (and scintillation) are amplitude and phase changes caused by the ionospheric turbulence. The self-calibration technique, developed in the radio astronomy community, manages to reduce these aberrations significantly. In this technique, a radio map of known radio objects (Cass-A, Crab, Cygn-A, *etc.*) are used as calibrators to iteratively adapt parameters of a model so that the radio interferences are eliminated.

The correction algorithm used at LOFAR (and tried at LOIS), is based on an empirical model where correction parameters are applied dynamically to the raw, sampled radio signals in a control loop manner to minimize the disturbances. The self-calibration process data can serve two purposes. It can be used for correcting signals received so that the information about distant objects is accurate. At the same time the self-calibration data provide valuable information about the dynamics of the Earth's ionosphere, which is of high interest in space physics research [50]. Similar self-calibration techniques can be envisioned to be used to compensate for imperfections in satellite-based navigation systems caused by ionospheric turbulence.

For low-frequency radio telescopes such as LOFAR, the self-calibration procedure must be repeated several times per minute. Figure 8.3 shows the difference between an un-calibrated (left) and calibrated (right) image of an astrophysical object observed at 74 MHz by the NRL-NROA VLA radio telescope: In the left panel the aberration of the ionospheric turbulence is clearly seen, while in the right panel, the self-calibration procedure has cleaned the radio image quite considerably.

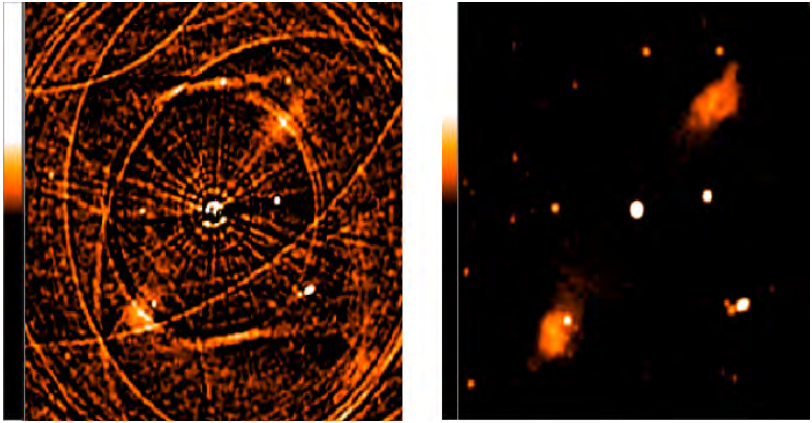

Figure 8.3: Example of how a self-calibration procedure can reduce radio image aberrations caused by ionospheric intensity and phase distortions. The left-hand panel shows the image before the self-calibration, the right-hand panel after. The radio image data were taken at the NRL-NRAO 74 MHz Very Large Array radio telescope near Socorro, NM. We expect that a self-calibration that also corrects for polarization and vorticity distortion, based on the LOIS vectorial radio field technique, utilizing OAM, will be able to improve the results considerably, particularly at low frequencies.

However, also the radio beam angular momentum (wave polarization and vorticity), which carries important information about the radio object under study, can be significantly influenced if the beam propagates through distorting media such as a turbulent ionosphere [21]. These aberrations have to be compensated for in order to obtain reliable data about the radio source under study. The vectorial sensing technique developed at LOIS and described in this thesis, makes it possible to extend the self-calibration technique to handle also vorticity distortion.

## 8.2 Communications

In optics, the transfer of information encoded in OAM has been demonstrated in practice. Using OAM one can use each state, in theory an infinite number of discrete states but in practice limited by the size of the sensor and actuator arrays and energy, to transfer information. In [22] *Gibson et al.* demonstrate the use of 8 OAM states information transfer. The communication can be made both classically and with single photons; in fact, Maxwell-Lorentz's equations in classical electrodynamics [31, 49] can be considered as the quantum theory for the single photon [1, 28]. If the receiving device had enough resolution to distinguish between 28 states, *i.e.*, in practical terms had enough vector sensing antenna units in the receiving array, one could encode the entire alphabet.

In the radio domain OAM can also be used to create multiple information channels, all using the same frequency but different $l$ so that they, ideally, do not interfere with each other (generalized Fresnel-Arago interference laws). This is achieved by demodulating different physical OAM states, using discrete Fourier transform techniques, just as one demodulates different frequencies in a normal signal. Theoretically this means that the OAM encoding technique allows for a much larger number of communication channels in each frequency band than today.

In optics, a beam carrying OAM has several restrictions for example the degrading of the beam over a distance and the beam width [22]. The degrading of the beam could be explained by atmospheric turbulence and could be mitigated by "adaptive optics" techniques of a similar kind as described in section 8.1.1. The beam width, in optics, scales with $\sqrt{l}$, according to [22]. This implies that the receiving aperture will have a physical limitation for how high OAM values it can utilize. In the radio domain, the radius seems to have a more linear relationship, and will result in a physical limitation for the arrays used, a similar restriction as for the optics case.

One advantage of using OAM in communications, is that it is resistant to eavesdropping [22]. The OAM data is extremely difficult to recover from the scattered signals originating from the beam. Misalignment with the axis of propagation also presents problems for the eavesdropper in the optical domain. In the radio domain the misalignment to the axis will yield less problem than in the optics case. If the direction of the of the beam is known the OAM should be obtainable even if one is misaligned from the axis of propagation.

## 8.3 Receiving

To receive a radio beam, and determine the variation of the instantaneous field vector properties across the beam in real time, is most conveniently done with an array of antennas. In this case, the array grid geometries shown in figure 6.1a, 6.1b and 6.1d on page 51 are not very useful, unless we know that the beam center will coincide with the center of the array or, at least partly, remain within the array grid. So for receiving a beam in general, a planar array such as the one shown in figure 6.1c, can be expected to yields an optimum result. The planar array geometry is quite simple and the calculations are therefore straightforward compared to the case of a circular grid, with the center of the beam displaced from the center of the array. If the centers would coincide between the array and the beam, a logarithmic spiral array, such as that shown in figure 6.1d, would increase the sensitivity of the array compared to the ERCG, shown in figure 6.1a. In the ERCG the elements are fanned out in sectors from the center of the array and the information of the field properties in between is lost. For the logarithmic spiral array the elements are deployed in a logarithmic spiral, so that sectors from the center of the array will less often contain no elements, giving a higher sensitivity for the array. The EDCG shown in figure 6.1b also give a better behavior for beams coinciding with the center of the array. So by comparing this array with the logarithmic spiral array it is hard to clearly see which one will the better one. Most likely the situation will become an important issue.

For receiving beams carrying orbital angular momentum, a planar array appears to be the best choice because of its simple composition of elements and geometry.

## 8.4 Outlook

In this thesis we have shown the importance of further improving, and developing, the use of the properties of the EM field. Today one are aware of these attributes of the EM field but there is little, if any, use of these in todays industry. In this thesis we have shown field gains that would have been hard to achieve without the knowledge of OAM from optics. This might be the key to the future.

We have presented results showing intensity patterns of beams with different OAM. By mixing several OAM states one can obtain very interesting and useful beam patterns and the ability to point beams simultaneously in several directions. This is likely to have useful application in, *e.g.*, radar and communication systems. By using radar systems with OAM, one could also detect rotations of the target since we know that the total angular momentum consisting of the sum of mechanical and electromagnetic angular momentum is conserved.

# Acknowledgements

First of all we would like to thank our supervisor Professor Bo Thidé for everything he has done for us. His encouragement and help in finding information is much appreciated. We are also grateful for all interesting discussions and his thorough proofreading. Secondly we would like to thank dr Jan Bergman for interesting inputs and help with finding information. Finally, a big thank you to all staff at IRFU and fellow thesis workers for making our stay here so pleasant.

# BIBLIOGRAPHY

[1]    A. I. AKHIEZER AND V. B. BERSTETSKII, *Quantum Electrodynamics*, Monographs on Physics & Astronomy. John Wiley & Sons, New York, 1965, ISBN 978-0470018484.

[2]    L. ALLEN, S. M. BARNETT, AND M. J. PADGETT, *Optical Angular Momentum*, IoP Publishing, Bristol, UK, 2003, ISBN 0 7503 0901 6.

[3]    L. ALLEN, M. W. BEIJERSBERGEN, R. J. C. SPREEUW, AND J. P. WOERDMAN, Orbital angular momentum of light and the transformation of Laguerre-Gaussian laser modes, *Physical Review A, 45*, 11 (2006), pp. 8185 – 8189.

[4]    ASTRON, Lunar Infrastructure for Exploration, Web site, `http://www.astron.nl/press/090306.htm`.

[5]    C. A. BALANIS, *Antenna Theory: Analysis and Design*, 3 ed., Wiley, 2005.

[6]    A. O. BARUT, *Electrodynamics and Classical Theory of Fields and Particles*, Dover Publications, Inc., New York, NY, 1980, ISBN 0-486-64038-8.

[7]    I. V. BASISTIY, M. S. SOSKIN, AND M. V. VASNETSOV, Optical wavefront dislocation and their properties, *Optics Communications, 119* (1995), pp. 604–612.

[8]    S. BATTERSBY, Twisting the light away, *NewScientist* (12 June 2004), pp. 36–40.

[9]    J. BERGMAN, Information Dense Antenna (IDA)—Technology White Paper.

[10]   M. V. BERRY, Paraxial beams of spinning light, In *Singular Optics* (August 1998), M. S. Soskin, Ed., vol. 3487, The International Society for Optical Engineering, SPIE, pp. 6–11.

[11]   R. A. BETH, Mechanical detection and measurement of the angular momentum of light, *Physical Review, 50* (1936), pp. 115–125.

[12]   T. H. BOYER, Illustrations of the relativistic conservation law for the center of energy, *American Journal of Physics, 73*, 10 (October 2005), pp. 953–961.

[13]   C. BROSSEAU, *Fundamentals of Polarized Light: A Statistical Optics Approach*, John Wiley & Sons, New York, 1998, Chapter 3.

[14] T. CAROZZI, R. KARLSSON, AND J. BERGMAN, Parameters characterizing electromagnetic wave polarization, *Physical Review E, 61* (2000), pp. 2024–2028, URL: `http://www.physics.irfu.se/Publications/Articles/CarozziEtAl:PRE:2000.pdf`.

[15] C. COHEN-TANNOUDJI, J. DUPONT-ROC, AND G. GRYNBERG, *Photons and Atoms: Introduction to Quantum Electrodynamics*, Wiley-Interscience, New York, NY, 1997, Chapter 1. ISBN 0-471-18433-0.

[16] C. N. COHEN-TANNOUDJI, Manipulating atoms with photons, *Journal de Physique et le Radium, 70*, 3 (July 1998), pp. 707–719, (Nobel Lecture).

[17] R. COMPTON, JR., The tripole antenna: An adaptive array with full polarization flexibility, *IEEE Transactions on Antennas and Propagation, 29*, 6 (1981), pp. 944–952, URL: `http://ieeexplore.ieee.org/xpl/freeabs_all.jsp?tp=&arnumber=1142690&isnumber=25617`.

[18] J. COURTIAL, K. DHOLAKIA, L. ALLEN, AND M. PADGETT, Gaussian beams with very high orbital angular momentum, *Optics Communications, 144* (1997), pp. 210–213.

[19] S. ERIKSSON, Study of tripole antenna arrays for space radio research, Masters thesis, Department of Astronomy and Space Physics, Uppsala University, Uppsala, Sweden, August 2003, UPTEC F03 062, URL: `http://www.physics.irfu.se/Publications/Theses/Eriksson:MSc:2003.pdf`.

[20] FOCUS, Breaking free of bits, Physical Review Letters Focus web site, URL: `http://focus.aps.org/v9/st29`.

[21] V. GARCÉS-CHAVÉZ, D. MCGLOIN, M. D. SUMMERS, A. FERNANDEZ-NIVES, G. C. SPALDING, G. CRISTOBAL, AND K. DHOLAKIA, The reconstruction of optical angular momentum after distortion in amplitude, phase and polarization, *Journal of Optics A: Pure and Applied Optics, 6* (2004), pp. S235–S238.

[22] G. GIBSON, J. COURTIAL, M. J. PADGETT, M. VASNETSOV, V. PAS'KO, S. M. BARNETT, AND S. FRANKE-ARNOLD, Free-space information transfer using light beams carrying orbital angular momentum, *Optics Express, 12*, 22 (2004), pp. 5448–5456.

[23] R. H. GOOD, JR. AND T. J. NELSON, *Classical Theory of Electric and Magnetic Fields*, Academic Press, New York and London, 1971, ISBN 0-12-290050-2.

[24] D. J. GRIFFITHS, *Introduction to Electrodynamics*, Prentice Hall, Upper Saddle River, NJ, 1999, Chapter 8.

[25] M. HARWIT, Photon orbital angular momentum in astrophysics, In *Bulletin of the American Astronomical Society* (Dec. 2003), vol. 35, The American Astronomical Society, p. 1246, AAS 203rd Meeting, Session 23, #23:07. URL: `http://adsabs.harvard.edu/cgi-bin/nph-bib_query?bibcode=2003AAS...203.2307H&db_key=AST`.

[26] M. HARWIT, Photon orbital angular momentum in astrophysics, *The Astrophysics Journal, 597*, 2 (10 November 2004), URL: `http://www.journals.uchicago.edu/ApJ/journal/issues/ApJ/v597n2/17268/brief/17268.abstract.html`.

[27] H. HE, M. E. J. FRIESE, N. R. HECKENBERG, AND H. RUBINSZTEIN-DUNLOP, Direct observation of transfer of angular momentum to absorptive particles from a laser beam with a phase singularity, *Physical Review Letters, 75*, 5 (1995), pp. 826–829.

[28] W. HEITLER, *The Quantum Theory of Radiation*, Dover Publications, New York, 1984, ISBN 978-0486645582.

[29] N. H. IBRAGIMOV, A new conservation theorem, *Journal of Mathematical Analysis and Applications, in press* (28 November 2006), doi:10.1016/j.jmaa.2006.10.078.

[30] ICT, NEC based antenna modeler and optimizer, Web site, URL: `http://home.ict.nl/~arivoors/Home.htm`.

[31] J. D. JACKSON, *Classical Electrodynamics*, 3 ed., John Wiley & Sons, New York, 1998, ISBN 978-0471427643.

[32] R. V. JONES, The Paraxial Wave Equation Gaussian Beams in Uniform Media, Lecture notes.

[33] A. KASTLER, The optical production and the optical detection of an inequality of population of the levels of spatial quantification of atoms. Application to the experiments of Stern and Gerlach and to magnetic resonance, *Reviews of Modern Physics, 11* (1950), pp. 255–265.

[34] M. KHOTYAINTSEV, *Radar Probing of the Sun*, Doctoral thesis, Uppsala University (AIM Graduate Programme), Uppsala, Sweden, 4 November 2006, Abstract: `http://publications.uu.se/abstract.xsql?dbid`, URL: `Theses/Khotyaintsev:PhD:2006.pdf`.

[35] M. V. KHOTYAINTSEV, V. N. MEL'NIK, B. THIDÉ, AND O. O. KONOVALENKO, Combination scattering by anisotropic Langmuir turbulence with

application to solar radar experiments, *Solar Physics, 234*, 1 (March 2006), pp. 169–186, URL: `http://www.springerlink.com/(32qia5nrry0y0t451pw0aj45)/app/home/contribution.asp?referrer=parent&backto=issue,9,10;journal,1,433;linkingpublicationresults,1:100339,1.`

[36] H. KOGELNIK AND T. LI, Laser beams and resonators, *Applied Optics, 5*, 10 (1966), pp. 1550–1567.

[37] LIFE, Lunar Infrastructure for Exploration, Web site, `http://www.beyondmoon.de.`

[38] LOFAR, Low Frequency Array, Web site, `http://www.lofar.org.`

[39] LOIS, LOFAR Outrigger in Scandinavia, Web site, `http://www.lois-space.net.`

[40] NEC2, Numerical Electromagnetic Code, Version 2, Web site, `http://www.nec2.org.`

[41] E. NOETHER, Invariante Variationsprobleme, *Nachrichten von der Gesellschaft der Wissenschaften zu Göttingen, 1*, 3 (1918), pp. 235–257, English transl.: *Invariant variation problems*, Transp. Th. Stat. Phys., **1**, 186–207 (1971).

[42] C. PATERSON, Atmospheric turbulence and orbital angular momentum of single photons for optical communication, *Physical Review Letters, 94*, 15 (2002), p. 153901.

[43] J. H. POYNTING, The wave motion of a revolving shaft, and a suggestion as to the angular momentum in a beam of circularly polarised light, *Proceedings of the Royal Society of London, 82*, 557 (31 July 1909), pp. 560–567, URL: `http://links.jstor.org/sici?sici=0950-1207(19090731)82%3A557%3C560%3ATWMOAR%3E2.0.CO%3B2-Z.`

[44] J. SCHWINGER, L. L. DERAAD, JR., K. A. MILTON, AND W. TSAI, *Classical Electrodynamics*, Perseus Books, Reading, MA, 1998, ISBN 0-7382-0056-5.

[45] A. SHADOWITZ, *The Electromagnetic Field*, McGraw-Hill, Tokyo, 1975, ISBN 0-07-056368-3.

[46] N. SIMPSON, K. DHOLAKI, L. ALLEN, AND M. J. PADGETT, Mechanical equivalence of spin and orbital angular momentum of light: an optical spanner, *Optics Letters, 22*, 1 (1997), pp. 52–54.

[47] M. Soljačić and M. Segev, Integer and fractional angular momentum borne on self-trapped necklace-ring beams, *Physical Review Letters, 86* (2001), p. 420.

[48] K. Sueda, N. M. G. Miyaji, and M. Nakatsuka, Laguerre-Gaussian beam generated with a multilevel spiral phase plate for high intensity laser pulses, *Optics Express, 12*, 15 (2004), pp. 3548–3553.

[49] B. Thidé, *Electromagnetic Field Theory*, Upsilon Books, Uppsala, 2006, URL: `http://www.plasma.uu.se/CED/Book`.

[50] B. Thidé, E. N. Sergeev, S. M. Grach, T. B. Leyser, and T. D. Carozzi, Competition between Langmuir and upper-hybrid turbulence in a high-frequency-pumped ionosphere, *Physical Review Letters, 95*, 25 (December 15 2005), p. 255002, URL: `http://link.aps.org/abstract/PRL/v95/e255002`, doi:10.1103/PhysRevLett.95.255002.

# Appendices

# A

## THE MATLAB CODES USED

### A.1 Phaseplanes_4_sub_.m

```matlab
% Plots the phase planes for four consecutive l values, in subfigures.
% LG beam.

clear;

%----------------constants-------------------

c=1;

%-------------things to change---------------

l_start=0;                                 %start OAM
lam=1.5;                                    %Wavelength
A=0.5;                                      %Amplitude
nog=100;                                    %
t_stop=10;                                  %Amount of timesteps
z_range=2*lam;                              %Length of z values

%-------------------------------------------

k=2*pi/lam;
w=k*c;
k_vector=[1 0 0]*k;
```

```
%---------vector x--------

z_value=linspace(0,z_range,nog);              %z on x-axis in plot
L_z=length(z_value);

r=0.5*ones(1,nog);

%----- --------- -------

t_steg=lam/(2*2*pi*c);                        %Not to big timesteps

t_vector=0:t_steg:t_stop;

%------------------------
z_antal=z_range/lam;
steg=zeros(1,z_antal);           %så att den plana vågen upprepas
                                 %då den passerar utanför bild.
t=1;
n=1;
l=l_start;
while t≤t_stop
      while l≤l_start+3

            if l==0 %Plane wave.
                clf;
                subplot(2,2,n);
                hold on;

                axis([0 z_range -2 2 -2 2]);
                view(51,30);

                axis fill;
                axis square;
                grid on;
                %light;

                z_value_tmp=z_value;                 %Z_value
                PHI_vector=linspace(0,2*pi,nog);
                L_PHI=length(PHI_vector);
                R=ones(1,L_PHI);
                [x_value y_value]=pol2cart(PHI_vector,R);
                temp=ones(1,L_z);
```

```
        z_value=temp*w*t_vector(t)/k;
        clear temp;

        for p=1:z_antal
            z_value_p=z_value+(p-1)*lam;
            if z_value_p≥(steg(1,p)+1)*z_range
                steg(1,p)=steg(1,p)+1;
            end

            z_value_p=z_value_p-2*steg(p)*lam;
            m=1;
            while m<L_z              %Plottar Phasplanes
                X=[z_value_p(m) ;z_value_p(m) ;z_value_p(m+1)...
                    ;z_value_p(m+1)      ];
                Y=[0             ;y_value(m)   ;y_value(m+1)...
                    ;0                         ];
                Z=[0             ;x_value(m)   ;x_value(m+1)...
                    ;0                         ];
                fill3(X,Y,Z,'y');
                m=m+1;
                alpha(1);
            end

        end   %for loop

        z_value=z_value_tmp;

else
        subplot(2,2,n);
        hold on;
        axis([0 z_range -2 2 -2 2]);
        view(51,30);

        axis fill;
        axis square;
        grid on;
        light;
        material shiny;
        l_steg=0;
        while l_steg<l

            temp=ones(1,L_z);
            PHI_start=l_steg*2*pi/l*temp;
            PHI_vector=PHI_start+k*z_value/l-w*t_vector(t)/l;
```

```
                     L_PHI=length(PHI_vector);
                     R=ones(1,L_PHI);

                     [x_value y_value]=pol2cart(PHI_vector,R);
                     clear temp;

                     m=1;
                     while m<L_z                   %%Plottar Phasplanes
                         X=[z_value(m)   ;z_value(m) ;z_value(m+1)...
                             ;z_value(m+1)   ];
                         Y=[0            ;y_value(m) ;y_value(m+1)...
                             ;0                 ];
                         Z=[0            ;x_value(m) ;x_value(m+1)...
                             ;0                 ];
                         fill3(X,Y,Z,'y')
                         m=m+1;

                     end
                 l_steg=l_steg+1;
               end   %while l_steg...
           end   %else satsen
           l=l+1;
           n=n+1;

     end  %while l.

     %Film(:,t)=getframe(gcf);
     hold off;
     n=1;            %Återställning av räknare
     t=t+1;          %Tidsstegning
     pause(0.1);
     l=l_start;

end

%movie2avi(Film,'phaseplanes_4.avi');   %For avi films.
%movieview('phaseplanes_4.avi')
```

## A.2  nec.m

```
% Written by Johan Sjöholm.
%
```

```matlab
% Returns the information, Theta, Phi, E_magn and phase) from 4nec2 data.
%

% The Data input file is obtained from the 4nec2 outputfile with
% modifications:
%   Only the values from "Radiation patterns" are kept.
%   "LINEAR", "RIGHT" and "LEFT" are replaced by appropriate amount of
%   spaces.

function [Phase_theta_10 Phase_phi_10]=nec_org(L,grad,step)

% L                % Data file from 4nec2
% grad_min         % Min meassurement theta angle.
% grad_max         % Max Meassurement theta angle (16 24 32 45 mainlobe)
% gradsteg         % Resolution in 4nec2.
% step             % Every step'th value.

p=1;o=1;
d=length(L);
for l=1:d                        % All values in L to be analyzed
    if L(l,1)==-grad             % If the theta angle equals grad.
        if p==1
            Phase_theta1=[L(l,1), L(l,2), L(l,8),L(l,9)];
            p=2;
        else
            Phase_theta1=[Phase_theta1;L(l,1), L(l,2),...
                L(l,8),L(l,9)];
        end
        if o==1
            Phase_phi1=[L(l,1), L(l,2), L(l,10), L(l,11)];
            o=2;
        else
            Phase_phi1=[Phase_phi1;L(l,1), L(l,2),...
                L(l,10), L(l,11)];
        end
    end
end

Phase_theta(:,:)=[Phase_theta1];        %[Theta       Phi
                                        %E_magn(theta)  E_phase(theta)]

Phase_phi(:,:)=[Phase_phi1];            %[Theta       Phi
                                        %E_magn(theta)  E_phase(theta)]
```

```
%________________________ Every "step" value ________________________
l_ph_theta=length(Phase_theta(:,1));
for ten=1:step:l_ph_theta
    if ten==1;
        Phase_theta_10(1,:)=Phase_theta(ten,:);
        Phase_phi_10(1,:)=Phase_phi(ten,:);
    else
        Phase_theta_10((ten-1)/step+1,:)=Phase_theta(ten,:);
        Phase_phi_10((ten-1)/step+1,:)=Phase_phi(ten,:);
    end
end
```

## A.3  Phase_arrows.m

```
% Written by Johan Sjöholm
%
% This program will plot what phase arrows from 4nec2 data.

%Source files from 4nec, modified. "LINEAR", "LEFT", "RIGHT" replaced
 %by appropriate amount of space. Radiation patterns values are kept.
  %Columns as follows:
        %( 1 - Theta               2 - Phi          ) Angles
        %( 3 - Vert. DB     4 - Hor. DB    5 - Tot. DB ) Power gains
        %( 6 - Axial Ratio        7 - Tilt deg       ) Polarization
        %( 8 - E_theta magnitude.   9 - Phase_theta    ) E(Theta)
        %(10 - E_phi magnitude.    11 - Phase_phi      ) E(Phi)

clear;
close all;

names_of_files=['ED_r10to20_lam11_dip_L00.dat';
                'ED_r10to20_lam11_dip_L01.dat';        %Names of 4nec2 files
                'ED_r10to20_lam11_dip_L02.dat';
                'ED_r10to20_lam11_dip_L04.dat'];

grad_min=[4 4 4 4];     % Min meassurement theta angle.
grad_max=[14 20 30 35]; % Max Meassurement theta angle (ex mainlobe)
grad_steg=2;            % Resolution in 4nec2 file.
step=4;                 % Every step'th value.
```

```
for m=1:length(names_of_files(:,1))
    Data=load(names_of_files(m,:));              %Data from file
    for grad=grad_min:grad_steg:grad_max         %Degrees for data.
        [Phase_theta_10 Phase_phi_10]=nec(Data,grad,step);
                        % nec obtains the data from the 4nec2 output file.

        % Ecos(theta)=-Ecos(theta+-180) same for phi. (For looks only)
        % No real change in the values of the E-field just easier to
        % visualize the phase.
        d=length(Phase_theta_10);
        for k=1:d
            if Phase_theta_10(k,2)≥180
                if Phase_theta_10(k,4)>0
                    Phase_theta_10(k,4)=Phase_theta_10(k,4)-180;
                else
                    Phase_theta_10(k,4)=Phase_theta_10(k,4)+180;
                end
                Phase_theta_10(k,3)=Phase_theta_10(k,3)*(-1);
            end
            if Phase_phi_10(k,2)≥90 & Phase_phi_10(k,2)<270
                if Phase_phi_10(k,4)>0
                    Phase_phi_10(k,3)=Phase_phi_10(k,3)*(-1);
                    Phase_phi_10(k,4)=Phase_phi_10(k,4)-180;
                else
                    Phase_phi_10(k,3)=Phase_phi_10(k,3)*(-1);
                    Phase_phi_10(k,4)=Phase_phi_10(k,4)+180;
                end
            end
        end

    end

% Distance between each ring:
    R=1;                    %Virtual distance just to get a feel for it.
    r=R.*tan(pi/180*grad).*ones(length(Phase_theta_10),1); %Placements
% Placement:
    [X Y]=pol2cart(pi/180.*Phase_theta_10(:,2),r);
        %Since phi=theta, in our case we can use theta or phi
        %for the phase.CAREFUL!!!!!
% Change of coordinates:
    [X_th Y_th]=pol2cart(pi/180.*Phase_theta_10(:,4),1);
    [X_ph Y_ph]=pol2cart(pi/180.*Phase_phi_10(:,4),1);
% PLOT:
    figure(m)
```

```
        quiver (X,Y,X_ph,Y_ph,0.5);                    %Vector arrows
        hold on;
        ax_values=(R*tan(pi/180*grad)+0.1);            %Axes values
        axis ([-ax_values ax_values -ax_values ax_values]);
        axis square;
        xlabel('X','fontsize',24);
        ylabel('Y','fontsize',24);
    end %GRAD
end %name_of_files
```

## A.4  E_field.m

```
% Written by Johan Sjöholm
%
% This program plots what the electric field looks like from
% 4nec2 data.

%Source files from 4nec, modified. "LINEAR", "LEFT", "RIGHT" replaced
  %by appropriate amount of space. Radiation patterns values are kept.
  %Columns as follows:
        %( 1 - Theta                2 - Phi                ) Angles
        %( 3 - Vert. DB     4 - Hor. DB     5 - Tot. DB    ) Power gains
        %( 6 - Axial Ratio          7 - Tilt deg           ) Polarization
        %( 8 - E_theta magnitude.   9 - Phase_theta        ) E(Theta)
        %(10 - E_phi magnitude.    11 - Phase_phi          ) E(Phi)

clear;
close all;

names_of_files=['ED_r10to20_lam11_dip_L00.dat';
                'ED_r10to20_lam11_dip_L01.dat';      %Names of 4nec2 files
                'ED_r10to20_lam11_dip_L02.dat';
                'ED_r10to20_lam11_dip_L04.dat'];

grad_min=[4 4 4 4];       % Min meassurement theta angle.
grad_max=[14 20 30 35]; % Max Meassurement theta angle (ex mainlobe)
grad_steg=2;              % Resolution in 4nec2 file.
step=2;                   % Every step'th value to be plotted

for m=1:length(names_of_files(:,1))
    Data=load(names_of_files(m,:));          %Data from file
    for grad=grad_min:grad_steg:grad_max     %Degrees for data.
```

```matlab
        [Phase_theta_10 Phase_phi_10]=nec(Data,grad,step);
                        % nec_org can also be used here, making it general.
                        % nec obtains the data from the 4nec2 output file.
    % E_phase:
        E_theta=pi/180.*Phase_theta_10(:,3).*...
            (cos(pi/180.*Phase_theta_10(:,4)));
        E_phi=pi/180.*Phase_phi_10(:,3).*...
            (cos(pi/180.*Phase_phi_10(:,4)));
    % Distance between each ring:
        R=1;
        z=0*ones(length(E_theta),1);
        rho=R.*tan(pi/180*grad).*ones(length(E_theta),1);
    %Placement:
        X=rho.*cos(pi/180.*Phase_theta_10(:,2));
        Y=rho.*sin(pi/180.*Phase_theta_10(:,2));
        Z=z;

        grad_theta=-pi/180.*Phase_theta_10(:,1);
        grad_phi=pi/180.*Phase_theta_10(:,2);
    % Change of coordinate system:
        EX=E_theta.*cos(grad_phi).*cos(grad_theta)-E_phi.*sin(grad_phi);
        EY=E_theta.*cos(grad_theta).*sin(grad_phi)+E_phi.*cos(grad_phi);
        EZ=-E_theta.*sin(grad_theta);
    % PLOT:
        figure(m)
        quiver3 (X,Y,Z,EX,EY,EZ,0.3);                  % Vector arrows.
        hold on;
        ax_values=(R.*tan(pi/180*grad)+0.1);       % Axes values.
        axis ([-ax_values ax_values -ax_values ax_values...
            -ax_values ax_values]);
        axis square;
        xlabel('X','fontsize',24);
        ylabel('Y','fontsize',24);
        zlabel('Z','fontsize',24);
    end %GRAD
end
```

## A.5   intensity.m

```matlab
% Written by Kristoffer Palmer 2007
% Plots intensities in beams created with NEC
% Input files should be with 5 degrees angular
% resolution and theta =< 90
```

```matlab
% degrees. Input file should have columns
% (theta, phi, vert. gain,
% hor. gain, tot. gain, AR, tilt angle, Ethetamagn,
% Ethetaphase, Ephimagn, Ephiphase)
clear;
c = 3e+8;
e0 = 8.854e-12;
% number of plots
korningar = input('number of different plots: ');
T40 = 0;   % needed?
RES = 0;   % needed
for korn = 1:korningar
clear T40;
clear RES;

file = input('filnamn: ','s');    %inpufile
RES = load(file);

R = 1; %may change
T40 = RES;

% change to radians and fix neg.
theta = pi/180*-1*T40(:,1);
phi = pi/180*T40(:,2);

% extract E magnitudes and phases
Ethetamagn = T40(:,8);
Ethetaphase = pi/180*T40(:,9);
Ephimagn = T40(:,10);
Ephiphase = pi/180*T40(:,11);

%size of E
Esqr = Ethetamagn.^2+Ephimagn.^2;

Int = c*e0/2*Esqr; %intesniy in each (theta,phi)

% sort values for ploting
for n = 0:72
    for m = 2:19
        INT(n*19+m) = Int((n+1)*19-(m-2));
    end
end

r = R*tan(theta);
```

```
    for n = 1:19
        rs(n) = r(n);
    end

    RS = sort(rs);

    for n = 1:18
        RS2(n)=RS(n);
    end

for n = 1:16
    RS3(n) = RS2(n);
end

    for m = 1:73
        for n = 1:16
            I0(n,m) = INT((m-1)*19+n);
        end
    end

%create mesh
[rr,tt] = meshgrid(RS3,linspace(-pi,pi,73));
xx = rr.*cos(tt);
yy = rr.*sin(tt);

%intensity in dB
I0dB = 10*log10(I0/(max(max(I0))));

%plots all beams in one plot
figure(4);
subplot(korningar,2,korn*2-1);
surf(xx,yy,I0'); view(2); axis([-4 4 -4 4]);
shading interp; axis image; axis off;
subplot(korningar,2,korn*2);
surf(xx,yy,I0dB'); view(2); axis([-4 4 -4 4]);
shading interp; axis image; axis off; colorbar;

%plots all beams in individual plots
figure(5+korn);
surf(xx,yy,I0'); view(2); axis([-4 4 -4 4]);
shading interp; axis image;
axis off;
figure(15+korn);
surf(xx,yy,I0dB'); view(2); axis([-4 4 -4 4]);
shading interp; axis image; axis off; colorbar;
```

```
end
```

## A.6 Phase.m

```matlab
% Written by Kristoffer Palmer 2007
% Plots phases of polarized beams in a cross section of beams
% created with NEC
% Input files should be with 5 degrees angular resolution and
% theta =< 90 degrees.
% Input file should have columns (theta, phi, vert. gain,
% hor. gain, tot. gain, AR, tilt angle, Ethetamagn, Ethetaphase,
% Ephimagn, Ephiphase)
clear;

korningar = input('number of different plots: ');% number of files
T40 = 0;
RES = 0;
for korn = 1:korningar
clear T40;
clear RES;

file = input('filnamn: ','s'); % load files
RES = load(file);

for g = 2:19 % skip values in center point

for n = 0:4:72 % to get square matrix
    T40(n/4+1,:) = RES(n*19+g,:);

end

% degrees to radians
theta = pi/180*-1*T40(:,1);
phi = pi/180*T40(:,2);

% extract E
Ethetamagn = T40(:,8);
Ethetaphase = pi/180*T40(:,9);
Ephimagn = T40(:,10);
Ephiphase = pi/180*T40(:,11);

% calculate theta- and phi-components with sign
```

```matlab
Etheta = Ethetamagn.*cos(Ethetaphase);
Ephi = Ephimagn.*cos(Ephiphase);

%change to carteisan coordinates
Ex = Etheta.*cos(theta).*cos(phi) - Ephi.*sin(phi);
Ey = Etheta.*cos(theta).*sin(phi) + Ephi.*cos(phi);
Ez = -Etheta.*cos(theta);

% calculate phase (vector direction) of electric field in xy-plane
for n = 1:19
if Ex(n) >= 0 && Ey(n) >= 0
    PHASE(g,n) = atan(Ey(n)/Ex(n));
elseif Ex(n) >= 0 && Ey(n) < 0
    PHASE(g,n) = atan(Ey(n)/Ex(n)) + 2*pi;
else
    PHASE(g,n) = (pi + atan(Ey(n)/Ex(n)));%*180/pi;
end
end

end

%resorting for plotting
for g = 19:-1:1
    PHASE2(g,:) = PHASE(20-g,:);
end
for g = 19:-1:1
    PHASE3(:,g) = PHASE(:,20-g);
end

rr(19) = tan(86*pi/180);
for n = 0:17   %17:-1:0
    rr(n+1) = tan(5*n*pi/180);
end

% build square mesh
[RR,TT] = meshgrid(rr,phi);
xx = RR.*cos(TT);
yy = RR.*sin(TT);

% plot phases in subplots
figure(4);
subplot(floor((-0.1+korningar)/2)+1,2,korn,'align');
surf(xx,yy,PHASE2'); view(2); axis image; axis off; colorbar
figure(5);
subplot(floor((-0.1+korningar)/2)+1,2,korn,'align');
```

```
surf(xx,yy,PHASE2'); view(2);
axis([-tan(60*pi/180) tan(60*pi/180) -tan(60*pi/180) tan(60*pi/180)]);
axis square; axis off; colorbar;
ARmean = sum(RES(:,6))/1387 %return mean value of AR
ARh=ARmean*2*pi*3e8*6.626e-34 %

% plot each beam in separate window
figure(5+korn);
surf(xx,yy,PHASE2');
axis([-tan(60*pi/180) tan(60*pi/180) -tan(60*pi/180) tan(60*pi/180)]);
axis square; axis off;  view(2);
%180 degree rotation somtimes needed if theta is negative
end
```

## A.7  Evectors.m

```
% By Kristoffer Palmer 2006
% Plots E vectors in xy plane. Assumes 5 degrees angular resolution.
% Input should look like input in Phase.m

clear;
lambda = 1;
c = 3e8;

tt = linspace(0,lambda/c,73);
dt= tt(2)-tt(1);
t=input('t: '); % can specify when to plot in one period (0<t<74)

file = input('filnamn: ','s');
RES = load(file);
%load data and store values to use

for g = 7:19  %between which theta?

for n = 0:72
    T40(n+1,:) = RES(n*19+g,:);

end
% extract data
theta = pi/180*-1*T40(:,1);
phi = pi/180*T40(:,2)+pi;

Ethetamagn = T40(:,8);
```

```matlab
Ethetaphase = pi/180*T40(:,9)+2*pi/lambda*c*tt(t);
Ephimagn = T40(:,10);
Ephiphase = pi/180*T40(:,11)+2*pi/lambda*c*tt(t);

%for plotting
for jogfed = 1:73
    ones(jogfed) = 1;
end
cx = tan(theta(1))*cos(linspace(0,2*pi,73));
cy = tan(theta(1))*sin(linspace(0,2*pi,73));
cz = 0*linspace(0,2*pi,73);

%calculate theta- and phi-components
Etheta = Ethetamagn.*cos(Ethetaphase);
Ephi = Ephimagn.*cos(Ephiphase);

%change to carteisan coordinates
Ex = Etheta.*cos(theta).*cos(phi) - Ephi.*sin(phi);
Ey = Etheta.*cos(theta).*sin(phi) + Ephi.*cos(phi);
Ez = -Etheta.*cos(theta);
figure(2);

%plot
for j = 1:2:72
    Ex(j) = 0;
    Ey(j) = 0;
    Ez(j) = 0;
end
Ex(73) = 0;
Ey(73) = 0;
Ez(73) = 0;

quiver3(cx,cy,cz,Ex,Ey,Ez,0.5); hold on
view(2);
%axis([-tan(60*pi/180) tan(60*pi/180) -tan(60*pi/180) tan(60*pi/180)]);
axis square; axis off;

end
```

## A.8 AngMom.m

```matlab
% Written by Kristoffer Palmer 2007
% calculates and plots Jz/U in beams created with NEC
```

```
% (but with unknown constant in the magnitude.
% Input files shold have only theta =< 90
% degrees. Input file should have columns (theta, phi, vert. gain,
% hor. gain, tot. gain, AR, tilt angle, Ethetamagn, Ethetaphase,
% Ephimagn, Ephiphase)
% For correct x axis, please give symmetric files (l=-4 to l=4
% in 9 files for instance)
clear;
N = input('Number of files: '); %Number of files
resolution = input('Angular resolution in input file: '); %angular
                                         % resolution in input file

for n = 1:N

    file = input('filnamn: ','s'); %give name of each file
    RES = load(file);

    JzUU = JzU(RES,resolution);  % Use function JzU

    JzUUU(n) = JzUU; %stores all Jz/U
end

l = linspace(-(N-1)/2,(N-1)/2,N); %creates l interval

plot(l,JzUUU,'b'); hold on; %plots result
```

## A.9 JzU.m

```
% By Kristoffer Palmer 2007
% function used by program AngMom.

function f = JzU(RES,resolution)

% constants and starting values
lambda = 1; %could make input lambda
c = 3e+8;
e0 = 8.854e-12;
Jxtot = 0;
Jytot = 0;
Jztot = 0;
Entot = 0;

% resolutiondependent parameters
```

```matlab
aa = 90/resolution +1;  %  g = 1:19
bb = 360/resolution;    %  n = 0:72

% timesteps
tt = linspace(0,lambda/c,73);
dt= tt(2)-tt(1);

for t = 0:dt:lambda/c %for one period
    for g = 1:aa
        for n = 0:bb
            T40(n+1,:) = RES(n*19+g,:);
        end

        theta = pi/180*-1*T40(:,1);
        phi = pi/180*T40(:,2);

        Ethetamagn = T40(:,8);
        Ethetaphase = pi/180*T40(:,9)+2*pi/lambda*c*t;
        Ephimagn = T40(:,10);
        Ephiphase = pi/180*T40(:,11)+2*pi/lambda*c*t;

        Etheta = Ethetamagn.*cos(Ethetaphase); %Etheta
        Ephi = Ephimagn.*cos(Ephiphase);        %Ephi

        Btheta = -1/c*Ephi; %create B from E
        Bphi = 1/c*Etheta;

        %change to carteisan coordinates
        Ex = Etheta.*cos(theta).*cos(phi) - Ephi.*sin(phi);
        Ey = Etheta.*cos(theta).*sin(phi) + Ephi.*cos(phi);
        Ez = -Etheta.*cos(theta);

        Bx = Btheta.*cos(theta).*cos(phi) - Bphi.*sin(phi);
        By = Btheta.*cos(theta).*sin(phi) + Bphi.*cos(phi);
        Bz = -Btheta.*cos(theta);

        % Vectors proportional to Poynting vector
        Sx = Ey.*Bz-Ez.*By;
        Sy = Ez.*Bx-Ex.*Bz;
        Sz = Ex.*By-Ey.*Bx;

        % for calculating J
        cx = cos(pi/2-theta(1))*cos(linspace(0,2*pi,bb+1));
        cy = cos(pi/2-theta(1))*sin(linspace(0,2*pi,bb+1));
        for n = 1:bb+1
```

```
            cz(n) = sin(pi/2-theta(1,1));
        end

        % Calculate J in one timestep
        Jx = e0*(cy.*Sz'-cz.*Sy');
        Jy = e0*(cz.*Sx'-cx.*Sz');
        Jz = e0*(cx.*Sy'-cy.*Sx');

        % Total J
        for n = 1:bb+1
            Jxtot = Jxtot + Jx(n);
            Jytot = Jytot + Jy(n);
            Jztot = Jztot + Jz(n);
        end

        Etot = sqrt(Ex.^2+Ey.^2+Ez.^2);
        Btot = sqrt(Bx.^2+By.^2+Bz.^2);

        % Total energy
        for n = 1:73
            Entot = Entot + e0/2*(Etot(n)^2 + c^2 * Btot(n)^2);
        end
    end
end
f = c/lambda*Jztot/Entot; %time average of Jz/U
end
```

## A.10 tripole.m

```
% Written by Kristoffer Palmer 2007.
% Calculates curents and phases for a circular tripole array.
% Works for all directions, polarizations and l numbers
%(if possible in theory). See report for explanation of parameters.
%The program also shows the total current vectors in a 3d plot.
% Red arrow indicates direction (theta0, phi0) of maximum (for l=0)
% or wanted symmetry axis. Assumes that the dipole antennas in each
% tripole is directed in one of x,y or z direction only.

clear;
runs = input('elements: '); %number of elements
theta0 = input('theta0: '); %beam symmetry axis theta direction
phi0 = input('phi0: ');     %beam symmetry axis phi direction
a = input('radius in wavelengths: '); %array radius
```

```matlab
%lambda = input('wavelenght: ');
l = input('l: ');           %l eigenvalue used
ax = input('ax: ');         %ax used

% calculate currents amplitudes and phases in one pointed antenna
J1 = 1/ax*cos(theta0*pi/180)*cos(phi0*pi/180) + i*sin(phi0*pi/180);
J2 = 1/ax*cos(theta0*pi/180)*sin(phi0*pi/180) - i*cos(phi0*pi/180);
J3 = -1/ax*sin(theta0*pi/180);
amp_x = abs(J1) % returns amplitude of x directed element
amp_y = abs(J2) % returns amplitude of y directed element
amp_z = abs(J3) % returns amplitude of z directed element

% calculate phase shift between elements in the array
for n = 1:runs
    phiN = input('phiN: '); %angular position of each element
    PHI = 180/pi*(l*phiN*pi/180-2*pi*a*sin(theta0*pi/180)...
        *cos(pi/180*phiN-pi/180*phi0));

    phase_x(n) = angle(J1)*180/pi + PHI;

    phase_y(n) = angle(J2)*180/pi + PHI;

    phase_z(n) = angle(J3)*180/pi + PHI;

    %%%%%%%%%%%%% for plotting "E" %%%%%%%%%%%%%%%%%%
    cx(n) = a*cos(pi/180*phiN);
    cy(n) = a*sin(pi/180*phiN);
    cz(n) = 0;
    %%%%%%%%%%%%%%%%%%%%%%%%%%%%%%%%%%%%%%%%%%%%%%%%%
end
phase_x % returns x phases
phase_y % returns y phases
phase_z % returns z phases

%%%%%%%%%%%%%%% plot "E" %%%%%%%%%%%%%%%%%%%%%%%%
Ex = amp_x.*cos(pi/180*phase_x);
Ey = amp_y.*cos(pi/180*phase_y);
Ez = amp_z.*cos(pi/180*phase_z);
E = sqrt(Ex(1)^2+Ey(1)^2+Ez(1)^2);

Refx = E*sin(theta0*pi/180)*cos(phi0*pi/180);
Refy = E*sin(theta0*pi/180)*sin(phi0*pi/180);
```

```
Refz = E*cos(theta0*pi/180);
figure(8);
quiver3(cx,cy,cz,Ex,Ey,Ez,0.3); hold on;
quiver3(0,0,0,Refx, Refy,Refz,0.3,'r'); hold off;
view(phi0+90,90-theta0)
axis equal;
```

## A.11  lg_beam.m

```
% By Kristoffer Palmer 2006
% Simulates intensity patterns of Laguerre-Gaussian beams.
% Can simulate any number of different beams interfering.

clear;
% number of patterns
korningar = input('give number of different plots: ');
u = 0;
U = 0;
for korn = 1:korningar
clear u;
clear U;

r = linspace(0,3,100); %may change 3 to other value
z = linspace(0,2*pi,100); %
phi = linspace(0,2*pi,100);

%number of beams interfering in each pattern
N = input('give number of beams: ');
lambda = input('give wavelength: '); %wavelength
w0 = input('give beamwaist: ');    %beamwaist

% constants and Rayleigh range
c = 3e+8;
e0 = 8.854e-12;
zr = (pi*(w0)^2)/lambda;

for q = 1:N

    p(q) = input('give p: ') % eigenvalue p. Return while runing
    l(q) = input('give l: ') % eigenvalue l. Return while runing
    %argument for rotation of beam. Return while running
    arg(q) = input('give argument in radians: ')
end
```

```
for q = 1:N
    for m = 1:100
        for n = 1:100
            % use function upl for scalar wave amplitude
            u(m,n,q) = upl(r(m),phi(n)+arg(q),z(1),p(q),l(q),lambda,w0);

        end
    end
end

U(100,100) = 0; %create U matrix

for q = 1:N %add all beams

    U = u(:,:,q)+U;
end

%calculate momentum
Uc = conj(U);
[Ux,Uy] = gradient(U,1/100,2*pi/100);
[Ucx,Ucy] = gradient(Uc,1/100,2*pi/100);
Px = i*2*pi*c/lambda*e0/2*(Uc.*Ux-U.*Ucx);
Py = i*2*pi*c/lambda*e0/2*(Uc.*Uy-U.*Ucy);
Pz = 2*pi*c/lambda*2*pi/lambda*e0*abs(U).^2; %neglect grad in z

% momentum vector length
Pabs = sqrt(Px.^2+Py.^2+Pz.^2);

% build mesh
[rr,tt] = meshgrid(linspace(0,3,100),linspace(0,2*pi,100));
xx = rr.*cos(tt);
yy = rr.*sin(tt);
% plot intesnity
figure(1)
subplot(korningar,2,korn);                              %
surf(xx,yy,Pabs'); view(2); axis image;; shading interp;%axis equal

%%%%%%%%%% change to max(max(Pabs)) if time, then this not needed
% for n = 1:100
%     for m = 1:100
%         if Pabs(n,m) == 0
%             Pabs(n,m) = pi;
%         end
%     end
% end
```

```
% for n = 1:100
%     for m = 1:100
%         if Pabs(n,m) == pi
%             Pabs(n,m) = min(min(Pabs));
%         end
%     end
% end

% plot linear and in db
figure(2)
subplot(korningar,2,korn);
PabsdB = 10*log10(Pabs/(max(max(Pabs))));  %change to max here
surf(xx,yy,PabsdB'); view(2); axis equal; shading interp;

figure(3);
subplot(korningar,2,korn*2-1);
surf(xx,yy,Pabs'); view(2); axis equal; shading interp;
subplot(korningar,2,korn*2);
surf(xx,yy,PabsdB'); view(2); axis equal; shading interp;

% plot in individual windows
figure(5+korn);
surf(xx,yy,Pabs'); view(2); axis image;
axis off; shading interp;
figure(15+korn);
surf(xx,yy,PabsdB'); view(2); axis image;
axis off;; shading interp;

end
```

## A.12  upl.m

```
% By Kristoffer Palmer 2006
% Function used for finding scalar amplitude u in LG beam

function f = upl(r,phi,z,p,l,lambda,w0)

k = 2*pi/lambda;
zr = (pi*(w0)^2)/lambda;
w = w0*sqrt(1+(z/zr).^2);
ll = abs(l); %so we can use negative l
maple('with','orthopoly'); % to use Laguerre polynomial
L = mfun('L',p,ll,2*r.^2/(w.^2)); % Laguerrepolynomial
```

```
f = sqrt(factorial(p)/factorial(p+ll)) * w0/(w*zr)*(r*sqrt(2)/w).^ll...
    *L*exp(-r.^2/w^2)*exp(-i*k*r.^2*z/(2*(z^2+zr^2)))
    *exp(-i*l*phi)*exp(i*(2*p+l+1)*atan(z/zr));
end
```

## A.13  makematrix.m

```
%Written by Kristoffer Palmer 2007
%If NEC output is a free space file, use this program to
%only save theta =< 90 degrees.
clear;
n=1;
file = input('filnamn: ','s'); %loads free space file
pellefant = load(file);
%load pellefant.dat
for m = 1:2701
    if pellefant(m,1) > -91
        A(n,:) = pellefant(m,:);
        n=n+1;
    end;
end;
clear m n pellefant.dat
utfil = input('utfil: ','s');  %give name of ouput file
save(utfil, 'A', '-ascii')     %saves output file
```

# B

## NEC INPUT FILES

## B.1 Electromagnetic beam with orbital angular momentum

### B.1.1 L0.txt

```
CE
GW    9   21   -7   6.5699   2.5 -7   7.5699   2.5 1.e-3
GW   10   21   -9.9289 -0.5      2.5 -9.9289 0.5 2.5 1.e-3
GW   11   21   -7  -7.572  2.5 -7  -6.572   2.5 1.e-3
GW   12   21   0.0711   -10.5    2.5 0.0711   -9.5     2.5 1.e-3
GW   13   21   7.1422   -7.57    2.5 7.1422   -6.57    2.5 1.e-3
GW   14   21   10.0711 -0.5      2.5 10.0711 0.5 2.5 1.e-3
GW   15   21   7.1422   6.572    2.5 7.1422   7.572    2.5 1.e-3
GW   33   21   -7.6537 17.977  2.5 -7.6537 18.977   2.5 1.e-3
GW   34   21   -14.1421      13.64    2.5 -14.1421     14.64    2.5 1.e-3
GW   35   21   -18.4776      7.15     2.5 -18.4776     8.15      2.5 1.e-3
GW   36   21   -20 -0.5      2.5 -20 0.5 2.5 1.e-3
GW   37   21   -18.5   -8.25    2.5 -18.5   -7.25    2.5 1.e-3
GW   38   21   -14.1645      -14.7385      2.5 -14.1645      -13.7385      2.5 1.e-3
GW   39   21   -7.65375      -18.9775      2.5 -7.65375      -17.9775      2.5 1.e-3
GW   40   21   0   -20.5    2.5 0   -19.5    2.5 1.e-3
GW   41   21   7.6537   -18.9776      2.5 7.6537   -17.9776      2.5 1.e-3
GW   42   21   14.1421 -14.6422      2.5 14.1421 -13.6422      2.5 1.e-3
GW   43   21   18.4776 -8.1536 2.5 18.4776 -7.1536 2.5 1.e-3
GW   44   21   20   -0.5      2.5 20  0.5 2.5 1.e-3
GW   45   21   18.5    7.25     2.5 18.5    8.25      2.5 1.e-3
GW   46   21   14.1421 13.6421 2.5 14.1421 14.6421 2.5 1.e-3
GW   47   21   7.6537   17.9776 2.5 7.6537   18.9776 2.5 1.e-3
```

```
GW   48   21   0    19.5     2.5 0    20.5     2.5 1.e-3
GW   16   21   0.0711  9.5 2.5 0.0711  10.5     2.5 1.e-3
GE   0
EX   6    9    11   0    1    0
EX   6    10   11   0    1    0
EX   6    11   11   0    1.002    0
EX   6    12   11   0    1    0
EX   6    13   11   0    1.004    0
EX   6    14   11   0    1    0
EX   6    15   11   0    1.003    0
EX   6    16   11   0    1.001    0
EX   6    33   11   0    1    0
EX   6    35   11   0    1    0
EX   6    36   11   0    1    0
EX   6    37   11   0    1    0
EX   6    38   11   0    1.002    0
EX   6    39   11   0    0.997    0
EX   6    40   11   0    1    0
EX   6    41   11   0    0.999    0
EX   6    42   11   0    1.004    0
EX   6    43   11   0    1.007    0
EX   6    44   11   0    1    0
EX   6    45   11   0    1.007    0
EX   6    46   11   0    1.002    0
EX   6    47   11   0    1    0
EX   6    48   11   0    1    0
EX   6    34   11   0    1    0
GN   1
FR   0    1    0    0    27.2545455   0
EN
```

## B.1.2  L1.txt

```
CE
GW   9    21   -7   6.5699   2.5 -7   7.5699   2.5 1.e-3
GW   10   21   -9.9289 -0.5     2.5 -9.9289 0.5 2.5 1.e-3
GW   11   21   -7   -7.572   2.5 -7   -6.572   2.5 1.e-3
GW   12   21   0.0711  -10.5    2.5 0.0711  -9.5    2.5 1.e-3
GW   13   21   7.1422  -7.57    2.5 7.1422  -6.57    2.5 1.e-3
GW   14   21   10.0711 -0.5     2.5 10.0711 0.5 2.5 1.e-3
GW   15   21   7.1422  6.572    2.5 7.1422  7.572    2.5 1.e-3
GW   33   21   -7.6537 17.977   2.5 -7.6537 18.977   2.5 1.e-3
GW   34   21   -14.1421    13.64    2.5 -14.1421    14.64    2.5 1.e-3
GW   35   21   -18.4776    7.15     2.5 -18.4776    8.15     2.5 1.e-3
```

```
GW  36  21  -20 -0.5   2.5 -20 0.5 2.5 1.e-3
GW  37  21  -18.5  -8.25   2.5 -18.5  -7.25  2.5 1.e-3
GW  38  21  -14.1645    -14.7385    2.5 -14.1645  -13.7385   2.5 1.e-3
GW  39  21  -7.65375    -18.9775    2.5 -7.65375   -17.9775   2.5 1.e-3
GW  40  21  0   -20.5   2.5 0   -19.5   2.5 1.e-3
GW  41  21  7.6537 -18.9776   2.5 7.6537 -17.9776   2.5 1.e-3
GW  42  21  14.1421 -14.6422   2.5 14.1421 -13.6422   2.5 1.e-3
GW  43  21  18.4776 -8.1536 2.5 18.4776 -7.1536 2.5 1.e-3
GW  44  21  20  -0.5   2.5 20  0.5 2.5 1.e-3
GW  45  21  18.5   7.25   2.5 18.5   8.25   2.5 1.e-3
GW  46  21  14.1421 13.6421 2.5 14.1421 14.6421 2.5 1.e-3
GW  47  21  7.6537 17.9776 2.5 7.6537 18.9776 2.5 1.e-3
GW  48  21  0   19.5   2.5 0   20.5   2.5 1.e-3
GW  16  21  0.0711 9.5 2.5 0.0711   10.5   2.5 1.e-3
GE  0
EX  6  9  11  0   -0.71  0.707
EX  6  10  11  0   -1  1.e-16
EX  6  11  11  0   -0.71  -0.71
EX  6  12  11  0   6.e-17  -1
EX  6  13  11  0   0.71   -0.71
EX  6  14  11  0   1   0
EX  6  15  11  0   0.707  0.707
EX  6  16  11  0   6.e-17  1.001
EX  6  33  11  0   -0.38  0.926
EX  6  35  11  0   -0.93  0.383
EX  6  36  11  0   -1  1.e-16
EX  6  37  11  0   -0.93  -0.38
EX  6  38  11  0   -0.71  -0.71
EX  6  39  11  0   -0.37  -0.92
EX  6  40  11  0   6.e-17  -1
EX  6  41  11  0   0.381  -0.92
EX  6  42  11  0   0.71   -0.71
EX  6  43  11  0   0.93   -0.38
EX  6  44  11  0   1   0
EX  6  45  11  0   0.928  0.384
EX  6  46  11  0   0.707  0.707
EX  6  47  11  0   0.383  0.926
EX  6  48  11  0   6.e-17  1
EX  6  34  11  0   -0.71  0.707
GN  1
FR  0  1  0  0   27.2545455  0
EN
```

### B.1.3 L2.txt

```
CE
GW   9    21    -7   6.5699   2.5 -7   7.5699   2.5 1.e-3
GW   10   21    -9.9289 -0.5     2.5 -9.9289 0.5 2.5 1.e-3
GW   11   21    -7   -7.572   2.5 -7   -6.572   2.5 1.e-3
GW   12   21    0.0711  -10.5    2.5 0.0711  -9.5     2.5 1.e-3
GW   13   21    7.1422  -7.57    2.5 7.1422  -6.57    2.5 1.e-3
GW   14   21    10.0711 -0.5     2.5 10.0711 0.5 2.5 1.e-3
GW   15   21    7.1422   6.572   2.5 7.1422   7.572   2.5 1.e-3
GW   33   21    -7.6537 17.977   2.5 -7.6537 18.977   2.5 1.e-3
GW   34   21    -14.1421    13.64    2.5 -14.1421    14.64    2.5 1.e-3
GW   35   21    -18.4776    7.15     2.5 -18.4776    8.15     2.5 1.e-3
GW   36   21    -20 -0.5     2.5 -20 0.5 2.5 1.e-3
GW   37   21    -18.5   -8.25    2.5 -18.5   -7.25    2.5 1.e-3
GW   38   21    -14.1645    -14.7385    2.5 -14.1645    -13.7385    2.5 1.e-3
GW   39   21    -7.65375    -18.9775    2.5 -7.65375    -17.9775    2.5 1.e-3
GW   40   21    0   -20.5    2.5 0   -19.5    2.5 1.e-3
GW   41   21    7.6537  -18.9776    2.5 7.6537  -17.9776    2.5 1.e-3
GW   42   21    14.1421 -14.6422    2.5 14.1421 -13.6422    2.5 1.e-3
GW   43   21    18.4776 -8.1536 2.5 18.4776 -7.1536 2.5 1.e-3
GW   44   21    20  -0.5     2.5 20  0.5 2.5 1.e-3
GW   45   21    18.5    7.25     2.5 18.5    8.25     2.5 1.e-3
GW   46   21    14.1421 13.6421 2.5 14.1421 14.6421 2.5 1.e-3
GW   47   21    7.6537  17.9776 2.5 7.6537  18.9776 2.5 1.e-3
GW   48   21    0   19.5     2.5 0   20.5     2.5 1.e-3
GW   16   21    0.0711  9.5 2.5 0.0711  10.5     2.5 1.e-3
GE   0
EX   6    9    11   0    6.e-17  -1
EX   6    10   11   0    1    0
EX   6    11   11   0    6.e-17  1
EX   6    12   11   0    -1   1.e-16
EX   6    13   11   0    6.e-17  -1
EX   6    14   11   0    1    0
EX   6    15   11   0    6.e-17  1
EX   6    16   11   0    -1   1.e-16
EX   6    33   11   0    -0.71   -0.71
EX   6    35   11   0    0.709   -0.71
EX   6    36   11   0    1    0
EX   6    37   11   0    0.71    0.71
EX   6    38   11   0    6.e-17  1
EX   6    39   11   0    -0.7    0.702
EX   6    40   11   0    -1   1.e-16
EX   6    41   11   0    -0.71   -0.71
EX   6    42   11   0    6.e-17  -1
EX   6    43   11   0    0.711   -0.71
EX   6    44   11   0    1    0
```

```
EX   6    45   11   0    0.71     0.71
EX   6    46   11   0    6.e-17  1
EX   6    47   11   0    -0.71    0.709
EX   6    48   11   0    -1 1.e-16
EX   6    34   11   0    6.e-17  -1
GN   1
FR   0    1    0    0    27.2545455  0
EN
```

## B.1.4  L4.txt

```
CE
GW   9    21   -7  6.5699   2.5 -7  7.5699   2.5 1.e-3
GW   10   21   -9.9289 -0.5    2.5 -9.9289 0.5 2.5 1.e-3
GW   11   21   -7  -7.572  2.5 -7  -6.572  2.5 1.e-3
GW   12   21   0.0711  -10.5   2.5 0.0711  -9.5    2.5 1.e-3
GW   13   21   7.1422  -7.57   2.5 7.1422  -6.57   2.5 1.e-3
GW   14   21   10.0711 -0.5    2.5 10.0711 0.5 2.5 1.e-3
GW   15   21   7.1422  6.572   2.5 7.1422  7.572   2.5 1.e-3
GW   33   21   -7.6537 17.977  2.5 -7.6537 18.977  2.5 1.e-3
GW   34   21   -14.1421   13.64   2.5 -14.1421   14.64   2.5 1.e-3
GW   35   21   -18.4776   7.15    2.5 -18.4776   8.15    2.5 1.e-3
GW   36   21   -20 -0.5    2.5 -20 0.5 2.5 1.e-3
GW   37   21   -18.5   -8.25   2.5 -18.5   -7.25   2.5 1.e-3
GW   38   21   -14.1645   -14.7385    2.5 -14.1645   -13.7385    2.5 1.e-3
GW   39   21   -7.65375   -18.9775    2.5 -7.65375   -17.9775    2.5 1.e-3
GW   40   21   0   -20.5   2.5 0   -19.5   2.5 1.e-3
GW   41   21   7.6537  -18.9776    2.5 7.6537  -17.9776    2.5 1.e-3
GW   42   21   14.1421 -14.6422    2.5 14.1421 -13.6422    2.5 1.e-3
GW   43   21   18.4776 -8.1536 2.5 18.4776 -7.1536 2.5 1.e-3
GW   44   21   20  -0.5    2.5 20  0.5 2.5 1.e-3
GW   45   21   18.5    7.25    2.5 18.5    8.25    2.5 1.e-3
GW   46   21   14.1421 13.6421 2.5 14.1421 14.6421 2.5 1.e-3
GW   47   21   7.6537  17.9776 2.5 7.6537  18.9776 2.5 1.e-3
GW   48   21   0   19.5    2.5 0   20.5    2.5 1.e-3
GW   16   21   0.0711  9.5 2.5 0.0711  10.5    2.5 1.e-3
GE   0
EX   6    9    11   0    -1 1.e-16
EX   6    10   11   0    1   0
EX   6    11   11   0    -1 1.e-16
EX   6    12   11   0    1   0
EX   6    13   11   0    -1 1.e-16
EX   6    14   11   0    1   0
EX   6    15   11   0    -1 1.e-16
```

```
EX  6   16  11  0   1.001    0
EX  6   33  11  0   6.e-17  1
EX  6   35  11  0   0   -1
EX  6   36  11  0   1    0
EX  6   37  11  0   6.e-17  1
EX  6   38  11  0   -1  1.e-16
EX  6   39  11  0   0   -1
EX  6   40  11  0   1    0
EX  6   41  11  0   6.e-17  0.999
EX  6   42  11  0   -1  1.e-16
EX  6   43  11  0   0   -1.01
EX  6   44  11  0   1    0
EX  6   45  11  0   6.e-17  1.007
EX  6   46  11  0   -1  1.e-16
EX  6   47  11  0   0   -1
EX  6   48  11  0   1    0
EX  6   34  11  0   -1  1.e-16
GN  1
FR  0   1   0   0   27.2545455   0
EN
```

# B.2  Multiple beam interference

## B.2.1  L01_L02.txt

```
CE
GW  9   21  -7   6.5699   2.5 -7   7.5699   2.5 1.e-3
GW  10  21  -9.9289 -0.5     2.5 -9.9289 0.5 2.5 1.e-3
GW  11  21  -7   -7.572   2.5 -7   -6.572   2.5 1.e-3
GW  12  21  0.0711  -10.5    2.5 0.0711  -9.5     2.5 1.e-3
GW  13  21  7.1422  -7.57    2.5 7.1422  -6.57    2.5 1.e-3
GW  14  21  10.0711 -0.5     2.5 10.0711 0.5 2.5 1.e-3
GW  15  21  7.1422  6.572    2.5 7.1422  7.572    2.5 1.e-3
GW  33  21  -7.6537 17.977   2.5 -7.6537 18.977   2.5 1.e-3
GW  34  21  -14.1421    13.64    2.5 -14.1421    14.64    2.5 1.e-3
GW  35  21  -18.4776    7.15     2.5 -18.4776    8.15     2.5 1.e-3
GW  36  21  -20 -0.5     2.5 -20 0.5 2.5 1.e-3
GW  37  21  -18.5   -8.25    2.5 -18.5   -7.25    2.5 1.e-3
GW  38  21  -14.1645    -14.7385    2.5 -14.1645    -13.7385    2.5 1.e-3
GW  39  21  -7.65375    -18.9775    2.5 -7.65375    -17.9775    2.5 1.e-3
GW  40  21  0   -20.5    2.5 0   -19.5    2.5 1.e-3
GW  41  21  7.6537  -18.9776    2.5 7.6537  -17.9776    2.5 1.e-3
GW  42  21  14.1421 -14.6422    2.5 14.1421 -13.6422    2.5 1.e-3
```

```
GW  43  21  18.4776 -8.1536 2.5 18.4776 -7.1536 2.5 1.e-3
GW  44  21  20  -0.5     2.5 20  0.5 2.5 1.e-3
GW  45  21  18.5    7.25     2.5 18.5    8.25     2.5 1.e-3
GW  46  21  14.1421 13.6421 2.5 14.1421 14.6421 2.5 1.e-3
GW  47  21  7.6537 17.9776 2.5 7.6537 18.9776 2.5 1.e-3
GW  48  21  0   19.5    2.5 0   20.5    2.5 1.e-3
GW  16  21  0.0711  9.5 2.5 0.0711  10.5    2.5 1.e-3
GE  0
EX  6   9   11  0   -0.71   0.707
EX  6   10  11  0   -1  1.e-16
EX  6   11  11  0   -0.71   -0.71
EX  6   12  11  0   6.e-17  -1
EX  6   13  11  0   0.71    -0.71
EX  6   14  11  0   1   0
EX  6   15  11  0   0.709   0.709
EX  6   16  11  0   6.e-17  1.001
EX  6   33  11  0   -0.71   -0.71
EX  6   35  11  0   0.705   -0.7
EX  6   36  11  0   1   0
EX  6   37  11  0   0.701   0.701
EX  6   38  11  0   6.e-17  1.004
EX  6   39  11  0   -0.7    0.701
EX  6   40  11  0   -1  1.e-16
EX  6   41  11  0   -0.7    -0.7
EX  6   42  11  0   6.e-17  -1
EX  6   43  11  0   0.713   -0.71
EX  6   44  11  0   1   0
EX  6   45  11  0   0.712   0.712
EX  6   46  11  0   6.e-17  1.003
EX  6   47  11  0   -0.71   0.707
EX  6   48  11  0   -1  1.e-16
EX  6   34  11  0   6.e-17  -1
GN  1
FR  0   1   0   0   14.99   0
EN
```

### B.2.2  L01_L04.txt

```
CE
GW  9   21  -7  6.5699  2.5 -7  7.5699  2.5 1.e-3
GW  10  21  -9.9289 -0.5     2.5 -9.9289 0.5 2.5 1.e-3
GW  11  21  -7  -7.572  2.5 -7  -6.572  2.5 1.e-3
GW  12  21  0.0711  -10.5   2.5 0.0711  -9.5     2.5 1.e-3
GW  13  21  7.1422  -7.57   2.5 7.1422  -6.57    2.5 1.e-3
```

```
GW   14   21   10.0711 -0.5     2.5 10.0711 0.5 2.5 1.e-3
GW   15   21   7.1422  6.572    2.5 7.1422  7.572   2.5 1.e-3
GW   33   21   -7.6537 17.977   2.5 -7.6537 18.977  2.5 1.e-3
GW   34   21   -14.1421    13.64    2.5 -14.1421    14.64    2.5 1.e-3
GW   35   21   -18.4776    7.15     2.5 -18.4776    8.15     2.5 1.e-3
GW   36   21   -20 -0.5    2.5 -20 0.5 2.5 1.e-3
GW   37   21   -18.5   -8.25    2.5 -18.5    -7.25    2.5 1.e-3
GW   38   21   -14.1645    -14.7385    2.5 -14.1645    -13.7385    2.5 1.e-3
GW   39   21   -7.65375    -18.9775    2.5 -7.65375    -17.9775    2.5 1.e-3
GW   40   21   0   -20.5   2.5 0   -19.5   2.5 1.e-3
GW   41   21   7.6537  -18.9776    2.5 7.6537  -17.9776    2.5 1.e-3
GW   42   21   14.1421 -14.6422    2.5 14.1421 -13.6422    2.5 1.e-3
GW   43   21   18.4776 -8.1536 2.5 18.4776 -7.1536 2.5 1.e-3
GW   44   21   20  -0.5    2.5 20  0.5 2.5 1.e-3
GW   45   21   18.5    7.25    2.5 18.5    8.25    2.5 1.e-3
GW   46   21   14.1421 13.6421 2.5 14.1421 14.6421 2.5 1.e-3
GW   47   21   7.6537  17.9776 2.5 7.6537  18.9776 2.5 1.e-3
GW   48   21   0   19.5    2.5 0   20.5    2.5 1.e-3
GW   16   21   0.0711  9.5 2.5 0.0711  10.5    2.5 1.e-3
GE   0
EX   6   9    11   0   -0.71   0.707
EX   6   10   11   0   -1  1.e-16
EX   6   11   11   0   -0.71   -0.71
EX   6   12   11   0   6.e-17  -1
EX   6   13   11   0   0.71    -0.71
EX   6   14   11   0   1   0
EX   6   15   11   0   0.709   0.709
EX   6   16   11   0   6.e-17  1.001
EX   6   33   11   0   6.e-17  1.004
EX   6   35   11   0   6.e-17  -0.99
EX   6   36   11   0   1   0
EX   6   37   11   0   6.e-17  0.991
EX   6   38   11   0   -1  1.e-16
EX   6   39   11   0   6.e-17  -0.99
EX   6   40   11   0   1   0
EX   6   41   11   0   6.e-17  0.99
EX   6   42   11   0   -1  1.e-16
EX   6   43   11   0   6.e-17  -1.01
EX   6   44   11   0   1   0
EX   6   45   11   0   6.e-17  1.007
EX   6   46   11   0   -1  1.e-16
EX   6   47   11   0   6.e-17  -1
EX   6   48   11   0   1   0
EX   6   34   11   0   -1  1.e-16
GN   1
```

```
FR  0   1   0   0   14.99   0
EN
```

### B.2.3 L02_L04.txt

```
CE
GW  9    21   -7  6.5699  2.5 -7  7.5699  2.5 1.e-3
GW  10   21   -9.9289 -0.5     2.5 -9.9289 0.5 2.5 1.e-3
GW  11   21   -7  -7.572  2.5 -7  -6.572  2.5 1.e-3
GW  12   21   0.0711  -10.5   2.5 0.0711  -9.5     2.5 1.e-3
GW  13   21   7.1422  -7.57   2.5 7.1422  -6.57    2.5 1.e-3
GW  14   21   10.0711 -0.5     2.5 10.0711 0.5 2.5 1.e-3
GW  15   21   7.1422  6.572   2.5 7.1422  7.572   2.5 1.e-3
GW  33   21   -7.6537 17.977   2.5 -7.6537 18.977   2.5 1.e-3
GW  34   21   -14.1421    13.64   2.5 -14.1421    14.64    2.5 1.e-3
GW  35   21   -18.4776    7.15    2.5 -18.4776    8.15     2.5 1.e-3
GW  36   21   -20 -0.5     2.5 -20 0.5 2.5 1.e-3
GW  37   21   -18.5   -8.25   2.5 -18.5   -7.25    2.5 1.e-3
GW  38   21   -14.1645    -14.7385    2.5 -14.1645    -13.7385    2.5 1.e-3
GW  39   21   -7.65375    -18.9775    2.5 -7.65375    -17.9775    2.5 1.e-3
GW  40   21   0   -20.5   2.5 0   -19.5   2.5 1.e-3
GW  41   21   7.6537  -18.9776    2.5 7.6537  -17.9776    2.5 1.e-3
GW  42   21   14.1421 -14.6422    2.5 14.1421 -13.6422    2.5 1.e-3
GW  43   21   18.4776 -8.1536 2.5 18.4776 -7.1536 2.5 1.e-3
GW  44   21   20  -0.5     2.5 20  0.5 2.5 1.e-3
GW  45   21   18.5    7.25    2.5 18.5    8.25    2.5 1.e-3
GW  46   21   14.1421 13.6421 2.5 14.1421 14.6421 2.5 1.e-3
GW  47   21   7.6537  17.9776 2.5 7.6537  18.9776 2.5 1.e-3
GW  48   21   0   19.5    2.5 0   20.5    2.5 1.e-3
GW  16   21   0.0711  9.5 2.5 0.0711  10.5    2.5 1.e-3
GE  0
EX  6    9    11  0   6.e-17  -1
EX  6    10   11  0   1   0
EX  6    11   11  0   6.e-17  1.004
EX  6    12   11  0   -1  1.e-16
EX  6    13   11  0   6.e-17  -1
EX  6    14   11  0   1   0
EX  6    15   11  0   6.e-17  1.003
EX  6    16   11  0   -1  1.e-16
EX  6    33   11  0   6.e-17  1.004
EX  6    35   11  0   6.e-17  -0.99
EX  6    36   11  0   1   0
EX  6    37   11  0   6.e-17  0.991
EX  6    38   11  0   -1  1.e-16
```

```
EX   6    39   11   0    6.e-17   -0.99
EX   6    40   11   0    1    0
EX   6    41   11   0    6.e-17   0.99
EX   6    42   11   0    -1   1.e-16
EX   6    43   11   0    6.e-17   -1.01
EX   6    44   11   0    1    0
EX   6    45   11   0    6.e-17   1.007
EX   6    46   11   0    -1   1.e-16
EX   6    47   11   0    6.e-17   -1
EX   6    48   11   0    1    0
EX   6    34   11   0    -1   1.e-16
GN   1
FR   0    1    0    0    14.99    0
EN
```

# B.3  Rotation of main lobe

## B.3.1  L01_L02_45deg.txt

```
CE
GW   9    21   -7   6.5699   2.5 -7   7.5699   2.5 1.e-3
GW   10   21   -9.9289 -0.5     2.5 -9.9289 0.5 2.5 1.e-3
GW   11   21   -7   -7.572   2.5 -7   -6.572   2.5 1.e-3
GW   12   21   0.0711   -10.5    2.5 0.0711   -9.5    2.5 1.e-3
GW   13   21   7.1422   -7.57    2.5 7.1422   -6.57    2.5 1.e-3
GW   14   21   10.0711  -0.5     2.5 10.0711  0.5 2.5 1.e-3
GW   15   21   7.1422   6.572    2.5 7.1422   7.572    2.5 1.e-3
GW   33   21   -7.6537 17.977    2.5 -7.6537 18.977    2.5 1.e-3
GW   34   21   -14.1421    13.64    2.5 -14.1421    14.64    2.5 1.e-3
GW   35   21   -18.4776    7.15     2.5 -18.4776    8.15     2.5 1.e-3
GW   36   21   -20 -0.5     2.5 -20 0.5 2.5 1.e-3
GW   37   21   -18.5    -8.25    2.5 -18.5    -7.25    2.5 1.e-3
GW   38   21   -14.1645    -14.7385    2.5 -14.1645    -13.7385    2.5 1.e-3
GW   39   21   -7.65375    -18.9775    2.5 -7.65375    -17.9775    2.5 1.e-3
GW   40   21   0    -20.5    2.5 0    -19.5    2.5 1.e-3
GW   41   21   7.6537   -18.9776    2.5 7.6537   -17.9776    2.5 1.e-3
GW   42   21   14.1421 -14.6422     2.5 14.1421 -13.6422     2.5 1.e-3
GW   43   21   18.4776 -8.1536 2.5 18.4776 -7.1536 2.5 1.e-3
GW   44   21   20   -0.5     2.5 20   0.5 2.5 1.e-3
GW   45   21   18.5     7.25     2.5 18.5     8.25     2.5 1.e-3
GW   46   21   14.1421 13.6421 2.5 14.1421 14.6421 2.5 1.e-3
GW   47   21   7.6537   17.9776 2.5 7.6537   18.9776 2.5 1.e-3
GW   48   21   0    19.5     2.5 0    20.5     2.5 1.e-3
```

```
GW  16   21   0.0711   9.5 2.5 0.0711  10.5    2.5 1.e-3
GE  0
EX  6    9    11  0    -0.71    0.707
EX  6    10   11  0    -1  1.e-16
EX  6    11   11  0    -0.71   -0.71
EX  6    12   11  0    6.e-17  -1
EX  6    13   11  0    0.71    -0.71
EX  6    14   11  0    1    0
EX  6    15   11  0    0.709    0.709
EX  6    16   11  0    6.e-17   1
EX  6    33   11  0    -1  1.e-16
EX  6    35   11  0    6.e-17  -0.99
EX  6    36   11  0    0.707   -0.71
EX  6    37   11  0    0.991    0
EX  6    38   11  0    0.71     0.71
EX  6    39   11  0    6.e-17   0.991
EX  6    40   11  0    -0.71    0.707
EX  6    41   11  0    -0.99   1.e-16
EX  6    42   11  0    -0.71   -0.71
EX  6    43   11  0    6.e-17  -1.01
EX  6    44   11  0    0.707   -0.71
EX  6    45   11  0    1.007    0
EX  6    46   11  0    0.709    0.709
EX  6    47   11  0    6.e-17   1.002
EX  6    48   11  0    -0.71    0.707
EX  6    34   11  0    -0.71   -0.71
GN  1
FR  0    1    0   0    14.99    0
EN
```

# B.4  Non linearly polarized beams

## B.4.1  Five element crossed dipole array

### Left-hand polarized, $l = -2$

```
CE
GW  1    3    0.2 0.03     0.01      0.2 -0.03    0.01     1.e-3
GW  2    3    0.06      0.22     0.01      0.06      0.16     0.01     1.e-3
GW  3    3    -0.17    0.14     0.01      -0.17    0.08     0.01     1.e-3
GW  4    3    -0.16    -0.09    0.01      -0.16    -0.15    0.01     1.e-3
GW  5    3    0.06     -0.16    0.01      0.06     -0.22    0.01     1.e-3
GW  6    3    0.17     -1.388e-17  9.e-3   0.23     0       9.e-3   1.e-3
```

```
GW   7   3    0.03      0.19     9.e-3    0.09     0.19     9.e-3    1.e-3
GW   8   3    -0.2      0.11     9.e-3    -0.14    0.11     9.e-3    1.e-3
GW   9   3    -0.19     -0.12    9.e-3    -0.13    -0.12    9.e-3    1.e-3
GW   10  3    0.03      -0.19    9.e-3    0.09     -0.19    9.e-3    1.e-3
GE   0
EX   0   1    2    0    1    0
EX   0   2    2    0    -0.81    -0.59
EX   0   3    2    0    0.309    0.95
EX   0   4    2    0    0.309    -0.95
EX   0   5    2    0    -0.81    0.589
EX   0   6    2    0    6.e-17   1
EX   0   7    2    0    0.589    -0.81
EX   0   8    2    0    -0.95    0.309
EX   0   9    2    0    0.95     0.309
EX   0   10   2    0    -0.59    -0.81
GN   1
FR   0   1    0    0    300 0
EN
```

Left-hand polarized, $l = -1$

```
CE
GW   1   3    0.2 0.03    0.01     0.2 -0.03   0.01     1.e-3
GW   2   3    0.06      0.22     0.01     0.06     0.16     0.01     1.e-3
GW   3   3    -0.17     0.14     0.01     -0.17    0.08     0.01     1.e-3
GW   4   3    -0.16     -0.09    0.01     -0.16    -0.15    0.01     1.e-3
GW   5   3    0.06      -0.16    0.01     0.06     -0.22    0.01     1.e-3
GW   6   3    0.17      -1.388e-17 9.e-3   0.23     0        9.e-3    1.e-3
GW   7   3    0.03      0.19     9.e-3    0.09     0.19     9.e-3    1.e-3
GW   8   3    -0.2      0.11     9.e-3    -0.14    0.11     9.e-3    1.e-3
GW   9   3    -0.19     -0.12    9.e-3    -0.13    -0.12    9.e-3    1.e-3
GW   10  3    0.03      -0.19    9.e-3    0.09     -0.19    9.e-3    1.e-3
GE   0
EX   0   1    2    0    1    0
EX   0   2    2    0    0.31     -0.95
EX   0   3    2    0    -0.81    -0.59
EX   0   4    2    0    -0.81    0.587
EX   0   5    2    0    0.31     0.953
EX   0   6    2    0    6.e-17   1
EX   0   7    2    0    0.953    0.31
EX   0   8    2    0    0.587    -0.81
EX   0   9    2    0    -0.59    -0.81
EX   0   10   2    0    -0.95    0.31
GN   1
```

```
FR  0   1   0   0   300 0
EN
```

Left-hand polarized, *l* = 0

```
CE
GW   1   3    0.2 0.03    0.01     0.2 −0.03    0.01       1.e−3
GW   2   3    0.06      0.22     0.01     0.06     0.16     0.01       1.e−3
GW   3   3    −0.17     0.14     0.01     −0.17    0.08     0.01       1.e−3
GW   4   3    −0.16     −0.09    0.01     −0.16    −0.15    0.01       1.e−3
GW   5   3    0.06      −0.16    0.01     0.06     −0.22    0.01       1.e−3
GW   6   3    0.17      −1.388e−17 9.e−3   0.23     0       9.e−3      1.e−3
GW   7   3    0.03      0.19     9.e−3    0.09     0.19     9.e−3      1.e−3
GW   8   3    −0.2      0.11     9.e−3    −0.14    0.11     9.e−3      1.e−3
GW   9   3    −0.19     −0.12    9.e−3    −0.13    −0.12    9.e−3      1.e−3
GW  10   3    0.03      −0.19    9.e−3    0.09     −0.19    9.e−3      1.e−3
GE   0
EX   0   1   2   0   1    0
EX   0   2   2   0   0.999    0
EX   0   3   2   0   1.002    0
EX   0   4   2   0   1    0
EX   0   5   2   0   1.002    0
EX   0   6   2   0   6.e−17   1
EX   0   7   2   0   6.e−17   1.002
EX   0   8   2   0   6.e−17   1
EX   0   9   2   0   6.e−17   1.002
EX   0  10   2   0   6.e−17   0.999
GN   1
FR   0   1   0   0   300 0
EN
```

Left-hand polarized, *l* = 1

```
CE
GW   1   3    0.2 0.03    0.01     0.2 −0.03    0.01       1.e−3
GW   2   3    0.06      0.22     0.01     0.06     0.16     0.01       1.e−3
GW   3   3    −0.17     0.14     0.01     −0.17    0.08     0.01       1.e−3
GW   4   3    −0.16     −0.09    0.01     −0.16    −0.15    0.01       1.e−3
GW   5   3    0.06      −0.16    0.01     0.06     −0.22    0.01       1.e−3
GW   6   3    0.17      −1.388e−17 9.e−3   0.23     0       9.e−3      1.e−3
GW   7   3    0.03      0.19     9.e−3    0.09     0.19     9.e−3      1.e−3
GW   8   3    −0.2      0.11     9.e−3    −0.14    0.11     9.e−3      1.e−3
GW   9   3    −0.19     −0.12    9.e−3    −0.13    −0.12    9.e−3      1.e−3
```

```
GW   10   3   0.03      -0.19    9.e-3    0.09     -0.19    9.e-3    1.e-3
GE   0
EX   0   1   2   0   1   0
EX   0   2   2   0   0.309    0.95
EX   0   3   2   0   -0.81    0.589
EX   0   4   2   0   -0.81    -0.59
EX   0   5   2   0   0.309    -0.95
EX   0   6   2   0   6.e-17   1
EX   0   7   2   0   -0.95    0.31
EX   0   8   2   0   -0.59    -0.81
EX   0   9   2   0   0.589    -0.81
EX   0   10  2   0   0.95     0.309
GN   1
FR   0   1   0   0   300 0
EN
```

Left-hand polarized, $l = 2$

```
CE
GW   1    3   0.2 0.03    0.01    0.2 -0.03   0.01    1.e-3
GW   2    3   0.06      0.22     0.01     0.06     0.16     0.01     1.e-3
GW   3    3   -0.17     0.14     0.01     -0.17    0.08     0.01     1.e-3
GW   4    3   -0.16     -0.09    0.01     -0.16    -0.15    0.01     1.e-3
GW   5    3   0.06      -0.16    0.01     0.06     -0.22    0.01     1.e-3
GW   6    3   0.17      -1.388e-17  9.e-3   0.23     0        9.e-3    1.e-3
GW   7    3   0.03      0.19     9.e-3    0.09     0.19     9.e-3    1.e-3
GW   8    3   -0.2      0.11     9.e-3    -0.14    0.11     9.e-3    1.e-3
GW   9    3   -0.19     -0.12    9.e-3    -0.13    -0.12    9.e-3    1.e-3
GW   10   3   0.03      -0.19    9.e-3    0.09     -0.19    9.e-3    1.e-3
GE   0
EX   0   1   2   0   1   0
EX   0   2   2   0   -0.81    0.586
EX   0   3   2   0   0.309    -0.95
EX   0   4   2   0   0.31     0.953
EX   0   5   2   0   -0.81    -0.59
EX   0   6   2   0   6.e-17   1
EX   0   7   2   0   -0.59    -0.81
EX   0   8   2   0   0.953    0.31
EX   0   9   2   0   -0.95    0.309
EX   0   10  2   0   0.586    -0.81
GN   1
FR   0   1   0   0   299.8    0
EN
```

Right-hand polarized, $l = -2$

```
CE
GW   1    3    0.2 0.03    0.01      0.2  -0.03    0.01    1.e-3
GW   2    3    0.06       0.22      0.01     0.06      0.16     0.01    1.e-3
GW   3    3    -0.17      0.14      0.01     -0.17     0.08     0.01    1.e-3
GW   4    3    -0.16      -0.09     0.01     -0.16     -0.15    0.01    1.e-3
GW   5    3    0.06       -0.16     0.01     0.06      -0.22    0.01    1.e-3
GW   6    3    0.17       -1.388e-17  9.e-3   0.23     0        9.e-3   1.e-3
GW   7    3    0.03       0.19      9.e-3    0.09      0.19     9.e-3   1.e-3
GW   8    3    -0.2       0.11      9.e-3    -0.14     0.11     9.e-3   1.e-3
GW   9    3    -0.19      -0.12     9.e-3    -0.13     -0.12    9.e-3   1.e-3
GW   10   3    0.03       -0.19     9.e-3    0.09      -0.19    9.e-3   1.e-3
GE   0
EX   0    1    2    0    6.e-17   1
EX   0    2    2    0    0.586    -0.81
EX   0    3    2    0    -0.95    0.31
EX   0    4    2    0    0.953    0.31
EX   0    5    2    0    -0.59    -0.81
EX   0    6    2    0    1    0
EX   0    7    2    0    -0.81    -0.59
EX   0    8    2    0    0.309    0.95
EX   0    9    2    0    0.31     -0.95
EX   0    10   2    0    -0.81    0.589
GN   1
FR   0    1    0    0    300 0
EN
```

Right-hand polarized, $l = -1$

```
CE
GW   1    3    0.2 0.03    0.01      0.2  -0.03    0.01    1.e-3
GW   2    3    0.06       0.22      0.01     0.06      0.16     0.01    1.e-3
GW   3    3    -0.17      0.14      0.01     -0.17     0.08     0.01    1.e-3
GW   4    3    -0.16      -0.09     0.01     -0.16     -0.15    0.01    1.e-3
GW   5    3    0.06       -0.16     0.01     0.06      -0.22    0.01    1.e-3
GW   6    3    0.17       -1.388e-17  9.e-3   0.23     0        9.e-3   1.e-3
GW   7    3    0.03       0.19      9.e-3    0.09      0.19     9.e-3   1.e-3
GW   8    3    -0.2       0.11      9.e-3    -0.14     0.11     9.e-3   1.e-3
GW   9    3    -0.19      -0.12     9.e-3    -0.13     -0.12    9.e-3   1.e-3
GW   10   3    0.03       -0.19     9.e-3    0.09      -0.19    9.e-3   1.e-3
GE   0
EX   0    1    2    0    6.e-17   1
```

```
EX  0   2    2   0   0.95     0.309
EX  0   3    2   0   0.589    -0.81
EX  0   4    2   0   -0.59    -0.81
EX  0   5    2   0   -0.95    0.31
EX  0   6    2   0   1   0
EX  0   7    2   0   0.31     -0.95
EX  0   8    2   0   -0.81    -0.59
EX  0   9    2   0   -0.81    0.589
EX  0   10   2   0   0.309    0.95
GN  1
FR  0   1    0   0   300 0
EN
```

Right-hand polarized, $l = 0$

```
CE
GW  1    3    0.2 0.03    0.01      0.2 -0.03    0.01     1.e-3
GW  2    3    0.06     0.22     0.01     0.06     0.16     0.01     1.e-3
GW  3    3    -0.17    0.14     0.01     -0.17    0.08     0.01     1.e-3
GW  4    3    -0.16    -0.09    0.01     -0.16    -0.15    0.01     1.e-3
GW  5    3    0.06     -0.16    0.01     0.06     -0.22    0.01     1.e-3
GW  6    3    0.17     -1.388e-17  9.e-3  0.23     0        9.e-3   1.e-3
GW  7    3    0.03     0.19     9.e-3    0.09     0.19     9.e-3    1.e-3
GW  8    3    -0.2     0.11     9.e-3    -0.14    0.11     9.e-3    1.e-3
GW  9    3    -0.19    -0.12    9.e-3    -0.13    -0.12    9.e-3    1.e-3
GW  10   3    0.03     -0.19    9.e-3    0.09     -0.19    9.e-3    1.e-3
GE  0
EX  0   1    2   0   6.e-17   1
EX  0   2    2   0   6.e-17   0.999
EX  0   3    2   0   6.e-17   1.002
EX  0   4    2   0   6.e-17   1
EX  0   5    2   0   6.e-17   1.002
EX  0   6    2   0   1   0
EX  0   7    2   0   1.002    0
EX  0   8    2   0   1   0
EX  0   9    2   0   1.002    0
EX  0   10   2   0   0.999    0
GN  1
FR  0   1    0   0   300 0
EN
```

Right-hand polarized, $l = 1$

```
CE
GW  1   3   0.2 0.03    0.01     0.2 -0.03   0.01     1.e-3
GW  2   3   0.06     0.22      0.01    0.06     0.16     0.01     1.e-3
GW  3   3   -0.17    0.14      0.01    -0.17    0.08     0.01     1.e-3
GW  4   3   -0.16    -0.09     0.01    -0.16    -0.15    0.01     1.e-3
GW  5   3   0.06     -0.16     0.01    0.06     -0.22    0.01     1.e-3
GW  6   3   0.17     -1.388e-17  9.e-3   0.23     0       9.e-3    1.e-3
GW  7   3   0.03     0.19      9.e-3   0.09     0.19     9.e-3    1.e-3
GW  8   3   -0.2     0.11      9.e-3   -0.14    0.11     9.e-3    1.e-3
GW  9   3   -0.19    -0.12     9.e-3   -0.13    -0.12    9.e-3    1.e-3
GW  10  3   0.03     -0.19     9.e-3   0.09     -0.19    9.e-3    1.e-3
GE  0
EX  0   1   2   0   6.e-17   1
EX  0   2   2   0   -0.95    0.31
EX  0   3   2   0   -0.59    -0.81
EX  0   4   2   0   0.586    -0.81
EX  0   5   2   0   0.953    0.31
EX  0   6   2   0   1   0
EX  0   7   2   0   0.31     0.953
EX  0   8   2   0   -0.81    0.587
EX  0   9   2   0   -0.81    -0.59
EX  0   10  2   0   0.31     -0.95
GN  1
FR  0   1   0   0   300 0
EN
```

Right-hand polarized, $l = 2$

```
CE
GW  1   3   0.2 0.03    0.01     0.2 -0.03   0.01     1.e-3
GW  2   3   0.06     0.22      0.01    0.06     0.16     0.01     1.e-3
GW  3   3   -0.17    0.14      0.01    -0.17    0.08     0.01     1.e-3
GW  4   3   -0.16    -0.09     0.01    -0.16    -0.15    0.01     1.e-3
GW  5   3   0.06     -0.16     0.01    0.06     -0.22    0.01     1.e-3
GW  6   3   0.17     -1.388e-17  9.e-3   0.23     0       9.e-3    1.e-3
GW  7   3   0.03     0.19      9.e-3   0.09     0.19     9.e-3    1.e-3
GW  8   3   -0.2     0.11      9.e-3   -0.14    0.11     9.e-3    1.e-3
GW  9   3   -0.19    -0.12     9.e-3   -0.13    -0.12    9.e-3    1.e-3
GW  10  3   0.03     -0.19     9.e-3   0.09     -0.19    9.e-3    1.e-3
GE  0
EX  0   1   2   0   6.e-17   1
EX  0   2   2   0   -0.59    -0.81
EX  0   3   2   0   0.953    0.31
EX  0   4   2   0   -0.95    0.309
```

```
EX   0   5    2    0     0.589     -0.81
EX   0   6    2    0     1    0
EX   0   7    2    0    -0.81      0.589
EX   0   8    2    0     0.309    -0.95
EX   0   9    2    0     0.31      0.953
EX   0   10   2    0    -0.81     -0.59
GN   1
FR   0   1    0    0     300  0
EN
```

## B.4.2  Ten element crossed dipole array

Left-hand polarized, $l = -4$, $a = \lambda$

```
CE
GW   1    5    0.9  0.31     0.1  1     0.31      0.1 1.e-3
GW   2    5    0.9 -0.31     0.1  1    -0.31      0.1 1.e-3
GW   3    5    0.54      -0.81   0.1  0.64     -0.81    0.1 1.e-3
GW   4    5   -0.05      -1    0.1  0.05      -1    0.1 1.e-3
GW   5    5   -0.64      -0.81   0.1 -0.54     -0.81    0.1 1.e-3
GW   6    5   -1    -0.31     0.1 -0.9     -0.31    0.1 1.e-3
GW   7    5   -1    0.31      0.1 -0.9      0.31     0.1 1.e-3
GW   8    5   -0.64      0.81    0.1 -0.54      0.81     0.1 1.e-3
GW   9    5   -0.05      1    0.1  0.05      1     0.1 1.e-3
GW   10   5    0.54      0.81    0.1  0.64      0.81     0.1 1.e-3
GW   11   5    0    1.05      0.099   0     0.95      0.099    1.e-3
GW   12   5    0.59      0.86     0.099   0.59      0.76     0.099    1.e-3
GW   13   5    0.95      0.36     0.099   0.95      0.26     0.099    1.e-3
GW   14   5    0.95     -0.26     0.099   0.95     -0.36     0.099    1.e-3
GW   15   5    0.59     -0.76     0.099   0.59     -0.86     0.099    1.e-3
GW   16   5    0   -0.95     0.099   0    -1.05     0.099    1.e-3
GW   17   5   -0.59     -0.76     0.099  -0.59     -0.86     0.099    1.e-3
GW   18   5   -0.95     -0.26     0.099  -0.95     -0.36     0.099    1.e-3
GW   19   5   -0.95      0.36     0.099  -0.95      0.26     0.099    1.e-3
GW   20   5   -0.59      0.86     0.099  -0.59      0.76     0.099    1.e-3
GE   0
EK
EX   0   1    3    0     0.95      0.309
EX   0   10   3    0    -0.59     -0.81
EX   0   9    3    0     6.e-16    1
EX   0   8    3    0     0.589    -0.81
EX   0   7    3    0    -0.95      0.309
EX   0   6    3    0     0.95      0.309
EX   0   5    3    0    -0.59     -0.81
```

```
EX  0   4   3   0   6.e-17  1
EX  0   3   3   0   0.589   -0.81
EX  0   2   3   0   -0.95   0.309
EX  0   13  3   0   0.309   -0.95
EX  0   12  3   0   -0.81   0.589
EX  0   11  3   0   1   0
EX  0   20  3   0   -0.81   -0.59
EX  0   19  3   0   0.309   0.95
EX  0   18  3   0   0.309   -0.95
EX  0   17  3   0   -0.81   0.589
EX  0   16  3   0   1   0
EX  0   15  3   0   -0.81   -0.59
EX  0   14  3   0   0.309   0.95
GN  1
FR  0   1   0   0   300 0
EN
```

Left-hand polarized, $l = -3$, $a = \lambda$

```
CE
GW  1   5   0.9 0.31    0.1 1   0.31    0.1 1.e-3
GW  2   5   0.9 -0.31   0.1 1   -0.31   0.1 1.e-3
GW  3   5   0.54    -0.81   0.1 0.64    -0.81   0.1 1.e-3
GW  4   5   -0.05   -1  0.1 0.05    -1  0.1 1.e-3
GW  5   5   -0.64   -0.81   0.1 -0.54   -0.81   0.1 1.e-3
GW  6   5   -1  -0.31   0.1 -0.9    -0.31   0.1 1.e-3
GW  7   5   -1  0.31    0.1 -0.9    0.31    0.1 1.e-3
GW  8   5   -0.64   0.81    0.1 -0.54   0.81    0.1 1.e-3
GW  9   5   -0.05   1   0.1 0.05    1   0.1 1.e-3
GW  10  5   0.54    0.81    0.1 0.64    0.81    0.1 1.e-3
GW  11  5   0   1.05    0.099   0   0.95    0.099   1.e-3
GW  12  5   0.59    0.86    0.099   0.59    0.76    0.099   1.e-3
GW  13  5   0.95    0.36    0.099   0.95    0.26    0.099   1.e-3
GW  14  5   0.95    -0.26   0.099   0.95    -0.36   0.099   1.e-3
GW  15  5   0.59    -0.76   0.099   0.59    -0.86   0.099   1.e-3
GW  16  5   0   -0.95   0.099   0   -1.05   0.099   1.e-3
GW  17  5   -0.59   -0.76   0.099   -0.59   -0.86   0.099   1.e-3
GW  18  5   -0.95   -0.26   0.099   -0.95   -0.36   0.099   1.e-3
GW  19  5   -0.95   0.36    0.099   -0.95   0.26    0.099   1.e-3
GW  20  5   -0.59   0.86    0.099   -0.59   0.76    0.099   1.e-3
GE  0
EK
EX  0   1   3   0   0.809   0.588
EX  0   10  3   0   0.309   -0.95
```

```
EX  0   9   3   0   -1  6.e-16
EX  0   8   3   0   0.31    0.953
EX  0   7   3   0   0.811   -0.59
EX  0   6   3   0   -0.81   -0.59
EX  0   5   3   0   -0.31   0.953
EX  0   4   3   0   1   0
EX  0   3   3   0   -0.31   -0.95
EX  0   2   3   0   -0.81   0.587
EX  0   13  3   0   0.587   -0.81
EX  0   12  3   0   -0.95   -0.31
EX  0   11  3   0   6.e-17  1
EX  0   20  3   0   0.953   -0.31
EX  0   19  3   0   -0.59   -0.81
EX  0   18  3   0   -0.59   0.811
EX  0   17  3   0   0.953   0.31
EX  0   16  3   0   0   -1
EX  0   15  3   0   -0.95   0.309
EX  0   14  3   0   0.586   0.81
GN  1
FR  0   1   0   0   300 0
EN
```

Left-hand polarized, $l = -2$, $a = \lambda$

```
CE
GW  1   5   0.9 0.31    0.1 1   0.31    0.1 1.e-3
GW  2   5   0.9 -0.31   0.1 1   -0.31   0.1 1.e-3
GW  3   5   0.54    -0.81   0.1 0.64    -0.81   0.1 1.e-3
GW  4   5   -0.05   -1  0.1 0.05    -1  0.1 1.e-3
GW  5   5   -0.64   -0.81   0.1 -0.54   -0.81   0.1 1.e-3
GW  6   5   -1  -0.31   0.1 -0.9    -0.31   0.1 1.e-3
GW  7   5   -1  0.31    0.1 -0.9    0.31    0.1 1.e-3
GW  8   5   -0.64   0.81    0.1 -0.54   0.81    0.1 1.e-3
GW  9   5   -0.05   1   0.1 0.05    1   0.1 1.e-3
GW  10  5   0.54    0.81    0.1 0.64    0.81    0.1 1.e-3
GW  11  5   0   1.05    0.099   0   0.95    0.099   1.e-3
GW  12  5   0.59    0.86    0.099   0.59    0.76    0.099   1.e-3
GW  13  5   0.95    0.36    0.099   0.95    0.26    0.099   1.e-3
GW  14  5   0.95    -0.26   0.099   0.95    -0.36   0.099   1.e-3
GW  15  5   0.59    -0.76   0.099   0.59    -0.86   0.099   1.e-3
GW  16  5   0   -0.95   0.099   0   -1.05   0.099   1.e-3
GW  17  5   -0.59   -0.76   0.099   -0.59   -0.86   0.099   1.e-3
GW  18  5   -0.95   -0.26   0.099   -0.95   -0.36   0.099   1.e-3
GW  19  5   -0.95   0.36    0.099   -0.95   0.26    0.099   1.e-3
```

```
GW  20  5   -0.59   0.86    0.099   -0.59   0.76    0.099   1.e-3
GE  0
EK
EX  0   1   3   0   0.589   0.811
EX  0   10  3   0   0.95    -0.31
EX  0   9   3   0   0   -1
EX  0   8   3   0   -0.95   -0.31
EX  0   7   3   0   -0.59   0.811
EX  0   6   3   0   0.589   0.811
EX  0   5   3   0   0.95    -0.31
EX  0   4   3   0   0   -1
EX  0   3   3   0   -0.95   -0.31
EX  0   2   3   0   -0.59   0.81
EX  0   13  3   0   0.81    -0.59
EX  0   12  3   0   -0.31   -0.95
EX  0   11  3   0   -1  4.e-16
EX  0   20  3   0   -0.31   0.95
EX  0   19  3   0   0.811   0.59
EX  0   18  3   0   0.811   -0.59
EX  0   17  3   0   -0.31   -0.95
EX  0   16  3   0   -1  1.e-16
EX  0   15  3   0   -0.31   0.95
EX  0   14  3   0   0.81    0.588
GN  1
FR  0   1   0   0   300 0
EN
```

Left-hand polarized, $l = 0$, $a = \lambda$

```
CE
GW  1   5   0.9 0.31    0.1 1   0.31    0.1 1.e-3
GW  2   5   0.9 -0.31   0.1 1   -0.31   0.1 1.e-3
GW  3   5   0.54    -0.81   0.1 0.64    -0.81   0.1 1.e-3
GW  4   5   -0.05   -1  0.1 0.05    -1  0.1 1.e-3
GW  5   5   -0.64   -0.81   0.1 -0.54   -0.81   0.1 1.e-3
GW  6   5   -1  -0.31   0.1 -0.9    -0.31   0.1 1.e-3
GW  7   5   -1  0.31    0.1 -0.9    0.31    0.1 1.e-3
GW  8   5   -0.64   0.81    0.1 -0.54   0.81    0.1 1.e-3
GW  9   5   -0.05   1   0.1 0.05    1   0.1 1.e-3
GW  10  5   0.54    0.81    0.1 0.64    0.81    0.1 1.e-3
GW  11  5   0   1.05    0.099   0   0.95    0.099   1.e-3
GW  12  5   0.59    0.86    0.099   0.59    0.76    0.099   1.e-3
GW  13  5   0.95    0.36    0.099   0.95    0.26    0.099   1.e-3
GW  14  5   0.95    -0.26   0.099   0.95    -0.36   0.099   1.e-3
```

```
GW  15   5    0.59    -0.76   0.099   0.59    -0.86   0.099   1.e-3
GW  16   5    0    -0.95   0.099   0    -1.05   0.099   1.e-3
GW  17   5   -0.59    -0.76   0.099   -0.59   -0.86   0.099   1.e-3
GW  18   5   -0.95    -0.26   0.099   -0.95   -0.36   0.099   1.e-3
GW  19   5   -0.95    0.36    0.099   -0.95   0.26    0.099   1.e-3
GW  20   5   -0.59    0.86    0.099   -0.59   0.76    0.099   1.e-3
GE  0
EK
EX  0    1    3    0    6.e-17  1
EX  0    10   3    0    6.e-17  0.999
EX  0    9    3    0    6.e-17  1
EX  0    8    3    0    6.e-17  1
EX  0    7    3    0    6.e-17  0.999
EX  0    6    3    0    6.e-17  0.999
EX  0    5    3    0    6.e-17  1.002
EX  0    4    3    0    6.e-17  1
EX  0    3    3    0    6.e-17  1
EX  0    2    3    0    6.e-17  1.001
EX  0    13   3    0    1.001   0
EX  0    12   3    0    1    0
EX  0    11   3    0    1    0
EX  0    20   3    0    1.002   0
EX  0    19   3    0    0.999   0
EX  0    18   3    0    0.999   0
EX  0    17   3    0    1    0
EX  0    16   3    0    1    0
EX  0    15   3    0    0.999   0
EX  0    14   3    0    1    0
GN  1
FR  0    1    0    0    300 0
EN
```

Left-hand polarized, $l = 1$, $a = \lambda$

```
CE
GW  1    5    0.9 0.31    0.1 1    0.31    0.1 1.e-3
GW  2    5    0.9 -0.31   0.1 1    -0.31   0.1 1.e-3
GW  3    5    0.54    -0.81   0.1 0.64    -0.81   0.1 1.e-3
GW  4    5    -0.05   -1   0.1 0.05    -1   0.1 1.e-3
GW  5    5    -0.64   -0.81   0.1 -0.54   -0.81   0.1 1.e-3
GW  6    5    -1   -0.31   0.1 -0.9   -0.31   0.1 1.e-3
GW  7    5    -1   0.31    0.1 -0.9   0.31    0.1 1.e-3
GW  8    5    -0.64   0.81    0.1 -0.54   0.81    0.1 1.e-3
GW  9    5    -0.05   1    0.1 0.05    1    0.1 1.e-3
```

```
GW  10   5     0.54      0.81     0.1 0.64     0.81     0.1 1.e-3
GW  11   5     0    1.05       0.099   0     0.95      0.099  1.e-3
GW  12   5     0.59      0.86      0.099    0.59     0.76      0.099   1.e-3
GW  13   5     0.95      0.36      0.099    0.95     0.26      0.099   1.e-3
GW  14   5     0.95     -0.26      0.099    0.95    -0.36      0.099   1.e-3
GW  15   5     0.59     -0.76      0.099    0.59    -0.86      0.099   1.e-3
GW  16   5     0    -0.95    0.099   0    -1.05     0.099  1.e-3
GW  17   5    -0.59     -0.76      0.099   -0.59    -0.86      0.099   1.e-3
GW  18   5    -0.95     -0.26      0.099   -0.95    -0.36      0.099   1.e-3
GW  19   5    -0.95      0.36      0.099   -0.95     0.26      0.099   1.e-3
GW  20   5    -0.59      0.86      0.099   -0.59     0.76      0.099   1.e-3
GE   0
EK
EX   0   1   3   0     0.31      0.953
EX   0   10  3   0     0.809     0.588
EX   0   9   3   0     1    0
EX   0   8   3   0     0.811    -0.59
EX   0   7   3   0     0.309    -0.95
EX   0   6   3   0    -0.31     -0.95
EX   0   5   3   0    -0.81     -0.59
EX   0   4   3   0    -1    1.e-16
EX   0   3   3   0    -0.81      0.587
EX   0   2   3   0    -0.31      0.953
EX   0   13  3   0     0.953    -0.31
EX   0   12  3   0     0.587    -0.81
EX   0   11  3   0     0    -1
EX   0   20  3   0    -0.59     -0.81
EX   0   19  3   0    -0.95     -0.31
EX   0   18  3   0    -0.95      0.309
EX   0   17  3   0    -0.59      0.811
EX   0   16  3   0     6.e-17    1
EX   0   15  3   0     0.586     0.81
EX   0   14  3   0     0.952     0.309
GN   1
FR   0   1   0   0     300 0
EN
```

Left-hand polarized, $l = 2$, $a = \lambda$

```
CE
GW   1   5     0.9 0.31      0.1 1     0.31      0.1 1.e-3
GW   2   5     0.9 -0.31     0.1 1    -0.31      0.1 1.e-3
GW   3   5     0.54     -0.81     0.1 0.64     -0.81     0.1 1.e-3
GW   4   5    -0.05     -1   0.1 0.05     -1   0.1 1.e-3
```

```
GW   5    5    -0.64   -0.81   0.1 -0.54   -0.81   0.1 1.e-3
GW   6    5    -1  -0.31   0.1 -0.9    -0.31   0.1 1.e-3
GW   7    5    -1  0.31    0.1 -0.9    0.31    0.1 1.e-3
GW   8    5    -0.64   0.81    0.1 -0.54   0.81    0.1 1.e-3
GW   9    5    -0.05   1   0.1 0.05    1    0.1 1.e-3
GW   10   5    0.54    0.81    0.1 0.64    0.81    0.1 1.e-3
GW   11   5    0   1.05    0.099   0   0.95    0.099   1.e-3
GW   12   5    0.59    0.86    0.099   0.59    0.76    0.099   1.e-3
GW   13   5    0.95    0.36    0.099   0.95    0.26    0.099   1.e-3
GW   14   5    0.95    -0.26   0.099   0.95    -0.36   0.099   1.e-3
GW   15   5    0.59    -0.76   0.099   0.59    -0.86   0.099   1.e-3
GW   16   5    0   -0.95   0.099   0   -1.05   0.099   1.e-3
GW   17   5    -0.59   -0.76   0.099   -0.59   -0.86   0.099   1.e-3
GW   18   5    -0.95   -0.26   0.099   -0.95   -0.36   0.099   1.e-3
GW   19   5    -0.95   0.36    0.099   -0.95   0.26    0.099   1.e-3
GW   20   5    -0.59   0.86    0.099   -0.59   0.76    0.099   1.e-3
GE   0
EK
EX   0    1    3    0    -0.59   0.809
EX   0    10   3    0    -0.95   -0.31
EX   0    9    3    0    0   -1
EX   0    8    3    0    0.953   -0.31
EX   0    7    3    0    0.59    0.811
EX   0    6    3    0    -0.59   0.811
EX   0    5    3    0    -0.95   -0.31
EX   0    4    3    0    0   -1
EX   0    3    3    0    0.95    -0.31
EX   0    2    3    0    0.588   0.809
EX   0    13   3    0    0.809   0.588
EX   0    12   3    0    -0.31   0.95
EX   0    11   3    0    -1  1.e-16
EX   0    20   3    0    -0.31   -0.95
EX   0    19   3    0    0.811   -0.59
EX   0    18   3    0    0.811   0.59
EX   0    17   3    0    -0.31   0.953
EX   0    16   3    0    -1  4.e-16
EX   0    15   3    0    -0.31   -0.95
EX   0    14   3    0    0.809   -0.59
GN   1
FR   0    1    0    0    300 0
EN
```

Left-hand polarized, $l = 3$, $a = \lambda$

```
CE
GW  1   5   0.9 0.31    0.1 1    0.31    0.1 1.e-3
GW  2   5   0.9 -0.31   0.1 1    -0.31   0.1 1.e-3
GW  3   5   0.54    -0.81   0.1 0.64    -0.81   0.1 1.e-3
GW  4   5   -0.05   -1  0.1 0.05    -1  0.1 1.e-3
GW  5   5   -0.64   -0.81   0.1 -0.54   -0.81   0.1 1.e-3
GW  6   5   -1  -0.31   0.1 -0.9    -0.31   0.1 1.e-3
GW  7   5   -1  0.31    0.1 -0.9    0.31    0.1 1.e-3
GW  8   5   -0.64   0.81    0.1 -0.54   0.81    0.1 1.e-3
GW  9   5   -0.05   1   0.1 0.05    1   0.1 1.e-3
GW  10  5   0.54    0.81    0.1 0.64    0.81    0.1 1.e-3
GW  11  5   0   1.05    0.099   0   0.95    0.099   1.e-3
GW  12  5   0.59    0.86    0.099   0.59    0.76    0.099   1.e-3
GW  13  5   0.95    0.36    0.099   0.95    0.26    0.099   1.e-3
GW  14  5   0.95    -0.26   0.099   0.95    -0.36   0.099   1.e-3
GW  15  5   0.59    -0.76   0.099   0.59    -0.86   0.099   1.e-3
GW  16  5   0   -0.95   0.099   0   -1.05   0.099   1.e-3
GW  17  5   -0.59   -0.76   0.099   -0.59   -0.86   0.099   1.e-3
GW  18  5   -0.95   -0.26   0.099   -0.95   -0.36   0.099   1.e-3
GW  19  5   -0.95   0.36    0.099   -0.95   0.26    0.099   1.e-3
GW  20  5   -0.59   0.86    0.099   -0.59   0.76    0.099   1.e-3
GE  0
EK
EX  0   1   3   0   -0.81   0.586
EX  0   10  3   0   -0.31   -0.95
EX  0   9   3   0   1   0
EX  0   8   3   0   -0.31   0.953
EX  0   7   3   0   -0.81   -0.59
EX  0   6   3   0   0.808   -0.59
EX  0   5   3   0   0.31    0.953
EX  0   4   3   0   -1  6.e-16
EX  0   3   3   0   0.31    -0.95
EX  0   2   3   0   0.808   0.587
EX  0   13  3   0   0.587   0.808
EX  0   12  3   0   -0.95   0.31
EX  0   11  3   0   0   -1
EX  0   20  3   0   0.953   0.31
EX  0   19  3   0   -0.59   0.808
EX  0   18  3   0   -0.59   -0.81
EX  0   17  3   0   0.953   -0.31
EX  0   16  3   0   6.e-17  1
EX  0   15  3   0   -0.95   -0.31
EX  0   14  3   0   0.587   -0.81
GN  1
FR  0   1   0   0   300 0
```

```
EN
```

Left-hand polarized, $l = 4$, $a = \lambda$

```
CE
GW   1   5    0.9 0.31     0.1 1     0.31     0.1 1.e-3
GW   2   5    0.9 -0.31    0.1 1    -0.31     0.1 1.e-3
GW   3   5    0.54       -0.81   0.1 0.64    -0.81    0.1 1.e-3
GW   4   5   -0.05    -1   0.1 0.05    -1   0.1 1.e-3
GW   5   5   -0.64    -0.81   0.1 -0.54   -0.81   0.1 1.e-3
GW   6   5   -1   -0.31    0.1 -0.9    -0.31   0.1 1.e-3
GW   7   5   -1   0.31     0.1 -0.9    0.31    0.1 1.e-3
GW   8   5   -0.64    0.81    0.1 -0.54    0.81    0.1 1.e-3
GW   9   5   -0.05    1    0.1 0.05    1    0.1 1.e-3
GW  10   5    0.54    0.81    0.1 0.64    0.81    0.1 1.e-3
GW  11   5    0    1.05    0.099    0    0.95    0.099    1.e-3
GW  12   5    0.59     0.86     0.099    0.59     0.76     0.099    1.e-3
GW  13   5    0.95     0.36     0.099    0.95     0.26     0.099    1.e-3
GW  14   5    0.95     -0.26    0.099    0.95    -0.36     0.099    1.e-3
GW  15   5    0.59     -0.76    0.099    0.59    -0.86     0.099    1.e-3
GW  16   5    0    -0.95   0.099    0    -1.05    0.099    1.e-3
GW  17   5   -0.59     -0.76    0.099   -0.59    -0.86     0.099    1.e-3
GW  18   5   -0.95     -0.26    0.099   -0.95    -0.36     0.099    1.e-3
GW  19   5   -0.95     0.36     0.099   -0.95     0.26     0.099    1.e-3
GW  20   5   -0.59     0.86     0.099   -0.59     0.76     0.099    1.e-3
GE   0
EK
EX   0   1    3    0   -0.95    0.309
EX   0   10   3    0    0.587   -0.81
EX   0   9    3    0    6.e-17   1
EX   0   8    3    0   -0.59    -0.81
EX   0   7    3    0    0.95     0.309
EX   0   6    3    0   -0.95     0.309
EX   0   5    3    0    0.589   -0.81
EX   0   4    3    0    6.e-16   1
EX   0   3    3    0   -0.59    -0.81
EX   0   2    3    0    0.952    0.309
EX   0   13   3    0    0.309    0.952
EX   0   12   3    0   -0.81    -0.59
EX   0   11   3    0    1    0
EX   0   20   3    0   -0.81    0.589
EX   0   19   3    0    0.309   -0.95
EX   0   18   3    0    0.309    0.95
EX   0   17   3    0   -0.81    -0.59
```

```
EX  0   16  3   0    1   0
EX  0   15  3   0   -0.81   0.587
EX  0   14  3   0    0.309  -0.95
GN  1
FR  0   1   0   0    300 0
EN
```

Right-hand polarized, $l = 0$, $a = \lambda/2$

```
CE
GW  1   5   0.425    0.15     0.01      0.525    0.15     0.01     1.e-3
GW  2   5   0.425   -0.15     0.01      0.525   -0.15     0.01     1.e-3
GW  3   5   0.225   -0.4      0.01      0.325   -0.4      0.01     1.e-3
GW  4   5  -0.05    -0.5      0.01      0.05    -0.5      0.01     1.e-3
GW  5   5  -0.35    -0.4      0.01     -0.25    -0.4      0.01     1.e-3
GW  6   5  -0.525   -0.15     0.01     -0.425   -0.15     0.01     1.e-3
GW  7   5  -0.525    0.15     0.01     -0.425    0.15     0.01     1.e-3
GW  8   5  -0.35     0.4 0.01          -0.25     0.4 0.01         1.e-3
GW  9   5  -0.05     0.5 0.01           0.05     0.5 0.01         1.e-3
GW  10  5   0.25     0.4 0.01           0.35     0.4 0.01         1.e-3
GW  11  5   0        0.55     9.e-3     0        0.45     9.e-3    1.e-3
GW  12  5   0.3 0.45          9.e-3     0.3 0.35          9.e-3    1.e-3
GW  13  5   0.475    0.2 9.e-3          0.475    0.1 9.e-3        1.e-3
GW  14  5   0.475   -0.1      9.e-3     0.475   -0.2      9.e-3    1.e-3
GW  15  5   0.275   -0.35     9.e-3     0.275   -0.45     9.e-3    1.e-3
GW  16  5   0       -0.45     9.e-3     0       -0.55     9.e-3    1.e-3
GW  17  5  -0.3     -0.35     9.e-3    -0.3     -0.45     9.e-3    1.e-3
GW  18  5  -0.475   -0.1      9.e-3    -0.475   -0.2      9.e-3    1.e-3
GW  19  5  -0.475    0.2 9.e-3         -0.475    0.1 9.e-3        1.e-3
GW  20  5  -0.3      0.45     9.e-3    -0.3      0.35     9.e-3    1.e-3
GE  0
EX  0   1   3   0    1   0
EX  0   10  3   0    1   0
EX  0   9   3   0    1   0
EX  0   8   3   0    1.002   0
EX  0   7   3   0    0.999   0
EX  0   6   3   0    0.999   0
EX  0   5   3   0    1.003   0
EX  0   4   3   0    1   0
EX  0   3   3   0    1   0
EX  0   2   3   0    1.001   0
EX  0   13  3   0    6.e-17  1.001
EX  0   12  3   0    6.e-17  1
EX  0   11  3   0    6.e-17  1
```

```
EX   0   20   3   0    6.e-17   1.003
EX   0   19   3   0    6.e-17   0.999
EX   0   18   3   0    6.e-17   0.999
EX   0   17   3   0    6.e-17   1.002
EX   0   16   3   0    6.e-17   1
EX   0   15   3   0    6.e-17   1
EX   0   14   3   0    6.e-17   1
GN   1
FR   0   1   0   0    300 0
EN
```

Right-hand polarized, $l = 1$, $a = \lambda/2$

```
CE
GW   1    5   0.42     0.15    0.01    0.52     0.15    0.01    1.e-3
GW   2    5   0.42    -0.15    0.01    0.52    -0.15    0.01    1.e-3
GW   3    5   0.24    -0.4     0.01    0.34    -0.4     0.01    1.e-3
GW   4    5  -0.06    -0.5     0.01    0.04    -0.5     0.01    1.e-3
GW   5    5  -0.35    -0.4     0.01   -0.25    -0.4     0.01    1.e-3
GW   6    5  -0.535   -0.15    0.01   -0.435   -0.15    0.01    1.e-3
GW   7    5  -0.535    0.15    0.01   -0.435    0.15    0.01    1.e-3
GW   8    5  -0.35     0.4 0.01   -0.25    0.4 0.01    1.e-3
GW   9    5  -0.06     0.5 0.01    0.04    0.5 0.01    1.e-3
GW   10   5   0.24     0.4 0.01    0.34    0.4 0.01    1.e-3
GW   11   5  -0.01     0.55    9.e-3   -0.01    0.45    9.e-3   1.e-3
GW   12   5   0.29     0.45    9.e-3    0.29    0.35    9.e-3   1.e-3
GW   13   5   0.47     0.2 9.e-3    0.47     0.1 9.e-3   1.e-3
GW   14   5   0.47    -0.1     9.e-3    0.47    -0.2     9.e-3   1.e-3
GW   15   5   0.29    -0.35    9.e-3    0.29    -0.45    9.e-3   1.e-3
GW   16   5  -0.01    -0.45    9.e-3   -0.01    -0.55    9.e-3   1.e-3
GW   17   5  -0.3     -0.35    9.e-3   -0.3     -0.45    9.e-3   1.e-3
GW   18   5  -0.485   -0.1     9.e-3   -0.485   -0.2     9.e-3   1.e-3
GW   19   5  -0.485    0.2 9.e-3   -0.485    0.1 9.e-3   1.e-3
GW   20   5  -0.3      0.45    9.e-3   -0.3      0.35    9.e-3   1.e-3
GE   0
EX   0   1    3   0    0.951    0.309
EX   0   10   3   0    0.587    0.808
EX   0   9    3   0    6.e-17   1
EX   0   8    3   0   -0.59     0.808
EX   0   7    3   0   -0.95     0.309
EX   0   6    3   0   -0.95    -0.31
EX   0   5    3   0   -0.59    -0.81
EX   0   4    3   0    0       -1
EX   0   3    3   0    0.587   -0.81
```

```
EX  0   2    3   0    0.952   -0.31
EX  0   13   3   0    -0.31   0.952
EX  0   12   3   0    -0.81   0.587
EX  0   11   3   0    -1  1.e-16
EX  0   20   3   0    -0.81   -0.59
EX  0   19   3   0    -0.31   -0.95
EX  0   18   3   0    0.309   -0.95
EX  0   17   3   0    0.808   -0.59
EX  0   16   3   0    1   0
EX  0   15   3   0    0.808   0.587
EX  0   14   3   0    0.588   0.809
GN  1
FR  0   1    0   0    300 0
EN
```

Right-hand polarized, $l = 2$, $a = \lambda/2$

```
CE
GW  1    5   0.42    0.15    0.01    0.52    0.15    0.01    1.e-3
GW  2    5   0.43    -0.15   0.01    0.53    -0.15   0.01    1.e-3
GW  3    5   0.24    -0.4    0.01    0.34    -0.4    0.01    1.e-3
GW  4    5   -0.05   -0.5    0.01    0.05    -0.5    0.01    1.e-3
GW  5    5   -0.35   -0.41   0.01    -0.25   -0.41   0.01    1.e-3
GW  6    5   -0.53   -0.15   0.01    -0.43   -0.15   0.01    1.e-3
GW  7    5   -0.53   0.15    0.01    -0.43   0.15    0.01    1.e-3
GW  8    5   -0.34   0.41    0.01    -0.24   0.41    0.01    1.e-3
GW  9    5   -0.05   0.5 0.01    0.05    0.5 0.01    1.e-3
GW  10   5   0.24    0.4 0.01    0.34    0.4 0.01    1.e-3
GW  11   5   0   0.55    9.e-3   0   0.45    9.e-3   1.e-3
GW  12   5   0.29    0.45    9.e-3   0.29    0.35    9.e-3   1.e-3
GW  13   5   0.47    0.2 9.e-3   0.47    0.1 9.e-3   1.e-3
GW  14   5   0.48    -0.1    9.e-3   0.48    -0.2    9.e-3   1.e-3
GW  15   5   0.29    -0.35   9.e-3   0.29    -0.45   9.e-3   1.e-3
GW  16   5   0   -0.45   9.e-3   0   -0.55   9.e-3   1.e-3
GW  17   5   -0.3    -0.36   9.e-3   -0.3    -0.46   9.e-3   1.e-3
GW  18   5   -0.48   -0.1    9.e-3   -0.48   -0.2    9.e-3   1.e-3
GW  19   5   -0.48   0.2 9.e-3   -0.48   0.1 9.e-3   1.e-3
GW  20   5   -0.29   0.46    9.e-3   -0.29   0.36    9.e-3   1.e-3
GE  0
EX  0   1    3   0    0.809   0.588
EX  0   10   3   0    -0.31   0.95
EX  0   9    3   0    -1  1.e-16
EX  0   8    3   0    -0.31   -0.95
EX  0   7    3   0    0.809   -0.59
```

```
EX   0    6    3    0     0.811    0.589
EX   0    5    3    0    -0.31     0.953
EX   0    4    3    0    -1  4.e-16
EX   0    3    3    0    -0.31    -0.95
EX   0    2    3    0     0.809   -0.59
EX   0   13    3    0    -0.59     0.809
EX   0   12    3    0    -0.95    -0.31
EX   0   11    3    0     0   -1
EX   0   20    3    0     0.953   -0.31
EX   0   19    3    0     0.589    0.811
EX   0   18    3    0    -0.59     0.809
EX   0   17    3    0    -0.95    -0.31
EX   0   16    3    0     0   -1
EX   0   15    3    0     0.95    -0.31
EX   0   14    3    0     0.588    0.809
GN   1
FR   0    1    0    0    300 0
EN
```

Right-hand polarized, $l = 3$, $a = \lambda/2$

```
CE
GW   1    5    0.43     0.15     0.01     0.53     0.15     0.01     1.e-3
GW   2    5    0.43    -0.15     0.01     0.53    -0.15     0.01     1.e-3
GW   3    5    0.24    -0.4      0.01     0.34    -0.4      0.01     1.e-3
GW   4    5   -0.05    -0.5      0.01     0.05    -0.5      0.01     1.e-3
GW   5    5   -0.34    -0.4      0.01    -0.24    -0.4      0.01     1.e-3
GW   6    5   -0.53    -0.16     0.01    -0.43    -0.16     0.01     1.e-3
GW   7    5   -0.53     0.16     0.01    -0.43     0.16     0.01     1.e-3
GW   8    5   -0.35     0.41     0.01    -0.25     0.41     0.01     1.e-3
GW   9    5   -0.05     0.5 0.01      0.05     0.5 0.01     1.e-3
GW  10    5    0.25     0.41     0.01     0.35     0.41     0.01     1.e-3
GW  11    5    0    0.55     9.e-3     0    0.45     9.e-3     1.e-3
GW  12    5    0.3 0.46      9.e-3    0.3 0.36      9.e-3     1.e-3
GW  13    5    0.48     0.2 9.e-3     0.48     0.1 9.e-3     1.e-3
GW  14    5    0.48    -0.1      9.e-3     0.48    -0.2      9.e-3     1.e-3
GW  15    5    0.29    -0.35     9.e-3     0.29    -0.45     9.e-3     1.e-3
GW  16    5    0   -0.45     9.e-3     0   -0.55     9.e-3     1.e-3
GW  17    5   -0.29    -0.35     9.e-3    -0.29    -0.45     9.e-3     1.e-3
GW  18    5   -0.48    -0.11     9.e-3    -0.48    -0.21     9.e-3     1.e-3
GW  19    5   -0.48     0.21     9.e-3    -0.48     0.11     9.e-3     1.e-3
GW  20    5   -0.3      0.46     9.e-3    -0.3      0.36     9.e-3     1.e-3
GE   0
EX   0    1    3    0     0.588    0.809
```

```
EX  0  10  3  0  -0.95   0.31
EX  0  9   3  0  0    -1
EX  0  8   3  0  0.952   0.309
EX  0  7   3  0  -0.59   0.808
EX  0  6   3  0  -0.59   -0.81
EX  0  5   3  0  0.953   -0.31
EX  0  4   3  0  6.e-17  1
EX  0  3   3  0  -0.95   -0.31
EX  0  2   3  0  0.587   -0.81
EX  0  13  3  0  -0.81   0.587
EX  0  12  3  0  -0.31   -0.95
EX  0  11  3  0  1    0
EX  0  20  3  0  -0.31   0.953
EX  0  19  3  0  -0.81   -0.59
EX  0  18  3  0  0.808   -0.59
EX  0  17  3  0  0.309   0.951
EX  0  16  3  0  -1   6.e-16
EX  0  15  3  0  0.31   -0.95
EX  0  14  3  0  0.809   0.588
GN  1
FR  0  1   0  0  300 0
EN
```

## Right-hand polarized, $l = 4$, $a = \lambda/2$

```
CE
GW  1   5   0.43    0.15    0.01    0.53    0.15    0.01    1.e-3
GW  2   5   0.43    -0.15   0.01    0.53    -0.15   0.01    1.e-3
GW  3   5   0.24    -0.4    0.01    0.34    -0.4    0.01    1.e-3
GW  4   5   -0.05   -0.5    0.01    0.05    -0.5    0.01    1.e-3
GW  5   5   -0.35   -0.41   0.01    -0.25   -0.41   0.01    1.e-3
GW  6   5   -0.53   -0.15   0.01    -0.43   -0.15   0.01    1.e-3
GW  7   5   -0.53   0.16    0.01    -0.43   0.16    0.01    1.e-3
GW  8   5   -0.35   0.41    0.01    -0.25   0.41    0.01    1.e-3
GW  9   5   -0.05   0.5  0.01      0.05    0.5  0.01    1.e-3
GW  10  5   0.25    0.4  0.01      0.35    0.4  0.01    1.e-3
GW  11  5   0    0.55    9.e-3     0    0.45    9.e-3    1.e-3
GW  12  5   0.3  0.45    9.e-3     0.3  0.35    9.e-3    1.e-3
GW  13  5   0.48    0.2  9.e-3     0.48    0.1  9.e-3    1.e-3
GW  14  5   0.48    -0.1    9.e-3   0.48    -0.2    9.e-3    1.e-3
GW  15  5   0.29    -0.35   9.e-3   0.29    -0.45   9.e-3    1.e-3
GW  16  5   0    -0.45   9.e-3     0    -0.55   9.e-3    1.e-3
GW  17  5   -0.3    -0.36   9.e-3   -0.3    -0.46   9.e-3    1.e-3
GW  18  5   -0.48   -0.1    9.e-3   -0.48   -0.2    9.e-3    1.e-3
```

```
GW   19   5    -0.48    0.21     9.e-3    -0.48    0.11     9.e-3    1.e-3
GW   20   5    -0.3     0.46     9.e-3    -0.3     0.36     9.e-3    1.e-3
GE   0
EX   0    1    3    0    0.309    0.951
EX   0    10   3    0    -0.81    -0.59
EX   0    9    3    0    1    0
EX   0    8    3    0    -0.81    0.588
EX   0    7    3    0    0.309    -0.95
EX   0    6    3    0    0.31     0.953
EX   0    5    3    0    -0.81    -0.59
EX   0    4    3    0    1    0
EX   0    3    3    0    -0.81    0.588
EX   0    2    3    0    0.309    -0.95
EX   0    13   3    0    -0.95    0.309
EX   0    12   3    0    0.587    -0.81
EX   0    11   3    0    6.e-17   1
EX   0    20   3    0    -0.59    -0.81
EX   0    19   3    0    0.953    0.31
EX   0    18   3    0    -0.95    0.309
EX   0    17   3    0    0.587    -0.81
EX   0    16   3    0    6.e-17   1
EX   0    15   3    0    -0.59    -0.81
EX   0    14   3    0    0.951    0.309
GN   1
FR   0    1    0    0    300 0
EN
```

### B.4.3  Tripole compared to crossed dipole

Crossed dipole array, $\theta_0 = 45°$, $\varphi_0 = 0°$

```
CE
GW   1    3    0.2 0.03     0.6 0.2  -0.03    0.6 1.e-3
GW   2    3    0.06     0.22     0.6 0.06     0.16     0.6 1.e-3
GW   3    3    -0.16    0.15     0.6 -0.16    0.09     0.6 1.e-3
GW   4    3    -0.16    -0.09    0.6 -0.16    -0.15    0.6 1.e-3
GW   5    3    0.06     -0.16    0.6 0.06     -0.22    0.6 1.e-3
GW   6    3    0.17     0    0.599    0.23     0    0.599    1.e-3
GW   7    3    0.03     0.19     0.599    0.09     0.19     0.599    1.e-3
GW   8    3    -0.19    0.12     0.599    -0.13    0.12     0.599    1.e-3
GW   9    3    -0.19    -0.12    0.599    -0.13    -0.12    0.599    1.e-3
GW   10   3    0.03     -0.19    0.599    0.09     -0.19    0.599    1.e-3
GW   11   3    0.201    -1.e-3   0.57     0.201    -1.e-3   0.63     1.e-3
GW   12   3    0.061    0.191    0.57     0.061    0.191    0.63     1.e-3
```

```
GW  13  3   -0.161  0.121   0.57    -0.161  0.121   0.63    1.e-3
GW  14  3   -0.161  -0.121  0.57    -0.161  -0.121  0.63    1.e-3
GW  15  3   0.061   -0.191  0.57    0.061   -0.191  0.63    1.e-3
GE  0
EX  0   1   2   0   0.633   -0.78
EX  0   2   2   0   0.969   -0.27
EX  0   3   2   0   0.759   0.664
EX  0   4   2   0   0.75    0.656
EX  0   5   2   0   0.969   -0.27
EX  0   6   2   0   0.551   0.448
EX  0   7   2   0   0.192   0.684
EX  0   8   2   0   -0.47   0.533
EX  0   9   2   0   -0.47   0.533
EX  0   10  2   0   0.192   0.684
GN  -1
FR  0   1   0   0   299.8   0
EN
```

Single tripole, $\theta_0 = 45°$, $\varphi_0 = 0°$

```
CE
GW  1   3   0   0.03    0   0   -0.03   0   1.e-3
GW  6   3   -0.03   0   -1.e-3  0.03    0   -1.e-3  1.e-3
GW  11  3   1.e-3   -1.e-3  -0.03   1.e-3   -1.e-3  0.03    1.e-3
GE  0
EX  0   1   2   0   -0.43   -0.91
EX  0   6   2   0   0.643   -0.31
EX  0   11  2   0   -0.64   0.304
GN  -1
FR  0   1   0   0   299.8   0
EN
```

Tripole array, $\theta_0 = 45°$, $\varphi_0 = 0°$

```
CE
GW  1   3   0.19    0.09    0.09    0.19    0.03    0.09    1.e-3
GW  2   3   0   0.23    0.09    0   0.17    0.09    1.e-3
GW  3   3   -0.19   0.09    0.09    -0.19   0.03    0.09    1.e-3
GW  4   3   -0.12   -0.13   0.09    -0.12   -0.19   0.09    1.e-3
GW  5   3   0.12    -0.13   0.09    0.12    -0.19   0.09    1.e-3
GW  6   3   0.16    0.06    0.089   0.22    0.06    0.089   1.e-3
GW  7   3   -0.03   0.2 0.089   0.03    0.2 0.089   1.e-3
GW  8   3   -0.22   0.06    0.089   -0.16   0.06    0.089   1.e-3
```

```
GW   9    3    -0.15    -0.16    0.089   -0.09    -0.16    0.089   1.e-3
GW  10    3     0.09    -0.16    0.089    0.15    -0.16    0.089   1.e-3
GW  11    3     0.191    0.059   0.06     0.191    0.059   0.12    1.e-3
GW  12    3     1.e-3    0.201   0.06     1.e-3    0.201   0.12    1.e-3
GW  13    3    -0.191    0.061   0.06    -0.191    0.061   0.12    1.e-3
GW  14    3    -0.121   -0.161   0.06    -0.121   -0.161   0.12    1.e-3
GW  15    3     0.121   -0.161   0.06     0.121   -0.161   0.12    1.e-3
GE   0
EX   0    1    2    0    -0.75    -0.66
EX   0    2    2    0     6.e-17  -1
EX   0    3    2    0     0.753   -0.67
EX   0    4    2    0     0.467   -0.88
EX   0    5    2    0    -0.5     -0.87
EX   0    6    2    0     0.469   -0.53
EX   0    7    2    0     0.706    0
EX   0    8    2    0     0.472    0.532
EX   0    9    2    0     0.613    0.352
EX   0   10    2    0     0.614   -0.35
EX   0   11    2    0    -0.47     0.526
EX   0   12    2    0    -0.71     9.e-17
EX   0   13    2    0    -0.47    -0.53
EX   0   14    2    0    -0.61    -0.35
EX   0   15    2    0    -0.61     0.353
GN  -1
FR   0    1    0    0     299.8    0
EN
```

### B.4.4  Pointed tripoles

Five element, $l = 0$, $\theta_0 = 10°$, $\varphi_0 = 40°$

```
CE
GW   1    3     0.19     0.09    0.09     0.19     0.03    0.09    1.e-3
GW   2    3     0    0.23    0.09     0    0.17     0.09    1.e-3
GW   3    3    -0.19     0.09    0.09    -0.19     0.03    0.09    1.e-3
GW   4    3    -0.12    -0.13    0.09    -0.12    -0.19    0.09    1.e-3
GW   5    3     0.12    -0.13    0.09     0.12    -0.19    0.09    1.e-3
GW   6    3     0.16     0.06    0.089    0.22     0.06    0.089   1.e-3
GW   7    3    -0.03     0.2  0.089    0.03     0.2  0.089   1.e-3
GW   8    3    -0.22     0.06    0.089   -0.16     0.06    0.089   1.e-3
GW   9    3    -0.15    -0.16    0.089   -0.09    -0.16    0.089   1.e-3
GW  10    3     0.09    -0.16    0.089    0.15    -0.16    0.089   1.e-3
GW  11    3     0.191    0.059   0.06     0.191    0.059   0.12    1.e-3
GW  12    3     1.e-3    0.201   0.06     1.e-3    0.201   0.12    1.e-3
```

```
GW   13   3     -0.191    0.061    0.06       -0.191    0.061    0.12     1.e-3
GW   14   3     -0.121    -0.161   0.06       -0.121    -0.161   0.12     1.e-3
GW   15   3     0.121     -0.161   0.06       0.121     -0.161   0.12     1.e-3
GE   0
EX   0    1    2    0    -0.61     0.78
EX   0    2    2    0    -0.66     0.742
EX   0    3    2    0    -0.83     0.553
EX   0    4    2    0    -0.88     0.473
EX   0    5    2    0    -0.76     0.631
EX   0    6    2    0    -0.77     -0.62
EX   0    7    2    0    -0.73     -0.67
EX   0    8    2    0    -0.54     -0.83
EX   0    9    2    0    -0.46     -0.88
EX   0    10   2    0    -0.62     -0.77
EX   0    11   2    0    0.171     -0.04
EX   0    12   2    0    0.171     -0.02
EX   0    13   2    0    0.172     0.02
EX   0    14   2    0    0.174     0.02
EX   0    15   2    0    0.174     3.e-3
GN   2    0    0    0    13   5.e-3
FR   0    1    0    0    299.8     0
EN
```

Five element, $l = 1$, $\theta_0 = 10°$, $\varphi_0 = 40°$

```
CE
GW   1    3    0.19      0.09     0.09      0.19      0.03     0.09      1.e-3
GW   2    3    0    0.23    0.09    0    0.17    0.09    1.e-3
GW   3    3    -0.19     0.09     0.09      -0.19     0.03     0.09      1.e-3
GW   4    3    -0.12     -0.13    0.09      -0.12     -0.19    0.09      1.e-3
GW   5    3    0.12      -0.13    0.09      0.12      -0.19    0.09      1.e-3
GW   6    3    0.16      0.06     0.089     0.22      0.06     0.089     1.e-3
GW   7    3    -0.03     0.2  0.089    0.03     0.2  0.089    1.e-3
GW   8    3    -0.22     0.06     0.089     -0.16     0.06     0.089     1.e-3
GW   9    3    -0.15     -0.16    0.089     -0.09     -0.16    0.089     1.e-3
GW   10   3    0.09      -0.16    0.089     0.15      -0.16    0.089     1.e-3
GW   11   3    0.191     0.059    0.06      0.191     0.059    0.12      1.e-3
GW   12   3    1.e-3     0.201    0.06      1.e-3     0.201    0.12      1.e-3
GW   13   3    -0.191    0.061    0.06      -0.191    0.061    0.12      1.e-3
GW   14   3    -0.121    -0.161   0.06      -0.121    -0.161   0.12      1.e-3
GW   15   3    0.121     -0.161   0.06      0.121     -0.161   0.12      1.e-3
GE   0
EX   0    1    2    0    -0.54     -0.83
EX   0    2    2    0    0.67      -0.73
```

```
EX  0   3   2   0    0.772    0.626
EX  0   4   2   0   −0.45     0.891
EX  0   5   2   0   −0.99     0.047
EX  0   6   2   0   −0.82     0.555
EX  0   7   2   0   −0.74    −0.66
EX  0   8   2   0    0.612   −0.78
EX  0   9   2   0    0.89     0.428
EX  0  10   2   0    0.061    0.988
EX  0  11   2   0    0.174    0.02
EX  0  12   2   0    0.024    0.171
EX  0  13   2   0   −0.17     0.034
EX  0  14   2   0   −0.07    −0.16
EX  0  15   2   0    0.104   −0.14
GN  1
FR  0   1   0   0    299.8    0
EN
```

Five element, $l = 2$, $\theta_0 = 10°$, $\varphi_0 = 40°$

```
CE
GW   1   3   0.19     0.09     0.09     0.19    0.03    0.09    1.e−3
GW   2   3   0      0.23     0.09      0     0.17    0.09    1.e−3
GW   3   3  −0.19     0.09     0.09    −0.19    0.03    0.09    1.e−3
GW   4   3  −0.12    −0.13     0.09    −0.12   −0.19    0.09    1.e−3
GW   5   3   0.12    −0.13     0.09     0.12   −0.19    0.09    1.e−3
GW   6   3   0.16     0.06     0.089    0.22    0.06    0.089   1.e−3
GW   7   3  −0.03     0.2 0.089    0.03     0.2 0.089   1.e−3
GW   8   3  −0.22     0.06     0.089   −0.16    0.06    0.089   1.e−3
GW   9   3  −0.15    −0.16     0.089   −0.09   −0.16    0.089   1.e−3
GW  10   3   0.09    −0.16     0.089    0.15   −0.16    0.089   1.e−3
GW  11   3   0.191    0.059    0.06     0.191   0.059   0.12    1.e−3
GW  12   3   1.e−3    0.201    0.06     1.e−3   0.201   0.12    1.e−3
GW  13   3  −0.191    0.061    0.06    −0.191   0.061   0.12    1.e−3
GW  14   3  −0.121   −0.161    0.06    −0.121  −0.161   0.12    1.e−3
GW  15   3   0.121   −0.161    0.06     0.121  −0.161   0.12    1.e−3
GE   0
EX  0   1   2   0   −0.26    −0.96
EX  0   2   2   0    0.733    0.67
EX  0   3   2   0   −0.93    −0.36
EX  0   4   2   0    0.985   −0.16
EX  0   5   2   0   −0.54     0.826
EX  0   6   2   0   −0.95     0.273
EX  0   7   2   0    0.657   −0.74
EX  0   8   2   0   −0.34     0.929
```

```
EX  0   9   2   0   -0.18   -0.98
EX  0   10  2   0   0.835   0.531
EX  0   11  2   0   0.16    0.073
EX  0   12  2   0   -0.17   0.024
EX  0   13  2   0   0.149   -0.08
EX  0   14  2   0   -0.09   0.154
EX  0   15  2   0   -0.05   -0.17
GN  2   0   0   0   13  5.e-3
FR  0   1   0   0   299.8   0
EN
```

Five element in free space, $l = 2$, $\theta_0 = 10°$, $\varphi_0 = 40°$

```
CE
GW  1   3   0.19    0.09    0.09    0.19    0.03    0.09    1.e-3
GW  2   3   0   0.23    0.09    0   0.17    0.09    1.e-3
GW  3   3   -0.19   0.09    0.09    -0.19   0.03    0.09    1.e-3
GW  4   3   -0.12   -0.13   0.09    -0.12   -0.19   0.09    1.e-3
GW  5   3   0.12    -0.13   0.09    0.12    -0.19   0.09    1.e-3
GW  6   3   0.16    0.06    0.089   0.22    0.06    0.089   1.e-3
GW  7   3   -0.03   0.2 0.089   0.03    0.2 0.089   1.e-3
GW  8   3   -0.22   0.06    0.089   -0.16   0.06    0.089   1.e-3
GW  9   3   -0.15   -0.16   0.089   -0.09   -0.16   0.089   1.e-3
GW  10  3   0.09    -0.16   0.089   0.15    -0.16   0.089   1.e-3
GW  11  3   0.191   0.059   0.06    0.191   0.059   0.12    1.e-3
GW  12  3   1.e-3   0.201   0.06    1.e-3   0.201   0.12    1.e-3
GW  13  3   -0.191  0.061   0.06    -0.191  0.061   0.12    1.e-3
GW  14  3   -0.121  -0.161  0.06    -0.121  -0.161  0.12    1.e-3
GW  15  3   0.121   -0.161  0.06    0.121   -0.161  0.12    1.e-3
GE  0
EX  0   1   2   0   -0.26   -0.96
EX  0   2   2   0   0.733   0.67
EX  0   3   2   0   -0.93   -0.36
EX  0   4   2   0   0.985   -0.16
EX  0   5   2   0   -0.54   0.826
EX  0   6   2   0   -0.95   0.273
EX  0   7   2   0   0.657   -0.74
EX  0   8   2   0   -0.34   0.929
EX  0   9   2   0   -0.18   -0.98
EX  0   10  2   0   0.835   0.531
EX  0   11  2   0   0.16    0.073
EX  0   12  2   0   -0.17   0.024
EX  0   13  2   0   0.149   -0.08
EX  0   14  2   0   -0.09   0.154
```

```
EX  0   15  2   0    -0.05   -0.17
GN  -1
FR  0   1   0   0    299.8   0
EN
```

Ten element, $l = 1$, $a = \lambda/5$, $\theta_0 = 45°$, $\varphi_0 = 0°$

```
CE
GW  1   3   0.19     0.09     0.125    0.19      0.03      0.125    1.e-3
GW  2   3   0.1175   0.19     0.125    0.1175    0.13      0.125    1.e-3
GW  3   3   0    0.23     0.125    0    0.17      0.125    1.e-3
GW  4   3   -0.1175  0.1925   0.125    -0.1175   0.1325    0.125    1.e-3
GW  5   3   -0.19    0.0925   0.125    -0.19     0.0325    0.125    1.e-3
GW  6   3   0.16     0.06     0.124    0.22      0.06      0.124    1.e-3
GW  7   3   0.0875   0.16     0.124    0.1475    0.16      0.124    1.e-3
GW  8   3   -0.03    0.2  0.124    0.03     0.2  0.124    1.e-3
GW  9   3   -0.1475  0.1625   0.124    -0.0875   0.1625    0.124    1.e-3
GW  10  3   -0.22    0.0625   0.124    -0.16     0.0625    0.124    1.e-3
GW  11  3   0.191    0.059    0.095    0.191     0.059     0.155    1.e-3
GW  12  3   0.1185   0.161    0.095    0.1185    0.161     0.155    1.e-3
GW  13  3   -1.e-3   0.201    0.095    -1.e-3    0.201     0.155    1.e-3
GW  14  3   -0.1185  0.1615   0.095    -0.1185   0.1615    0.155    1.e-3
GW  15  3   -0.189   0.0615   0.095    -0.189    0.0615    0.155    1.e-3
GW  16  3   -0.19    -0.03    0.125    -0.19     -0.09     0.125    1.e-3
GW  17  3   -0.22    -0.06    0.124    -0.16     -0.06     0.124    1.e-3
GW  18  3   -0.189   -0.061   0.095    -0.189    -0.061    0.155    1.e-3
GW  19  3   -0.1175  -0.13    0.125    -0.1175   -0.19     0.125    1.e-3
GW  20  3   -0.15    -0.16    0.124    -0.09     -0.16     0.124    1.e-3
GW  21  3   -0.1165  -0.161   0.095    -0.1165   -0.161    0.155    1.e-3
GW  22  3   0    -0.17    0.125    0    -0.23     0.125    1.e-3
GW  23  3   -0.03    -0.2     0.124    0.03      -0.2      0.124    1.e-3
GW  24  3   1.e-3    -0.201   0.095    1.e-3     -0.201    0.155    1.e-3
GW  25  3   0.1175   -0.1325  0.125    0.1175    -0.1925   0.125    1.e-3
GW  26  3   0.0875   -0.1625  0.124    0.1475    -0.1625   0.124    1.e-3
GW  27  3   0.1185   -0.1635  0.095    0.1185    -0.1635   0.155    1.e-3
GW  28  3   0.19     -0.03    0.125    0.19      -0.09     0.125    1.e-3
GW  29  3   0.16     -0.06    0.124    0.22      -0.06     0.124    1.e-3
GW  30  3   0.191    -0.061   0.095    0.191     -0.061    0.155    1.e-3
GE  0
EK
EX  0   1   2   0    -0.51   -0.87
EX  0   2   2   0    0.405   -0.91
EX  0   3   2   0    1.006   0
EX  0   4   2   0    0.404   0.905
```

```
EX   0   5    2   0    -0.51   0.868
EX   0   6    2   0    0.611   -0.36
EX   0   7    2   0    0.645   0.288
EX   0   8    2   0    4.e-17  0.711
EX   0   9    2   0    -0.64   0.288
EX   0   10   2   0    -0.61   -0.36
EX   0   11   2   0    -0.61   0.356
EX   0   12   2   0    -0.64   -0.29
EX   0   13   2   0    0    -0.71
EX   0   14   2   0    0.647   -0.29
EX   0   15   2   0    0.611   0.358
EX   0   16   2   0    -0.92   0.403
EX   0   17   2   0    -0.29   -0.65
EX   0   18   2   0    0.285   0.653
EX   0   19   2   0    -1  0.106
EX   0   20   2   0    -0.07   -0.7
EX   0   21   2   0    0.075   0.709
EX   0   22   2   0    -1  1.e-16
EX   0   23   2   0    0.71    0
EX   0   24   2   0    4.e-17  0.71
EX   0   25   2   0    -1  -0.11
EX   0   26   2   0    0.075   -0.71
EX   0   27   2   0    -0.08   0.706
EX   0   28   2   0    -0.92   -0.4
EX   0   29   2   0    0.286   -0.66
EX   0   30   2   0    -0.28   0.651
GN   -1
FR   0   1    0   0    300 0
EN
```

Ten element, $l = 2$, $a = \lambda/5$, $\theta_0 = 45°$, $\varphi_0 = 0°$

```
CE
GW   1    3   0.19    0.09    0.125   0.19    0.03    0.125   1.e-3
GW   2    3   0.115   0.19    0.125   0.115   0.13    0.125   1.e-3
GW   3    3   0   0.23    0.125   0   0.17    0.125   1.e-3
GW   4    3   -0.115  0.19    0.125   -0.115  0.13    0.125   1.e-3
GW   5    3   -0.19   0.09    0.125   -0.19   0.03    0.125   1.e-3
GW   6    3   0.16    0.06    0.124   0.22    0.06    0.124   1.e-3
GW   7    3   0.085   0.16    0.124   0.145   0.16    0.124   1.e-3
GW   8    3   -0.03   0.2 0.124   0.03    0.2 0.124   1.e-3
GW   9    3   -0.145  0.16    0.124   -0.085  0.16    0.124   1.e-3
GW   10   3   -0.22   0.06    0.124   -0.16   0.06    0.124   1.e-3
GW   11   3   0.191   0.059   0.095   0.191   0.059   0.155   1.e-3
```

```
GW   12   3    0.116     0.161    0.095     0.116     0.161    0.155    1.e-3
GW   13   3   -1.e-3     0.201    0.095    -1.e-3     0.201    0.155    1.e-3
GW   14   3   -0.116     0.159    0.095    -0.116     0.159    0.155    1.e-3
GW   15   3   -0.189     0.059    0.095    -0.189     0.059    0.155    1.e-3
GW   16   3   -0.19     -0.035    0.125    -0.19     -0.095    0.125    1.e-3
GW   17   3   -0.22     -0.065    0.124    -0.16     -0.065    0.124    1.e-3
GW   18   3   -0.189    -0.066    0.095    -0.189    -0.066    0.155    1.e-3
GW   19   3   -0.115    -0.135    0.125    -0.115    -0.195    0.125    1.e-3
GW   20   3   -0.145    -0.165    0.124    -0.085    -0.165    0.124    1.e-3
GW   21   3   -0.114    -0.166    0.095    -0.114    -0.166    0.155    1.e-3
GW   22   3    0        -0.17     0.125     0        -0.23     0.125    1.e-3
GW   23   3   -0.03     -0.2      0.124     0.03     -0.2      0.124    1.e-3
GW   24   3    1.e-3    -0.201    0.095     1.e-3    -0.201    0.155    1.e-3
GW   25   3    0.115    -0.13     0.125     0.115    -0.19     0.125    1.e-3
GW   26   3    0.085    -0.16     0.124     0.145    -0.16     0.124    1.e-3
GW   27   3    0.116    -0.161    0.095     0.116    -0.161    0.155    1.e-3
GW   28   3    0.19     -0.035    0.125     0.19     -0.095    0.125    1.e-3
GW   29   3    0.16     -0.065    0.124     0.22     -0.065    0.124    1.e-3
GW   30   3    0.191    -0.066    0.095     0.191    -0.066    0.155    1.e-3
GE    0
EK
EX    0    1    2    0   -0.22     -0.98
EX    0    2    2    0    0.973    -0.21
EX    0    3    2    0    6.e-17    1.006
EX    0    4    2    0   -0.97     -0.2
EX    0    5    2    0    0.217    -0.98
EX    0    6    2    0    0.691    -0.15
EX    0    7    2    0    0.146     0.691
EX    0    8    2    0   -0.71      9.e-17
EX    0    9    2    0    0.146    -0.69
EX    0   10    2    0    0.691     0.152
EX    0   11    2    0   -0.69      0.151
EX    0   12    2    0   -0.15     -0.69
EX    0   13    2    0    0.71      0
EX    0   14    2    0   -0.15      0.694
EX    0   15    2    0   -0.69     -0.15
EX    0   16    2    0    1.002    -0.1
EX    0   17    2    0    0.069     0.71
EX    0   18    2    0   -0.07     -0.71
EX    0   19    2    0    0.672     0.744
EX    0   20    2    0   -0.52      0.474
EX    0   21    2    0    0.529    -0.48
EX    0   22    2    0    3.e-16    1
EX    0   23    2    0   -0.71      3.e-16
EX    0   24    2    0    0.71      0
```

```
EX  0   25  2   0    -0.67   0.743
EX  0   26  2   0    -0.53   -0.48
EX  0   27  2   0    0.527   0.476
EX  0   28  2   0    -1  -0.1
EX  0   29  2   0    0.07    -0.71
EX  0   30  2   0    -0.07   0.707
GN  -1
FR  0   1   0   0    300 0
EN
```

Ten element, $l = 1$, $a = \lambda/2$, $\theta_0 = 45°$, $\varphi_0 = 0°$

```
CE
GW  1   3   0.475   0.185   0.125   0.475   0.125   0.125   1.e-3
GW  2   3   0.295   0.435   0.125   0.295   0.375   0.125   1.e-3
GW  3   3   0   0.53    0.125   0   0.47    0.125   1.e-3
GW  4   3   -0.295  0.435   0.125   -0.295  0.375   0.125   1.e-3
GW  5   3   -0.475  0.185   0.125   -0.475  0.125   0.125   1.e-3
GW  6   3   0.445   0.155   0.124   0.505   0.155   0.124   1.e-3
GW  7   3   0.265   0.405   0.124   0.325   0.405   0.124   1.e-3
GW  8   3   -0.03   0.5 0.124   0.03    0.5 0.124   1.e-3
GW  9   3   -0.325  0.405   0.124   -0.265  0.405   0.124   1.e-3
GW  10  3   -0.505  0.155   0.124   -0.445  0.155   0.124   1.e-3
GW  11  3   0.476   0.154   0.095   0.476   0.154   0.155   1.e-3
GW  12  3   0.296   0.406   0.095   0.296   0.406   0.155   1.e-3
GW  13  3   -1.e-3  0.501   0.095   -1.e-3  0.501   0.155   1.e-3
GW  14  3   -0.296  0.404   0.095   -0.296  0.404   0.155   1.e-3
GW  15  3   -0.474  0.154   0.095   -0.474  0.154   0.155   1.e-3
GW  16  3   -0.475  -0.125  0.125   -0.475  -0.185  0.125   1.e-3
GW  17  3   -0.505  -0.155  0.124   -0.445  -0.155  0.124   1.e-3
GW  18  3   -0.474  -0.156  0.095   -0.474  -0.156  0.155   1.e-3
GW  19  3   -0.295  -0.375  0.125   -0.295  -0.435  0.125   1.e-3
GW  20  3   -0.325  -0.405  0.124   -0.265  -0.405  0.124   1.e-3
GW  21  3   -0.294  -0.406  0.095   -0.294  -0.406  0.155   1.e-3
GW  22  3   0   -0.47   0.125   0   -0.53   0.125   1.e-3
GW  23  3   -0.03   -0.5    0.124   0.03    -0.5    0.124   1.e-3
GW  24  3   1.e-3   -0.501  0.095   1.e-3   -0.501  0.155   1.e-3
GW  25  3   0.29    -0.375  0.125   0.29    -0.435  0.125   1.e-3
GW  26  3   0.26    -0.405  0.124   0.32    -0.405  0.124   1.e-3
GW  27  3   0.291   -0.406  0.095   0.291   -0.406  0.155   1.e-3
GW  28  3   0.475   -0.125  0.125   0.475   -0.185  0.125   1.e-3
GW  29  3   0.445   -0.155  0.124   0.505   -0.155  0.124   1.e-3
GW  30  3   0.476   -0.156  0.095   0.476   -0.156  0.155   1.e-3
GE  0
```

```
EK
EX  0   1   2   0   -1   -0.09
EX  0   2   2   0   0.544     -0.83
EX  0   3   2   0   6.e-17  1.006
EX  0   4   2   0   -0.54     -0.83
EX  0   5   2   0   1.003   -0.09
EX  0   6   2   0   0.061   -0.71
EX  0   7   2   0   0.591   0.386
EX  0   8   2   0   -0.71   9.e-17
EX  0   9   2   0   0.591   -0.39
EX  0   10  2   0   0.061   0.705
EX  0   11  2   0   -0.06   0.7
EX  0   12  2   0   -0.59   -0.39
EX  0   13  2   0   -0.59   -0.39
EX  0   14  2   0   -0.59   0.388
EX  0   15  2   0   -0.06   -0.71
EX  0   16  2   0   0.393   0.927
EX  0   17  2   0   -0.66   0.277
EX  0   18  2   0   0.657   -0.28
EX  0   19  2   0   -0.05   1.002
EX  0   20  2   0   -0.71   -0.03
EX  0   21  2   0   0.712   0.035
EX  0   22  2   0   3.e-16  1
EX  0   23  2   0   -0.71   3.e-16
EX  0   24  2   0   0.71    0
EX  0   25  2   0   0.049   1
EX  0   26  2   0   -0.71   0.035
EX  0   27  2   0   0.709   -0.03
EX  0   28  2   0   -0.39   0.924
EX  0   29  2   0   -0.66   -0.28
EX  0   30  2   0   0.654   0.277
GN  -1
FR  0   1   0   0   300 0
EN
```

Ten element, $l = 2$, $a = \lambda/2$, $\theta_0 = 45°$, $\varphi_0 = 0°$

```
CE
GW  1   3   0.475   0.185   0.125   0.475   0.125   0.125   1.e-3
GW  2   3   0.295   0.435   0.125   0.295   0.375   0.125   1.e-3
GW  3   3   0   0.53    0.125   0   0.47    0.125   1.e-3
GW  4   3   -0.295  0.435   0.125   -0.295  0.375   0.125   1.e-3
GW  5   3   -0.475  0.185   0.125   -0.475  0.125   0.125   1.e-3
GW  6   3   0.445   0.155   0.124   0.505   0.155   0.124   1.e-3
```

```
GW   7   3   0.265    0.405    0.124    0.325    0.405    0.124    1.e-3
GW   8   3   -0.03    0.5 0.124    0.03    0.5 0.124    1.e-3
GW   9   3   -0.325   0.405    0.124    -0.265   0.405    0.124    1.e-3
GW  10   3   -0.505   0.155    0.124    -0.445   0.155    0.124    1.e-3
GW  11   3   0.476    0.154    0.095    0.476    0.154    0.155    1.e-3
GW  12   3   0.296    0.406    0.095    0.296    0.406    0.155    1.e-3
GW  13   3   -1.e-3   0.501    0.095    -1.e-3   0.501    0.155    1.e-3
GW  14   3   -0.296   0.404    0.095    -0.296   0.404    0.155    1.e-3
GW  15   3   -0.474   0.154    0.095    -0.474   0.154    0.155    1.e-3
GW  16   3   -0.475   -0.125   0.125    -0.475   -0.185   0.125    1.e-3
GW  17   3   -0.505   -0.155   0.124    -0.445   -0.155   0.124    1.e-3
GW  18   3   -0.474   -0.156   0.095    -0.474   -0.156   0.155    1.e-3
GW  19   3   -0.295   -0.375   0.125    -0.295   -0.435   0.125    1.e-3
GW  20   3   -0.325   -0.405   0.124    -0.265   -0.405   0.124    1.e-3
GW  21   3   -0.294   -0.406   0.095    -0.294   -0.406   0.155    1.e-3
GW  22   3   0    -0.47    0.125    0    -0.53    0.125    1.e-3
GW  23   3   -0.03    -0.5     0.124    0.03    -0.5     0.124    1.e-3
GW  24   3   1.e-3    -0.501   0.095    1.e-3    -0.501   0.155    1.e-3
GW  25   3   0.29     -0.375   0.125    0.29     -0.435   0.125    1.e-3
GW  26   3   0.26     -0.405   0.124    0.32     -0.405   0.124    1.e-3
GW  27   3   0.291    -0.406   0.095    0.291    -0.406   0.155    1.e-3
GW  28   3   0.475    -0.125   0.125    0.475    -0.185   0.125    1.e-3
GW  29   3   0.445    -0.155   0.124    0.505    -0.155   0.124    1.e-3
GW  30   3   0.476    -0.156   0.095    0.476    -0.156   0.155    1.e-3
GE   0
EK
EX   0   1   2   0   -1   -0.09
EX   0   2   2   0   0.544    -0.83
EX   0   3   2   0   6.e-17   1.006
EX   0   4   2   0   -0.54    -0.83
EX   0   5   2   0   1.003    -0.09
EX   0   6   2   0   0.061    -0.71
EX   0   7   2   0   0.591    0.386
EX   0   8   2   0   -0.71    9.e-17
EX   0   9   2   0   0.591    -0.39
EX   0  10   2   0   0.061    0.705
EX   0  11   2   0   -0.06    0.7
EX   0  12   2   0   -0.59    -0.39
EX   0  13   2   0   0.707    0
EX   0  14   2   0   -0.59    0.386
EX   0  15   2   0   -0.06    -0.71
EX   0  16   2   0   0.393    0.927
EX   0  17   2   0   -0.66    0.277
EX   0  18   2   0   0.657    -0.28
EX   0  19   2   0   -0.05    1.002
```

```
EX   0   20   2   0   -0.71    -0.02
EX   0   21   2   0   0.712    0.035
EX   0   22   2   0   6.e-17   1
EX   0   23   2   0   -0.71    9.e-17
EX   0   24   2   0   0.71     0
EX   0   25   2   0   0.049    1
EX   0   26   2   0   -0.71    0.035
EX   0   27   2   0   0.709    -0.03
EX   0   28   2   0   -0.39    0.924
EX   0   29   2   0   -0.66    -0.28
EX   0   30   2   0   0.654    0.277
GN  -1
FR   0   1    0   0   300 0
EN
```

Ten element, $l = 3$, $a = \lambda/2$, $\theta_0 = 45°$, $\varphi_0 = 0°$

```
CE
GW   1    3   0.475    0.185    0.125    0.475    0.125    0.125    1.e-3
GW   2    3   0.295    0.435    0.125    0.295    0.375    0.125    1.e-3
GW   3    3   0    0.53     0.125    0    0.47     0.125    1.e-3
GW   4    3   -0.295   0.435    0.125    -0.295   0.375    0.125    1.e-3
GW   5    3   -0.475   0.185    0.125    -0.475   0.125    0.125    1.e-3
GW   6    3   0.445    0.155    0.124    0.505    0.155    0.124    1.e-3
GW   7    3   0.265    0.405    0.124    0.325    0.405    0.124    1.e-3
GW   8    3   -0.03    0.5 0.124    0.03     0.5 0.124    1.e-3
GW   9    3   -0.325   0.405    0.124    -0.265   0.405    0.124    1.e-3
GW   10   3   -0.505   0.155    0.124    -0.445   0.155    0.124    1.e-3
GW   11   3   0.476    0.154    0.095    0.476    0.154    0.155    1.e-3
GW   12   3   0.296    0.406    0.095    0.296    0.406    0.155    1.e-3
GW   13   3   -1.e-3   0.501    0.095    -1.e-3   0.501    0.155    1.e-3
GW   14   3   -0.296   0.404    0.095    -0.296   0.404    0.155    1.e-3
GW   15   3   -0.474   0.154    0.095    -0.474   0.154    0.155    1.e-3
GW   16   3   -0.475   -0.125   0.125    -0.475   -0.185   0.125    1.e-3
GW   17   3   -0.505   -0.155   0.124    -0.445   -0.155   0.124    1.e-3
GW   18   3   -0.474   -0.156   0.095    -0.474   -0.156   0.155    1.e-3
GW   19   3   -0.295   -0.375   0.125    -0.295   -0.435   0.125    1.e-3
GW   20   3   -0.325   -0.405   0.124    -0.265   -0.405   0.124    1.e-3
GW   21   3   -0.294   -0.406   0.095    -0.294   -0.406   0.155    1.e-3
GW   22   3   0    -0.47    0.125    0    -0.53    0.125    1.e-3
GW   23   3   -0.03    -0.5     0.124    0.03     -0.5     0.124    1.e-3
GW   24   3   1.e-3    -0.501   0.095    1.e-3    -0.501   0.155    1.e-3
GW   25   3   0.29     -0.375   0.125    0.29     -0.435   0.125    1.e-3
GW   26   3   0.26     -0.405   0.124    0.32     -0.405   0.124    1.e-3
```

```
GW  27  3    0.291   -0.406  0.095   0.291   -0.406  0.155   1.e-3
GW  28  3    0.475   -0.125  0.125   0.475   -0.185  0.125   1.e-3
GW  29  3    0.445   -0.155  0.124   0.505   -0.155  0.124   1.e-3
GW  30  3    0.476   -0.156  0.095   0.476   -0.156  0.155   1.e-3
GE  0
EK
EX  0   1    2   0   -0.93   -0.39
EX  0   2    2   0   0.993   -0.05
EX  0   3    2   0   -1.01   1.e-16
EX  0   4    2   0   0.99    0.049
EX  0   5    2   0   -0.93   0.394
EX  0   6    2   0   0.276   -0.65
EX  0   7    2   0   0.035   0.705
EX  0   8    2   0   4.e-17  -0.71
EX  0   9    2   0   -0.03   0.705
EX  0   10   2   0   -0.28   -0.65
EX  0   11   2   0   -0.27   0.647
EX  0   12   2   0   -0.04   -0.71
EX  0   13   2   0   2.e-16  0.71
EX  0   14   2   0   0.035   -0.71
EX  0   15   2   0   0.277   0.652
EX  0   16   2   0   -0.09   -1
EX  0   17   2   0   0.71    -0.06
EX  0   18   2   0   -0.71   0.062
EX  0   19   2   0   0.839   -0.55
EX  0   20   2   0   0.387   0.592
EX  0   21   2   0   -0.39   -0.6
EX  0   22   2   0   1   0
EX  0   23   2   0   4.e-16  0.71
EX  0   24   2   0   8.e-16  -0.71
EX  0   25   2   0   0.838   0.548
EX  0   26   2   0   -0.39   0.595
EX  0   27   2   0   0.391   -0.6
EX  0   28   2   0   -0.09   0.998
EX  0   29   2   0   -0.72   -0.06
EX  0   30   2   0   0.707   0.062
GN  -1
FR  0   1    0   0   300 0
EN
```

Ten element, $l = 4$, $a = \lambda/2$, $\theta_0 = 15°$, $\varphi_0 = 0°$

```
CE
GW  1   3    0.475   0.185   0.125   0.475   0.125   0.125   1.e-3
```

```
GW    2    3     0.295    0.435    0.125    0.295    0.375    0.125    1.e-3
GW    3    3     0     0.53     0.125    0     0.47     0.125    1.e-3
GW    4    3    -0.295    0.435    0.125   -0.295    0.375    0.125    1.e-3
GW    5    3    -0.475    0.185    0.125   -0.475    0.125    0.125    1.e-3
GW    6    3     0.445    0.155    0.124    0.505    0.155    0.124    1.e-3
GW    7    3     0.265    0.405    0.124    0.325    0.405    0.124    1.e-3
GW    8    3    -0.03     0.5  0.124    0.03     0.5  0.124    1.e-3
GW    9    3    -0.325    0.405    0.124   -0.265    0.405    0.124    1.e-3
GW   10    3    -0.505    0.155    0.124   -0.445    0.155    0.124    1.e-3
GW   11    3     0.476    0.154    0.095    0.476    0.154    0.155    1.e-3
GW   12    3     0.296    0.406    0.095    0.296    0.406    0.155    1.e-3
GW   13    3    -1.e-3    0.501    0.095   -1.e-3    0.501    0.155    1.e-3
GW   14    3    -0.296    0.404    0.095   -0.296    0.404    0.155    1.e-3
GW   15    3    -0.474    0.154    0.095   -0.474    0.154    0.155    1.e-3
GW   16    3    -0.475   -0.125    0.125   -0.475   -0.185    0.125    1.e-3
GW   17    3    -0.505   -0.155    0.124   -0.445   -0.155    0.124    1.e-3
GW   18    3    -0.474   -0.156    0.095   -0.474   -0.156    0.155    1.e-3
GW   19    3    -0.295   -0.375    0.125   -0.295   -0.435    0.125    1.e-3
GW   20    3    -0.325   -0.405    0.124   -0.265   -0.405    0.124    1.e-3
GW   21    3    -0.294   -0.406    0.095   -0.294   -0.406    0.155    1.e-3
GW   22    3     0    -0.47     0.125    0    -0.53     0.125    1.e-3
GW   23    3    -0.03    -0.5     0.124    0.03    -0.5     0.124    1.e-3
GW   24    3     1.e-3   -0.501    0.095    1.e-3   -0.501    0.155    1.e-3
GW   25    3     0.29    -0.375    0.125    0.29    -0.435    0.125    1.e-3
GW   26    3     0.26    -0.405    0.124    0.32    -0.405    0.124    1.e-3
GW   27    3     0.291   -0.406    0.095    0.291   -0.406    0.155    1.e-3
GW   28    3     0.475   -0.125    0.125    0.475   -0.185    0.125    1.e-3
GW   29    3     0.445   -0.155    0.124    0.505   -0.155    0.124    1.e-3
GW   30    3     0.476   -0.156    0.095    0.476   -0.156    0.155    1.e-3
GE    0
EK
EX    0    1    2    0     0.468   -0.89
EX    0    2    2    0    -0.15     0.983
EX    0    3    2    0     3.e-16  -1.01
EX    0    4    2    0     0.148    0.98
EX    0    5    2    0    -0.47    -0.89
EX    0    6    2    0     0.856    0.449
EX    0    7    2    0    -0.96    -0.14
EX    0    8    2    0     0.97     2.e-16
EX    0    9    2    0    -0.96     0.145
EX    0   10    2    0     0.857   -0.45
EX    0   11    2    0    -0.23    -0.12
EX    0   12    2    0     0.257    0.039
EX    0   13    2    0    -0.26     0
EX    0   14    2    0     0.257   -0.04
```

```
EX  0   15  2   0   -0.23   0.12
EX  0   16  2   0   0.9 0.445
EX  0   17  2   0   -0.43   0.865
EX  0   18  2   0   0.116   -0.23
EX  0   19  2   0   -0.9    0.449
EX  0   20  2   0   -0.43   -0.86
EX  0   21  2   0   0.117   0.233
EX  0   22  2   0   0   -1
EX  0   23  2   0   0.966   0
EX  0   24  2   0   -0.26   1.e-16
EX  0   25  2   0   0.895   0.448
EX  0   26  2   0   -0.43   0.864
EX  0   27  2   0   0.117   -0.23
EX  0   28  2   0   -0.9    0.444
EX  0   29  2   0   -0.43   -0.86
EX  0   30  2   0   0.115   0.232
GN  -1
FR  0   1   0   0   300 0
EN
```

## B.5  The LOIS Test Station

### B.5.1  LOIS_L0.txt

```
CE
GW  1   11  6.75    4.5 2   8.872   4.5 3.502   1.e-3
GW  2   11  8.34236644  3.5820664   2.0002664   7.281   5.4188  3.5018  1.e-3
GW  3   11  8.3415  5.4188  1.9994  7.281   3.582   3.5018  1.e-3
GW  10  11  6.75    -4.5    2   8.872   -4.5    3.502   1.e-3
GW  11  11  8.34236644  -5.4179336  2.0002664   7.281   -3.5812 3.5018  1.e-3
GW  12  11  8.3415  -3.5812 1.9994  7.281   -5.418  3.5018  1.e-3
GW  13  11  -1.122  -9  2   1   -9  3.502   1.e-3
GW  14  11  0.47036644  -9.9179336  2.0002664   -0.591  -8.0812 3.5018  1.e-3
GW  15  11  0.4695  -8.0812 1.9994  -0.591  -9.918  3.5018  1.e-3
GW  16  11  -8.872  -4.5    2   -6.75   -4.5    3.502   1.e-3
GW  17  11  -7.2796336  -5.4179336  2.0002664   -8.341  -3.5812 3.5018  1.e-3
GW  18  11  -7.2805 -3.5812 1.9994  -8.341  -5.418  3.5018  1.e-3
GW  19  11  -8.872  4.5 2   -6.75   4.5 3.502   1.e-3
GW  20  11  -7.2796336  3.5820664   2.0002664   -8.341  5.4188  3.5018  1.e-3
GW  21  11  -7.2805 5.4188  1.9994  -8.341  3.582   3.5018  1.e-3
GW  22  11  -1.122  9   2   1   9   3.502   1.e-3
GW  23  11  0.47036644  8.0820664   2.0002664   -0.591  9.9188  3.5018  1.e-3
GW  24  11  0.4695  9.9188  1.9994  -0.591  8.082   3.5018  1.e-3
```

```
GE   0
EX   6   1    6    0    -2   2.e-16
EX   6   2    6    0    1    0
EX   6   3    6    0    1    0
EX   6   10   6    0    -2   2.e-16
EX   6   11   6    0    1    0
EX   6   12   6    0    1    0
EX   6   13   6    0    -2   2.e-16
EX   6   14   6    0    1    0
EX   6   15   6    0    1    0
EX   6   16   6    0    -2   2.e-16
EX   6   17   6    0    1    0
EX   6   18   6    0    1    0
EX   6   19   6    0    -2   2.e-16
EX   6   20   6    0    1    0
EX   6   21   6    0    1    0
EX   6   22   6    0    -2   2.e-16
EX   6   23   6    0    1    0
EX   6   24   6    0    1    0
GN   1
FR   0   1    0    0    9.99333333   0
EN
```

## B.5.2  LOIS_L1.txt

```
CE
GW   1    11   6.75        4.5 2      8.872       4.5 3.502    1.e-3
GW   2    11   8.34236644  3.5820664  2.0002664   7.281    5.4188  3.5018  1.e-3
GW   3    11   8.3415   5.4188   1.9994   7.281    3.582    3.5018  1.e-3
GW   10   11   6.75     -4.5     2    8.872    -4.5     3.502    1.e-3
GW   11   11   8.34236644  -5.4179336  2.0002664   7.281    -3.5812 3.5018  1.e-3
GW   12   11   8.3415   -3.5812 1.9994   7.281    -5.418  3.5018  1.e-3
GW   13   11   -1.122   -9   2    1    -9   3.502    1.e-3
GW   14   11   0.47036644  -9.9179336  2.0002664   -0.591   -8.0812 3.5018  1.e-3
GW   15   11   0.4695   -8.0812 1.9994   -0.591   -9.918  3.5018  1.e-3
GW   16   11   -8.872   -4.5     2    -6.75    -4.5     3.502    1.e-3
GW   17   11   -7.2796336  -5.4179336  2.0002664   -8.341   -3.5812 3.5018  1.e-3
GW   18   11   -7.2805  -3.5812 1.9994   -8.341   -5.418  3.5018  1.e-3
GW   19   11   -8.872   4.5 2    -6.75    4.5 3.502    1.e-3
GW   20   11   -7.2796336  3.5820664  2.0002664   -8.341   5.4188  3.5018  1.e-3
GW   21   11   -7.2805  5.4188   1.9994   -8.341   3.582    3.5018  1.e-3
GW   22   11   -1.122   9   2    1    9    3.502    1.e-3
GW   23   11   0.47036644  8.0820664  2.0002664   -0.591   9.9188  3.5018  1.e-3
GW   24   11   0.4695   9.9188   1.9994   -0.591   8.082    3.5018  1.e-3
```

```
GE   0
EX   6   1    6    0    -1.73     -1
EX   6   2    6    0    0.866     0.5
EX   6   3    6    0    0.869     0.502
EX   6   10   6    0    -1.73     0.997
EX   6   11   6    0    0.866     -0.5
EX   6   12   6    0    0.869     -0.5
EX   6   13   6    0    1.e-16    2
EX   6   14   6    0    6.e-17    -1
EX   6   15   6    0    6.e-17    -1
EX   6   16   6    0    1.732     1
EX   6   17   6    0    -0.87     -0.5
EX   6   18   6    0    -0.87     -0.5
EX   6   19   6    0    1.734     -1
EX   6   20   6    0    -0.87     0.5
EX   6   21   6    0    -0.87     0.5
EX   6   22   6    0    1.e-16    -2
EX   6   23   6    0    6.e-17    1
EX   6   24   6    0    6.e-17    1
GN   1
FR   0   1    0    0    9.99333333   0
EN
```